\documentclass[rmp,aps,twocolumn,showpacs]{revtex4-1}
\usepackage[utf8]{inputenc}
\usepackage{amsmath,graphicx,float,latexsym,color}
\usepackage[normalem]{ulem}
\usepackage{subfigure}
\newcommand{\be}{\begin{equation}}
\newcommand{\ee}{\end{equation}}
\newcommand{\ba}{\begin{eqnarray}}
\newcommand{\ea}{\end{eqnarray}}
\newcommand{\figref}[1]{Fig.~\ref{#1}}
\newcommand{\tblref}[1]{Table~\ref{#1}}
\newcommand{\sectionref}[1]{Sec.~\ref{#1}}
\newcommand{\eqnref}[1]{Eq.~\eqref{#1}}
\begin{document}
\title{Experimental review of graphene}
\author{Daniel R. Cooper}
\thanks{Equal contributions from all authors}
\author{Benjamin D'Anjou}
\author{Nageswara Ghattamaneni}
\author{Benjamin Harack}
\author{Michael Hilke}
\author{Alexandre Horth}
\author{Norberto Majlis}
\author{Mathieu Massicotte}
\author{Leron Vandsburger}
\author{Eric Whiteway}
\author{Victor Yu}
\affiliation{McGill University, Montr\'eal, Canada, H3A 2T8}
\date{\today}
\begin{abstract}

This review examines the properties of graphene from an experimental perspective. The intent is to review the most important experimental results at a level of detail appropriate for new graduate students who are interested in a general overview of the fascinating properties of graphene. While some introductory theoretical concepts are provided, including a discussion of the electronic band structure and phonon dispersion, the main emphasis is on describing relevant experiments and important results as well as some of the novel applications of graphene. In particular, this review covers graphene synthesis and characterization, field-effect behavior, electronic transport properties, magneto-transport, integer and fractional quantum Hall effects, mechanical properties, transistors, optoelectronics, graphene-based sensors, and biosensors. This approach attempts to highlight both the means by which the current understanding of graphene has come about and some tools for future contributions. 
\end{abstract}
\maketitle
\tableofcontents

\section{Introduction}
Graphene is a single two-dimensional layer of carbon atoms bound in a hexagonal lattice structure. It has been extensively studied in the last several years even though it was only isolated for the first time in 2004 \cite{nov04}. Andre Geim and Konstantin Novoselov won the 2010 Nobel Prize in Physics for their groundbreaking work on graphene. The fast uptake of interest in graphene is due primarily to a number of exceptional properties that it has been found to possess.

There have been several reviews discussing the topic of graphene in recent years. Many are theoretically oriented, with Castro Neto \textit{et al.}'s review of the electronic properties as a prominent example \cite{CastroNeto2009} and a more focused review of the electronic transport properties \cite{Sarma2010}. Experimental reviews, to name only a few, include detailed discussions of synthesis \cite{choi10} and Raman characterization methods \cite{ni08}, of transport mechanisms \cite{Avouris10, Giannazzo11}, of relevant applications of graphene such as transistors and the related bandgap engineering \cite{Schwierz:2010ix}, and of graphene optoelectronic technologies \cite{Bonaccorso10}. We feel, however, that the literature is lacking a comprehensive overview of all major recent experimental results related to graphene and its applications. It is with the intent to produce such a document that we wrote this review. We gathered a great number of results from what we believe to be the most relevant fields in current graphene research in order to give a starting point to readers interested in expanding their knowledge on the topic. The review should be particularly well-suited to graduate students who desire an introduction to the study of graphene that will provide them with many references for further reading.

We have attempted to deliver an up-to-date account of most topics. For example, we present recent results on the fractional quantum Hall effect and some of the newest developments and device details in optoelectronics.  In addition, we include a summary of the work that has been done in the field of graphene biosensors.  As a practical tool, we also give comparative analyses of graphene substrate properties, of available bandgap engineering techniques and of photovoltaic devices in order to provide the researcher with useful and summarized laboratory references.

This review is structured as follows: We start by reviewing the electronic band structure and its associated properties in \sectionref{sec:BandStructure} before introducing the vibrational properties, including the phonon dispersion, in \sectionref{VibrationalProperties}. We then detail the various synthesis methods for graphene monolayers and their corresponding characterization. The main properties are then discussed, starting with the electric field effect, which originally spurred the intense activity in graphene research. This is then followed by a review of the magneto-transport properties, which includes the quantum Hall effect and recent results on the fractional quantum Hall effect. We conclude the review of the main properties with a discussion of the mechanical properties of graphene, before turning to some important applications. The transistor applications are discussed first, with particular emphasis on band gap engineering, before reviewing optoelectronic applications. We finish the applications section with a discussion of graphene-based sensors and biosensors.


\section{Electronic Structure}
\label{sec:BandStructure}


Graphene has a remarkable band structure thanks to its crystal structure. Carbon atoms form a hexagonal lattice on a two-dimensional plane. Each carbon atom is about $a=$ 1.42 \AA{} from its three neighbors, with each of which it shares one $\sigma$ bond. The fourth bond is a $\pi$-bond, which is oriented in the z-direction (out of the plane). One can visualize the $\pi$ orbital as a pair of symmetric lobes oriented along the z-axis and centered on the nucleus. Each atom has one of these $\pi$-bonds, which are then hybridized together to form what are referred to as the $\pi$-band and $\pi^*$-bands. These bands are responsible for most of the peculiar electronic properties of graphene.

\begin{figure}[htbp]
  	\centering \includegraphics[width=0.40\textwidth]{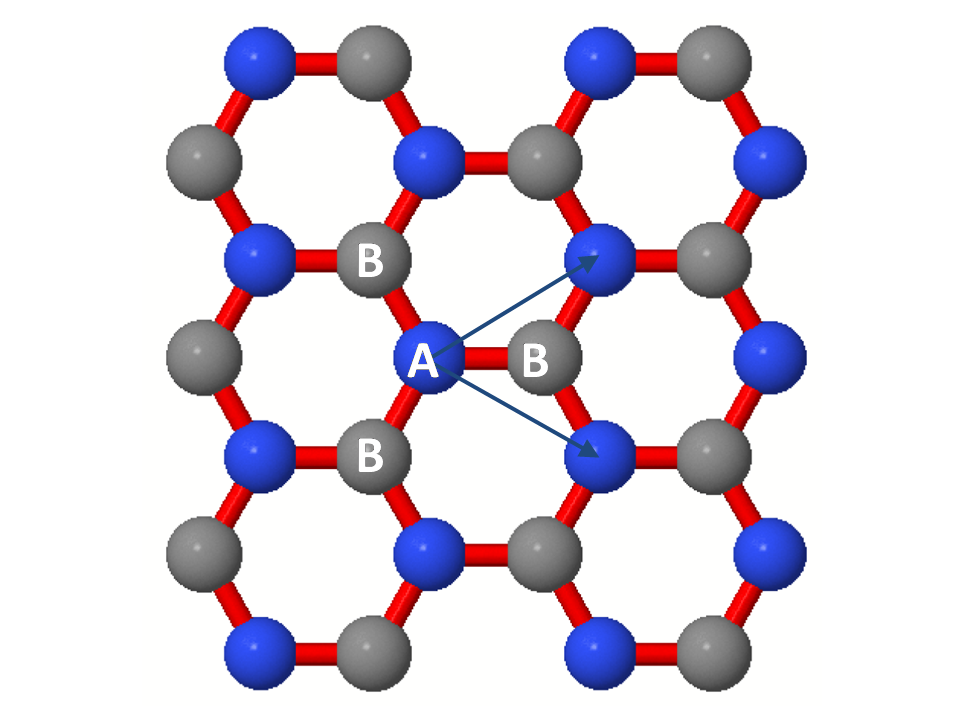}
  	\caption{\textbf{Triangular sublattices of graphene.} Each atom in one sublattice (A) has 3 nearest neighbors in sublattice (B) and vice-versa.}
  	\label{fig:GrapheneSubLattices}
\end{figure}

The hexagonal lattice of graphene can be regarded as two interleaving triangular lattices. This is illustrated in \figref{fig:GrapheneSubLattices}. This perspective was successfully used as far back as 1947 when Wallace calculated the band structure for a single graphite layer using a tight-binding approximation \cite{Wallace1947}.

\begin{figure}[htbp]
  	\centering
      	\includegraphics[width=0.5\textwidth]{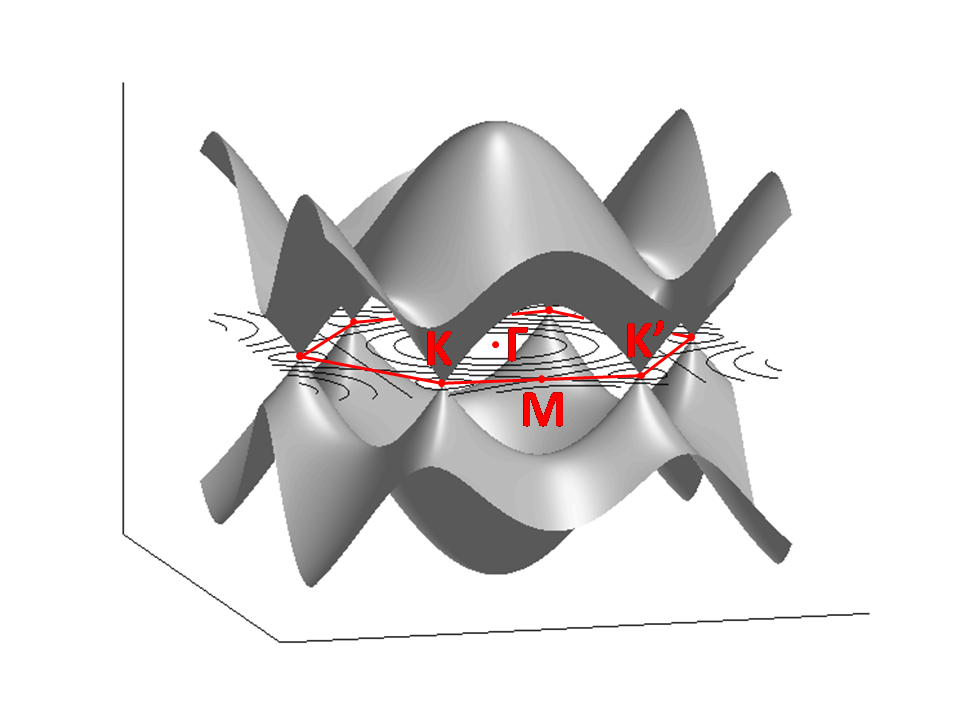}
  	\caption{\textbf{First Brillouin zone and band structure of graphene.} The vertical axis is energy, while the horizontal axes are momentum space on the graphene lattice. The first Brillouin zone of graphene is illustrated in the horizontal plane and labeled with some points of interest. K and K' are the two non-equivalent corners of the zone, and M is the midpoint between adjacent K and K' points. $\Gamma$ is the zone center. K and K' are also known as the Dirac points. The Dirac points are the transition between the valence band and the conduction band.}
  	\label{fig:BandStructure}
\end{figure}

Band structure is most often studied from a standpoint of the relationship between the energy and momentum of electrons within a given material. Since graphene constrains the motion of electrons to two dimensions, our momentum space is also constrained to two dimensions. A plot of the energy versus momentum dispersion relation for graphene can be found in \figref{fig:BandStructure} using the tight binding approximation discussed below.

Graphene is a zero-gap semiconductor because the conduction and valence bands meet at the Dirac points (see \figref{fig:BandStructure}). The Dirac points are locations in momentum space, on the edge of the Brillouin zone. There are two sets of three Dirac points. Each set is not equivalent with the other set of three. The two sets are labeled K and K'. The two sets of Dirac points give graphene a valley degeneracy of $g_v = 2$. The K and K' points are the primary points of interest when studying the electronic properties of graphene. This is noteworthy in comparison to traditional semiconductors where the primary point of interest is generally $\Gamma$, where momentum is zero.

Following the original work \cite{Wallace1947}, the tight binding Hamiltonian, written in a basis of sublattice positions A and B, is given by:

\begin{align}
\mathcal{H}_{\vec{k}} = \left( \begin{array}{cc}
\epsilon_A & te^{i\vec{k}\cdot\vec{a_1}}+te^{i\vec{k}\cdot\vec{a_2}}+te^{i\vec{k}\cdot\vec{a_3}} \\
c.c. & \epsilon_B \end{array} \right)
\label{Hk}
\end{align}

where $\epsilon_A = \epsilon_B = 0$ are the on-site energies of the carbon atoms on sites A and B, $t\simeq$ 2.7 eV is the next nearest hopping element (between A and B sites), $\vec{a_1} = a(1,0)$, $\vec{a_2} = a(-1/2,\sin[\pi/3])$, $\vec{a_3} = a(-1/2,-\sin[\pi/3])$ are the positions of the three nearest neighbors, and $c.c.$ the complex conjugate of the off-diagonal matrix element. The eigenvalues of this tight binding Hamiltonian are shown in \figref{fig:BandStructure} as a function of $\vec{k} = (k_x,k_y)$.

\subsection{Massless Dirac fermions}

Looking closely at the region near one of the Dirac points (K or K') in \figref{fig:BandStructure}, the cone-like linear dispersion relation is evident. The Fermi energy for neutral (or ideal) graphene is at the Dirac energy, which is the energy of the Dirac point. In graphene devices, the Fermi energy can be significantly different from the Dirac energy. 

Electrons within about 1 eV of the Dirac energy have a linear dispersion relation. The linear dispersion region is well-described by the Dirac equation for massless fermions. That is, the effective mass of the charge carriers in this region is zero. The dispersion relation near the K points is generally expressed as follows:

\begin{align}
E_{\pm} (k) \approx \pm \hbar v_F | k - K|
\end{align}

which corresponds to the spectrum of the Dirac-like Hamiltonian for low-energy massless Dirac fermions (again in the sublattice basis \{A,B\}):

\begin{align}
\mathcal{H}_K = \hbar v_F \left( \begin{array}{cc}
0 & k_x - ik_y \\
k_x+ik_y & 0 \end{array} \right) = \hbar v_F \vec{\sigma} \cdot \vec{ k}
\label{HK}
\end{align}

This Dirac Hamiltonian is simply the tight binding Hamiltonian from \eqnref{Hk} expanded close to K with $\hbar v_F = 3ta/2$. Close to K' the Hamiltonian becomes $\mathcal{H}_{K'} = \hbar v_F \vec{\sigma}^* \cdot \vec{ k} $, where $\vec{\sigma} = (\sigma_x, \sigma_y)$ is the 2D vector of the Pauli matrices (and $^*$ denotes the complex conjugate), $\vec{k}$ is the wavevector, and the Fermi velocity is $v_F \approx 10^6$ m/s, or 1/300th the speed of light in vacuum. Charge carriers in graphene behave like relativistic particles with an effective speed of light given by the Fermi velocity. This behavior is one of the most intriguing aspects about graphene, and is responsible for much of the research attention that graphene has received.

\subsection{Chirality}
Transport in graphene exhibits a novel chirality which we will now briefly describe. Each graphene sublattice can be regarded as being responsible for one branch of the dispersion. These dispersion branches interact very weakly with one another.

This chiral effect indicates the existence of a pseudospin quantum number for the charge carriers. This quantum number is analogous to spin but is completely independent of the `real' spin. The pseudospin lets us differentiate between contributions from each of the sublattices. This independence is called chirality because of the inability to transform one type of dispersion into another \cite{Sarma2010}. A typical example of chirality is that you cannot transform a right hand into a left hand with only translations, scalings, and rotations. The chirality of graphene can also be understood in terms of the Pauli matrix contributions in the Dirac-like Hamiltonian described in the previous section.

\subsection{Klein paradox}
A peculiar property of the Dirac Hamiltonian is that charge carriers cannot be confined by electrostatic potentials. In traditional semiconductors, if an electron strikes an electrostatic barrier that has a height above the electron's kinetic energy, the electron wavefunction will become evanescent within the barrier and exponentially decay with distance into the barrier. This means that the taller and wider a barrier is, the more the electron wavefunction will decay before reaching the other side. Thus, the taller and wider the barrier is, the lower the probability of the electron quantum tunneling through the barrier.

However, if the particles are governed by the Dirac equation, their transmission probability actually increases with increasing barrier height. A Dirac electron that hits a tall barrier will turn into a hole, and propagate through the barrier until it reaches the other side, where it will turn back into an electron. This phenomenon is called Klein tunneling.

An explanation for this phenomenon is that increasing barrier height leads to an increased degree of mode-matching between the wavefunctions of the holes within the barrier and the electrons outside of it. When the modes are perfectly matched (in the case of an infinitely tall barrier), we have perfect transmission through the barrier. In the case of graphene, the chirality discussed earlier leads to a varying transmission probability depending on the angle of incidence to the barrier \cite{Katsnelson2006}.

Some experimental results have been interpreted as evidence for Klein tunneling. Klein tunneling has been observed through electrostatic barriers, which were created by gate voltages \cite{Stander2009}. Similar effects have also been observed in narrow graphene resonant heterostructures \cite{Young2009}.

\subsection{Graphene vs traditional materials}
Here we summarize some of the interesting properties of graphene by comparing them with more traditional materials such as 2D semiconductors.

\begin{enumerate}
\item Traditional semiconductors have a finite band gap while graphene has a nominal gap of zero. Normally, the study of electron and hole motion through a semiconductor must be done with differently doped materials. However, in graphene the nature of a charge carrier changes at the Dirac point from an electron to a hole or vice-versa. On a related note, the Fermi level in graphene is always within the conduction or valence band while in traditional semiconductors the Fermi level often falls within the band gap when pinned by impurity states.
\item Dispersion in graphene is chiral. This is related to some very distinctive material behaviors like Klein tunneling.
\item Graphene has a linear dispersion relation while semiconductors tend to have quadratic dispersion. Many of the impressive physical and electronic properties of graphene can be considered to be consequences of this fact.
\item Graphene is much thinner than a traditional 2D electron gas (2DEG). A traditional 2DEG in a quantum well or heterostructure tends to have an effective thickness around 5-50 nm. This is due to the constraints on construction and the fact that the confined electron wavefunctions have an evanescent tail that stretches into the barriers. Graphene on the other hand is only a single layer of carbon atoms, generally regarded to have a thickness of about 3 \AA{}  (twice the carbon-carbon bond length). Electrons conducting through graphene are constrained in the z-axis to a much greater extent than those that conduct through a traditional 2DEG.
\item Graphene has been found to have a finite minimum conductivity, even in the case of vanishing charge carriers \cite{nov05, Tan2007}. This is an issue for the construction of field-effect transistors (FETs), as we will see in more detail in section \ref{ElectronicProperties}, since it contributes to relatively low on/off ratios for graphene-based transistors.
\end{enumerate}

Readers seeking further reading about distinctly electronic properties of graphene should refer to \sectionref{ElectronicProperties}. The next section will introduce and discuss the vibrational properties of graphene.


\section{Vibrational properties}
\label{VibrationalProperties}

While the electronic properties have attracted the lion's share of the interest in graphene, the vibrational properties are of great importance too. They are responsible for several fascinating properties such as record thermal conductivities. Since graphene is composed of a light atom, where the in-plane bonding is very strong, graphene exhibits a very high sound velocity. This large sound velocity is responsible for the very high thermal conductivity of graphene that is useful in many applications. Moreover, vibrational properties are instrumental in understanding other graphene attributes, including optical properties via phonon-photon scattering (e.g. in Raman scattering) and electronic properties via electron-phonon scattering.

\subsection{Phonon dispersion}

\begin{figure}[h!]
    \centering
        \includegraphics[width=0.5\textwidth]{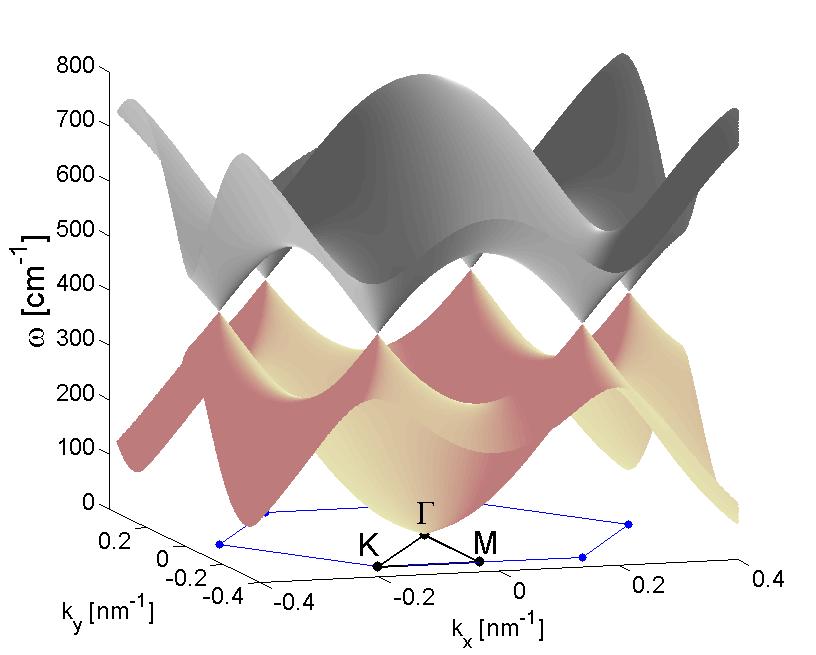}
    \caption{\textbf{First Brillouin zone and out-of-plane phonon modes.} The phonon dispersion relation of the ZA and ZO modes as a function of the in-plane reciprocal vector $\vec{k}$. The vertical axis is the phonon frequency, while the horizontal axes are momentum space on the graphene lattice. The dispersion relation is obtained using the second-nearest-neighbor model for graphene \cite{Falkovsky2007}. The gray surface corresponds to the ZO (optical) mode, whereas the pink surface shows the ZA (acoustic) mode. Also shown are the corresponding K, $\Gamma$, and M points of the Brillouin zone.}
    \label{PhononBS1}
\end{figure}

Most of the vibrational properties of graphene can be understood with the help of the phonon dispersion relation. Interestingly, the phonon dispersion has some similarity with the electronic band structure discussed in the previous section, which stems from the identical honeycomb structure (the out-of-plane modes are shown in \figref{PhononBS1}). In order to obtain the phonon dispersion it is necessary to consider the vibrational modes of the crystal in thermal equilibrium. This is done by considering the displacement of each atom from its equilibrium position, written $\vec{u}_n$ for the atom labeled $n$. Each atom is effectively coupled to its neighbors by some torsional and longitudinal force constants, which only depend on the relative positions of the atoms. This allows one to write the Newtonian coupled equation of motion in frequency space as:

\begin{align}
-\sum_m \Phi_{m,n}\vec{u}_n=\omega^2\vec{u}_n,
\end{align}

where in graphene the sum over $m$ is typically over the second or fourth nearest neighbors, with the corresponding coefficients $\Phi_{m,n}$, also known as the dynamical matrix.

In graphene, and similarly  to the electronic structure, the two sublattices A and B have to be considered explicitly to solve for the eigenspectrum of the dynamical matrix. However, the atoms can vibrate in all three dimensions, hence the dynamical matrix has to be written in terms of both the sublattices A and B as well as the 3 spatial dimensions. This leads to a dynamical matrix in reciprocal space $\Phi_{m,n}(\vec{k})$, which is given by a $6\times 6$ matrix when assuming $\vec{u}_n^{A,B}\sim e^{i\vec{k}\cdot\vec{R}^{A,B}_n}$. Here one applies Bloch's theorem, where $\vec{R}^{A,B}_n$ is the equilibrium position of atom $n$ in the sublattice A and B, respectively. Two of the eigenvalues correspond to the out-of-plane vibrations, ZA (acoustic) and ZO (optical), and the remaining 4 correspond to the in-plane vibrations: TA (transverse acoustic), TO (transverse optical), LA (longitudinal acoustic) and LO (longitudinal optical).

\begin{figure}[h!]
    \centering
        \includegraphics[width=0.45\textwidth]{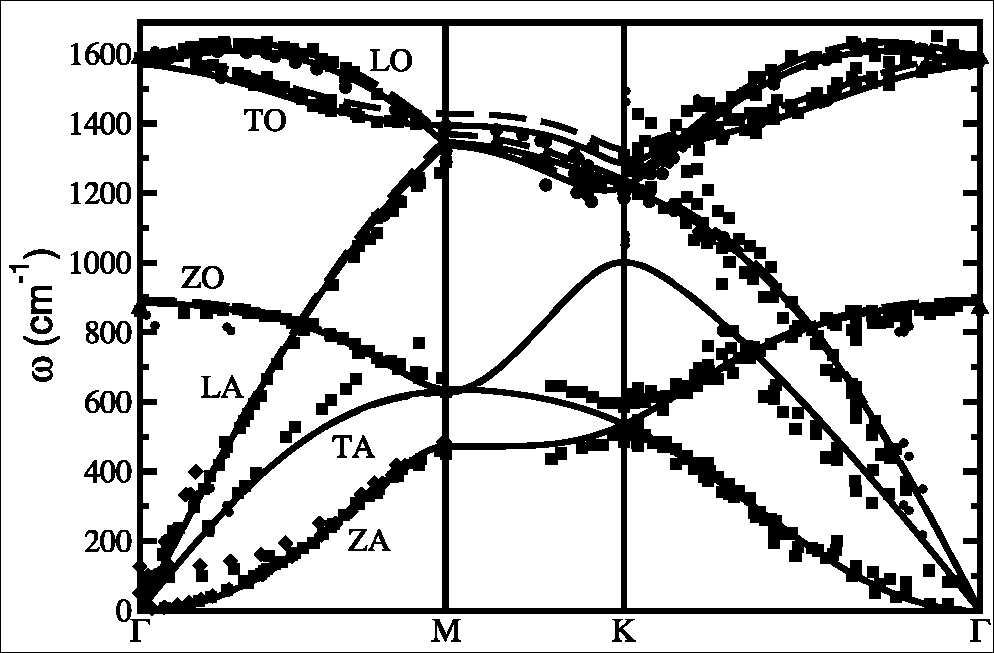}
    \caption{\textbf{Experimental phonon dispersion relation in graphite.} All the phonon modes are shown, including the in-plane and out-of-plane modes. The data (full symbols) is shown along the special symmetry points $\Gamma$, K, and M along with results from ab initio calculations (lines). Figure adapted from \cite{Wirtz2004}. }
    \label{PhononBS3}
\end{figure}

The ZA and ZO modes are often assumed to be decoupled from the in-plane modes \cite{Falkovsky2007}, which leads to a simple dispersion relation very similar to the electronic band structure as shown in \figref{PhononBS1}. It is interesting to note that the dispersion is quadratic at the $\Gamma$ point, which is unusual for acoustic modes. In contrast, a simple graphene phonon model based on atomic potentials containing only three parameters leads to a linear dispersion for the ZA mode \cite{adamyan}. However, the experimental data seems to be more consistent with a quadratic dispersion (see \figref{PhononBS3}) \cite{popov,dressel}. At the K and K' points we recover a cone structure similar to the Dirac cones in the electronic structure. However, the phonon density of states does not vanish at these points because of the presence of the in-plane modes.

\begin{figure}[h!]
    \centering
        \includegraphics[width=0.5\textwidth]{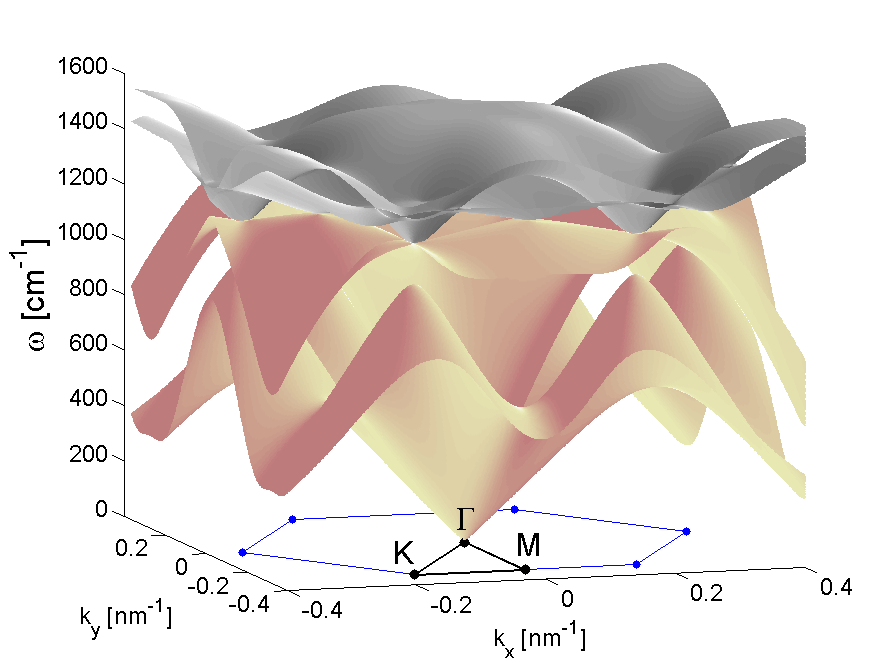}
    \caption{\textbf{First Brillouin zone and in-plane phonon modes.} The phonon dispersion relation of the TO and LO modes in gray and the TA and LA modes in pink as a function of the in-plane reciprocal vector $\vec{k}$. The longitudinal modes are on top of the transverse modes. The vertical axis is the phonon frequency, while the horizontal axes are the momentum space on the graphene lattice.}
    \label{PhononBS2}
\end{figure}

The in-plane modes are constituted by two acoustic modes and two optical modes. These modes can be obtained from the reduced in-plane dynamical matrix, which can be described by a $4\times4$ matrix, assuming no coupling to the out-of-plane modes. Using the parameters given by reference \cite{Falkovsky2007}, the in-plane modes are shown in \figref{PhononBS2}.

The in-plane modes show the expected linear dispersion at the $\Gamma$ point. The transverse modes closely follow the longitudinal modes but with a slightly lower frequency, for both the acoustic and optical modes. For comparison with experiments, it is more instructive to show the dispersion relation following surface cuts along the lines $\Gamma$ to M, M to K and K to $\Gamma$ in the Brillouin zone. An extensive collection of experimental data for graphite is shown in \figref{PhononBS3} along with ab initio calculations. The data was obtained using several techniques, including neutron scattering, electron energy loss spectroscopy, X-ray scattering, infrared absorption and double resonant Raman scattering experiments \cite{Wirtz2004}. There is no comparably extensive data on graphene yet, but it is expected to be very similar to graphite, since the coupling between planes is very weak.

\begin{figure}[h!]
    \centering
        \includegraphics[width=0.5\textwidth]{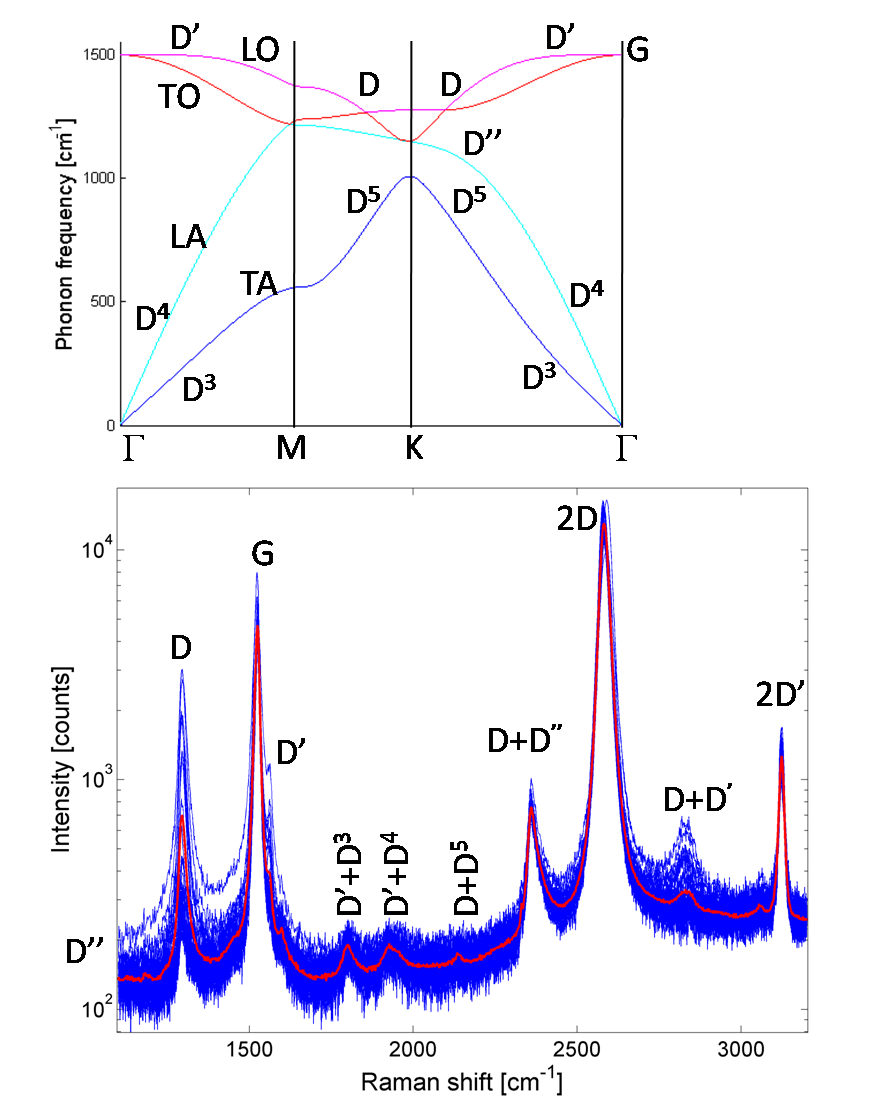}
    \caption{\textbf{Theoretical phonon dispersion of graphene and Raman spectrum of large scale graphene.} The labels (G, D, 2D, etc) identify the peaks from the Raman spectrum with the corresponding phonon energies \cite{Mauri2011,Rao2011}. Only the in-plane modes are shown since they are the only ones which are Raman active. Several Raman data sets of graphene $^{13}$C are shown in blue taken from different spots of the same sample at a laser wavelength of 514nm. The red line corresponds to the average over these data sets. The phonon dispersion is calculated using the mass of $^{13}$C.
    }
    \label{PhononBS4}
\end{figure}

Most of the data on graphene stems from Raman scattering, which allows for the determination of the phonon spectrum close to special symmetry points. This is illustrated in \figref{PhononBS4}, where the theoretical phonon dispersion with parameters from \cite{Falkovsky2007} is shown with the labels corresponding to the various Raman peaks. The Raman peaks determine the phonon energies for some values of the momentum. The Raman data was obtained from chemical vapor deposition (CVD)  $^{13}$C grown graphene. More conventional $^{12}$C graphene is very similar except for a rescaling of the Raman peaks due to the change of mass. Other aspects of the Raman spectrum are discussed in \sectionref{Characterization}.

An important consequence of the phonon dispersion relation in graphene is the very high value of the in-plane sound velocity, close to $c_{ph}\simeq 20$ km/s \cite{adamyan}, which leads to very high thermal conductivities.

\subsection{Thermal conductivity}

From the kinetic theory of gases, the thermal conductivity due to phonons is given by $\kappa\sim c_{ph} ~ C_{V}(T) ~ \lambda$, where $C_{V}(T)$ is the specific heat per unit volume and $\lambda$ is the phonon mean free path. This implies that since $c_{ph}$ is very large in graphene, one can expect a large thermal conductivity. Indeed, experiments at near room temperature obtain $\kappa\simeq$ 3080-5150 W/mK  and a phonon mean free path of $\lambda\simeq 775$ nm for a set of graphene flakes \cite{ghosh,balandin}.\\

These results indicate that graphene is a good candidate for applications to electronic devices, since a high thermal conductivity facilitates the diffusion of heat to the contacts and allows for more compact circuits. Phonons also play an important role in electronic transport via electron-phonon scattering, which is discussed in \sectionref{ElectronicProperties}. The mechanical properties of graphene are discussed in \sectionref{MechanicalProperties}, whereas the synthesis of graphene is treated in the next section.

\section{Synthesis}
\label{Synthesis}
Since graphene was isolated in 2004 by Geim and Novoselov using the now famous Scotch tape method, there have been many processes developed to produce few-to-single layer graphene. One of the primary concerns in graphene synthesis is producing samples with high carrier mobility and low density of defects. To date there is no method that can match mechanical exfoliation for producing high-quality, high-mobility graphene flakes. However, mechanical exfoliation is a time consuming process limited to small scale production. There is great interest in producing large scale graphene suitable for applications in flexible transparent electronics, transistors, etc. Some concerns in producing large scale graphene are the quality and consistency between samples as well as the cost and difficulty involved in the method.

\tblref{methods} shows a summary of some of the most important synthesis methods. The typical number of graphene layers produced as well as currently achievable dimensions are given. For comparison the mobilities listed are for graphene transferred to Si/SiO$_2$ wafers, since the electron mobility of graphene is heavily substrate dependent, as discussed in \sectionref{ElectronicProperties}.

\begin{table}[H]
\centering
\begin{tabular}{ l  l  l  c }
\hline Method & Layers & Size & Mobility (cm$^2$V$^{-1}$s$^{-1}$)\\
\hline
\hline Exfoliation & 1 to 10+ & 1 mm$^{a}$ & 15000$^{b}$\\
\hline Thermal SiC & 1 to 4 & 50 $\mu$m$^{c}$ & 2000$^{c}$ \\
\hline Ni-CVD & 1 to 4 & 1 cm$^{d}$ & 3700$^{d}$ \\
\hline Cu-CVD & 1 & 65 cm$^{e}$ & 16000$^{f}$ \\
\hline
\end{tabular}
\caption{\textbf{Comparison of graphene synthesis methods.} Shows typical number of layers produced, size of graphene layers (largest dimension) and mobility on Si/SiO$_2$. a:\cite{geim09}, b:\cite{nov05}, c:\cite{emts09}, d:\cite{Kim09}, e:\cite{bae10}, f:\cite{li10}.}
\label{methods}
\end{table}

\subsection{Mechanical exfoliation}
Developed by Geim and Novoselov, the exfoliation process uses HOPG (highly oriented pyrolitic graphite) as a precursor. The HOPG was subjected to an oxygen plasma etching to create 5 $\mu$m deep mesas and these mesas were then pressed into a layer of photoresist. The photoresist was baked and the HOPG was cleaved from the resist. Scotch tape was used to repeatedly peel flakes of graphite from the mesas. These thin flakes were then released in acetone and captured on the surface of a Si/SiO$_2$ wafer \cite{nov04}.

\begin{figure}[htbp]
\centering
\includegraphics[scale=0.3]{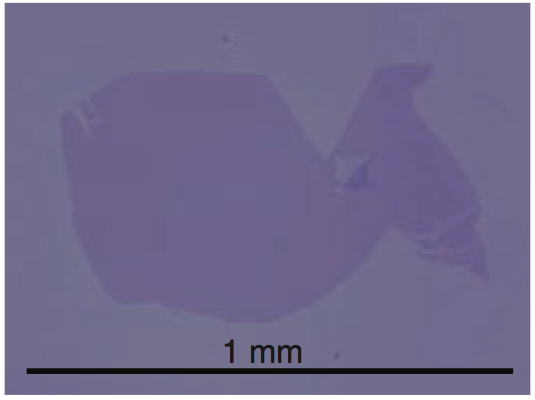}
\caption{\textbf{Monolayer graphene produced by mechanical exfoliation.} Large sample with length of 1 mm on Si/SiO$_2$.}
\label{1mmg}
\end{figure}

These few-layer graphene (FLG) flakes were identified using the contrast difference in an optical microscope and single layers using an SEM. Using this technique Geim and Novoselov were able to generate few- and single-layer graphene flakes with dimensions of up to 10 $\mu$m \cite{nov04}. The few-layer graphene flakes were found to have ballistic transport at room temperature and mobilities as high as 15000 cm$^2$V$^{-1}$s$^{-1}$ on Si/SiO$_2$ wafers \cite{nov05}. The scotch tape method can generate flakes with sides of up to 1 mm in length \cite{geim09}, of excellent quality and well suited for fundamental research. However, the process is limited to small sizes and cannot be scaled for industrial production.


\subsection{Thermal decomposition of SiC}
The thermal decomposition of silicon carbide is a technique that consists of heating SiC in ultra-high vacuum (UHV) to temperatures between 1000$^{\circ}$C and 1500$^{\circ}$C. This causes Si to sublimate from the material and leave behind a carbon rich surface. Low-energy electron microscopy (LEEM) studies indicate that this carbon layer is graphitic in nature, which suggests that the technique could be used to form graphene \cite{hass08}.

Berger and De Heer produced few-layer graphene by thermal decomposition of SiC. The Si face of a 6H-SiC single crystal was first prepared by oxidation or H$_2$ etching in order to improve surface quality. The sample was then heated by electron bombardment in UHV to 1000$^{\circ}$C to remove the oxide layer. Once the oxide was removed the samples were heated to 1250-1450$^{\circ}$C, resulting in the formation of thin graphitic layers. Typically between 1 and 3 layers were formed depending on the decomposition temperature. Using this method, devices were produced with mobilities of 1100 cm$^2$V$^{-1}$s$^{-1}$ \cite{berg04}.

This technique is capable of generating wafer-scale graphene layers and is potentially of interest to the semiconductor industry. Several issues still remain, notably controlling the number of layers produced, repeatability of large area growths and interface effects with the SiC substrate \cite{choi10}.

\begin{figure}[htbp]
\centering
\includegraphics[width=\columnwidth]{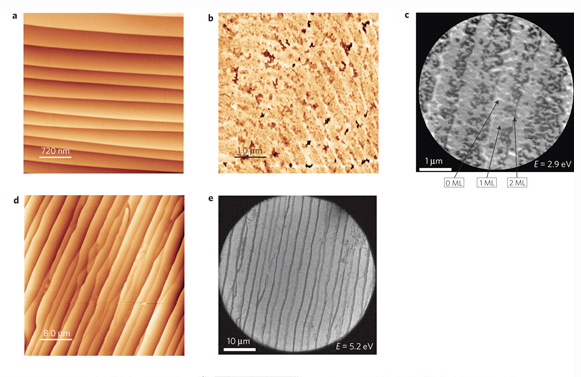}
\caption{\textbf{Graphene produced by thermal decomposition of SiC.} (a) AFM image of graphene growth on SiC annealed at UHV. (b) LEEM image of UHV grown graphene film. (c) AFM image of graphene annealed in Ar at 900 mbar. (d) LEEM image of graphene on Ar annealed SiC substrate showing terraces up to 50 $\mu$m in length \cite{emts09}. }
\label{sic}
\end{figure}

Emtsev \textit{et al.} found that by heating SiC in Ar at 900 mbar as opposed to UHV they were able to reduce surface roughness and produce much larger continuous graphene layers, up to 50 $\mu$m in length. The graphene on SiC was characterized using atomic force microscopy (AFM) and LEEM, as shown in \figref{sic}. They measured electron mobilities of up to 2000 cm$^2$V$^{-1}$s$^{-1}$ \cite{emts09}.

Juang \textit{et al.} synthesized millimeter size few- to single-layer graphene sheets using SiC substrate coated in a thin Ni film. 200 nm of Ni was evaporated onto the surface of the SiC and the sample was heated to 750$^{\circ}$C in vacuum. Graphene was found to segregate to the surface of the Ni on cooling. This gave a continuous graphene layer over the entire nickel surface \cite{juan09}.

Unarunotai \textit{et al.} have developed a technique to transfer graphene synthesized on SiC onto arbitrary insulating substrates. Graphene was first produced using a typical thermal decomposition of SiC technique. A bilayer film of gold/polyimide was deposited onto the SiC wafer and then peeled off. The gold/polyimide film was then transferred onto a Si/SiO$_2$ substrate and the gold/polyimide layers were removed using oxygen plasma reactive ion etching. This yielded single-layer graphene flakes with mm$^2$ areas \cite{unar09}.

\subsection{Chemical vapor deposition}
In contrast to the thermal decomposition of SiC, where carbon is already present in the substrate, in chemical vapor deposition (CVD), carbon is supplied in gas form and a metal is used as both catalyst and substrate to grow the graphene layer. 

\subsubsection{Growth on nickel}
Yu \textit{et al.} grew few-layer graphene sheets on polycrystalline Ni foils. The foils were first annealed in hydrogen and then exposed to a CH$_4$-Ar-H$_2$ environment at atmospheric pressure for 20 mins at a temperature of 1000$^{\circ}$C. The foils were then cooled at different rates between 20$^{\circ}$C/s and 0.1$^{\circ}$C/s. The thickness of the graphene layers was found to be dependent on the cooling rate, with few-layer graphene (typically 3-4 layers) being produced with a cooling rate of 10$^{\circ}$C/s. Faster cooling rates result in thicker graphite layers, whereas slower cooling prevents carbon from segregating to the surface of the Ni foil \cite{yu08}.

To transfer the graphene layers to an insulating substrate, the Ni foil with graphene was first coated in silicone rubber and covered with a glass slide then the Ni was etched in HNO$_3$.

\subsubsection{Growth on copper}
Li \textit{et al.} used a similar process to produce large scale monolayer graphene on copper foils. 25 $\mu$m thick copper foils were first heated to 1000$^{\circ}$C in a flow of 2 sccm (standard cubic centimeters per minute) hydrogen at low pressure and then exposed to methane flow of 35 sccm and pressure of 500 mTorr. Raman spectroscopy and SEM imaging confirm the graphene to be primarily monolayer independent of growth time. This indicates that the process is surface mediated and self limiting. They fabricated dual gated FETs using graphene and extracted a carrier mobility of 4050 cm$^2$V$^{-1}$s$^{-1}$ \cite{li09}.

Recently a roll-to-roll process was demonstrated to produce graphene layers with a diagonal of up to 30 inches as well as transfer them to transparent flexible substrates \cite{bae10}. Graphene was grown by CVD on copper, and a polymer support layer was adhered to the graphene-copper. The copper was then removed by chemical etching and the graphene film transferred to a polyethylene terephthalate (PET) substrate. These films demonstrate excellent sheet resistances of 125 $\Omega/\Box$ for a single layer. Using a repeated transfer process, doped 4-layer graphene sheets were produced with sheet resistances as low as 30 $\Omega/\Box$ and optical transmittance greater than 90\%. These 4-layer graphene sheets are superior to commercially available indium tin oxide (ITO) currently used in flat panel displays and touch screens in terms of sheet resistance ($\sim$100 $\Omega/\Box$ for ITO) and optical transmittance ($\sim$90\% for ITO).

\begin{figure}[H]
\centering
\includegraphics[width=\columnwidth]{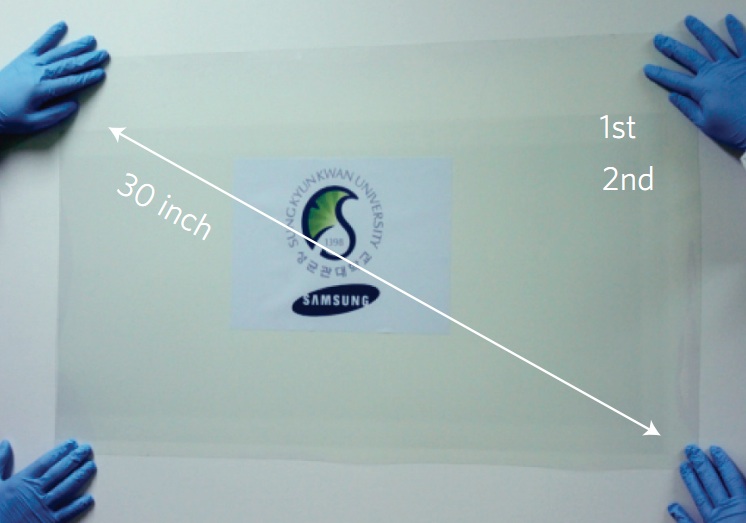}
\caption{\textbf{Multiple CVD graphene sheets transferred to PET.} A roll-to-roll process was used to produce graphene sheets with up to 30 inch diagonal \cite{bae10}. }
\label{30g}
\end{figure}

Li \textit{et al.} have shown the dependence on the size of graphene domains synthesized by CVD with temperature, methane flow and methane pressure. Performing the growth at 1035$^{\circ}$C with methane flow of 7 sccm and pressure 160 mTorr led to the largest graphene domains with average areas of 142 $\mu$m$^2$. A two-step process was used to first grow large graphene flakes and then by modifying the growth conditions to fill in the gaps in the graphene sheet. Using this technique they were able to produce samples with carrier mobility of up to 16000 cm$^2$V$^{-1}$s$^{-1}$ \cite{li10}. In general, the graphene layer is slightly strained on the copper foil due to the high temperature growth \cite{Yu2011}.

\begin{figure}[H]
\centering
\includegraphics[width=\columnwidth]{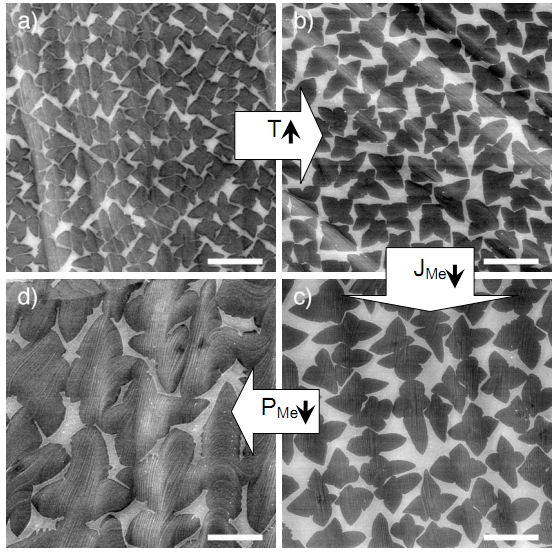}
\caption{\textbf{Controlling domain size in CVD graphene.} Effect of temperature, methane flow and methane partial pressure on the size of graphene domains in CVD growth, scale bars are 10 $\mu$m \cite{li10}. }
\label{dsize}
\end{figure}

Recently Lee \textit{et al.} have demonstrated a technique to produce uniform bilayer graphene by chemical vapor deposition on copper using a similar process but with modified growth conditions. They determined optimal bilayer growth conditions to be: 15 minutes at 1000$^{\circ}$C with methane flow of 70 sccm and pressure of 500 mTorr. The bilayer nature of the graphene was confirmed by Raman spectroscopy, AFM, and transmission electron microscopy (TEM). Electrical transport measurements in a dual gated device indicate that a band gap is opened in CVD bilayer graphene \cite{lee10}.

\begin{figure}[H]
\centering
\includegraphics[width=\columnwidth]{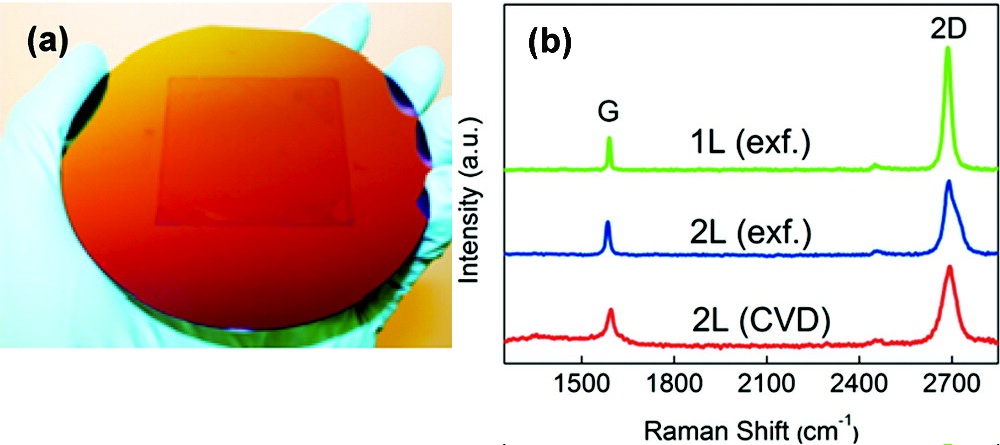}
\caption{\textbf{Bilayer CVD growth on copper.} (a) 2 x 2 inch bilayer graphene on Si/SiO$_2$. (b) Raman spectrum with 514 nm laser source of 1 and 2 layers of graphene produced by exfoliation and CVD \cite{lee10}.}
\label{blcvd}
\end{figure}

\subsection{Molecular beam deposition}

Zhan \textit{et al.} succeeded in layer-by-layer growth of graphene using a molecular beam deposition technique. Starting with an ethylene gas source, gas was broken down at 1200$^{\circ}$C using a thermal cracker and deposited on a nickel substrate. Large area, high quality graphene layers were produced at 800$^{\circ}$C. This technique is capable of forming one layer on top of another, allowing for synthesis of one to several layers of graphene. The number of graphene layers produced was found to be independent of cooling rate, indicating that carbon was not absorbed into the bulk of the Ni as in CVD growth on nickel. Results were confirmed using Raman spectroscopy and TEM \cite{zhan11}.

\begin{figure}[H]
\centering
\includegraphics[width=\columnwidth]{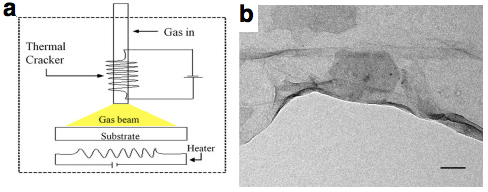}
\caption{\textbf{Molecular beam deposition produced graphene.} (a) Diagram of thermal cracker setup. (b) TEM image of graphene film, scale bar 100 nm \cite{zhan11}.}
\label{mbd}
\end{figure}

\subsection{Unzipping carbon nanotubes}

Multi-walled carbon nanotubes were cut longitudinally by first suspending them in sulphuric acid and then treating them with KMnO$_4$. This produced oxidized graphene nanoribbons which were subsequently reduced chemically. The resulting graphene nanoribbons were found to be conducting, but electronically inferior to large scale graphene sheets due to the presence of oxygen defect sites \cite{kosy09}.

\subsection{Sodium-ethanol pyrolysis}
Graphene was produced by heating sodium and ethanol at a 1:1 molar ratio in a sealed vessel. The product of this reaction is then pyrolized to produce a material consisting of fused graphene sheets, which can then be released by sonication. This yielded graphene sheets with dimensions of up to 10 $\mu$m. The individual layer, crystalline and graphitic nature of the samples was confirmed by TEM, selected area electron diffraction (SAED) and Raman spectroscopy \cite{chou09}.

\subsection{Other methods}
There are several other ways to produce graphene such as electron beam-irradiation of PMMA nanofibres \cite{duan08}, arc discharge of graphite \cite{sub09}, thermal fusion of PAHs \cite{wang082}, and conversion of nanodiamond \cite{sub08}.

\subsection{Graphene oxide}
\label{GOsynth}

Another approach to the production of graphene is sonication and reduction of graphene oxide (GO). The polar O and OH groups formed during the oxidation process render graphite oxide hydrophilic, and it can be chemically exfoliated in several solvents, including water \cite{zhu10}. The graphite oxide solution can then be sonicated in order to form GO nanoplatelets. The oxygen groups can then be removed in a reduction process involving one of several reducing agents. This method was used by Stankovich \textit{et al.} using a hydrazine reducing agent, but the reduction process was found to be incomplete, leaving some oxygen remaining \cite{Stankovich2007}.


Graphene oxide (GO) is produced as a precursor to graphene synthesis. GO is useful because its individual layers are hydrophilic, in contrast to graphite. GO is suspended in water by sonication \cite{McAllister2007, Paredes2008}, then deposited onto surfaces by spin coating or filtration to make single or double layer graphene oxide. Graphene films are then made by
reducing the graphene oxide either thermally or chemically \cite{Marcano2010}. The exact structure of graphene oxide is still a matter of debate, although there is considerable agreement as to the general types and proportion of oxygen bonds present in the graphene lattice \cite{He1996}.

\subsubsection{Wet chemical synthesis}

\begin{figure}[H]
\centering \includegraphics[width=9cm]{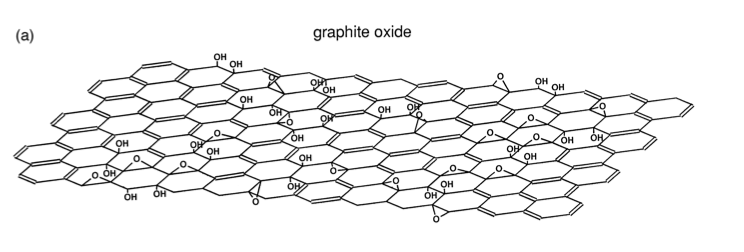}
\caption{\textbf{Structure of a monolayer of graphite oxide} \cite{He1998}.}
\label{GO}
\end{figure}

The chemical methods to produce GO were all developed before 1960. The most recent and most commonly employed is the Hummers procedure \cite{Hummers1958}. This process treats graphite in an anhydrous mixture of sulfuric acid, sodium nitrate, and potassium permanganate for several hours, followed by the addition of water. The resulting material is graphite oxide hydrate, which contains approximately 23\% water. Subsequent nuclear magnetic resonance and X-ray diffraction studies of the structure of GO have led to fairly detailed models based on a combination of hydroxide, carbonyl, carboxyl and epoxide groups covalently bonded to the graphene lattice. \figref{GO} shows a predicted structure for GO produced using the Hummers method \cite{He1998}.

\begin{figure}[H]
\centering \includegraphics[width=6.996cm,height=4.643cm]{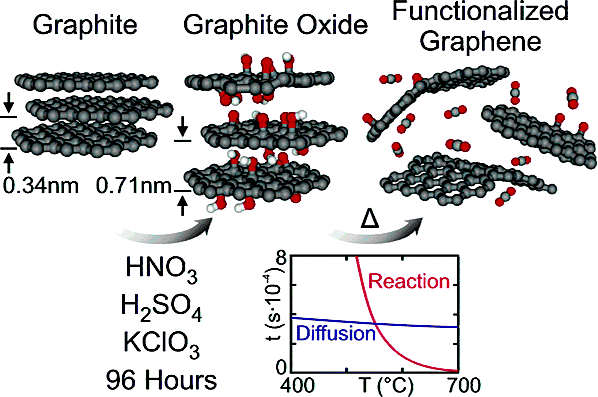}
\caption{\textbf{Summary of the Hummers method and thermal reduction.} Bulk graphite is oxidized then separated in water. Then it is thermally reduced to make single-layer graphene \cite{McAllister2007}.}
\label{Hummers}
\end{figure}

Understandably, the degree of oxidation strongly affects the in-plane electrical and thermal conductivity of graphene oxide. Increased introduction of oxygen groups into the graphene lattice interrupts the $sp^2$ hybridization of electron orbitals. Epoxide groups can be reduced by thermal treatment, or reaction with potassium iodide (KI), leading to a similar structure, where only hydroxyl groups are present. This leads to improved electrical conductivity and unaffected hydrophilicity. In each case the procedure requires tight temperature control and long reaction times of several hours. The basic process, including thermal treatment, is shown in \figref{Hummers}.

\subsubsection{Plasma functionalization}

\ \ Following the realization of the potential importance of graphene as a replacement for semiconductor materials and indium tin oxide (ITO), as discussed in \sectionref{TransparentConductingElectrodes}, alternative methods for graphene production have been explored. Other approaches have been sought to produce the same hydrophilicity in graphite without the time and material requirements of the Hummers method. Very recently, glow discharge treatment has been proven to introduce oxygen species into the lattices of all forms of graphitic materials (e.g. buckyballs, CNTs, graphene, carbon nanofibers, and graphite) \cite{Vandsburger2009}. The resulting graphene/graphite oxides have a structure very similar to Hummers GO, and can be thermally treated to selectively reduce epoxides. Unlike the Hummers method, plasma functionalization requires no strong acids, can proceed at room temperature and can be completed very quickly, often in a matter of seconds or minutes.

Aside from its potential to replace the Hummers method, plasma treatment in itself is interesting for altering the electrical conductivity of graphene or thin graphite. This allows for bandgap engineering as well as phenomena like photoluminescence (PL).

\subsubsection{RF plasma}

Radio frequency (RF) plasma refers to a processing technique whereby a capacitive plasma is ignited in an isolated volume. RF treatment is used almost exclusively for surface treatment of graphene, because ion bombardment is significantly reduced in RF treatments, as opposed to DC discharges. In RF plasma, electrodes need not be in contact with the plasma gas, current is supplied with an alternation frequency of 13.57 MHz, and power ranges from 10 W to 50
W. RF treatment has been shown to selectively affect the outermost surface of graphene \cite{Hazra2011}. Other work using only oxygen for RF treatment allows for layer-by-layer etching of a graphite surface, producing islands of GO \cite{You1993}.

\subsubsection{Photoluminescence}

Single- and dual-layer graphene do not exhibit photoluminescence, due mainly to the negligible bandgap of native graphene. GO does have a photoluminescent response, but the typical oxidation methods, sonication of bulk graphite oxide, are inappropriate for use in photoluminescence applications. For that reason, RF plasma oxidation has been the
subject of recent work at producing photoluminescent single layers of GO. The procedure for producing GO thin films from single layer graphene was reported in \cite{Gokus2009}. Rather than oxidizing bulk graphite to produce single layers of graphene, single- or few-layer graphene is oxidized after isolation. Typically, graphene is prepared by micro-cleavage using the scotch tape or other methods, and electrical contacts or other additions are put in place. Following installation, RF plasma treatment in Ar-O$_2$ mixes is applied in one to six second intervals. The plasma power reported was 10 W, the pressure 0.04 mbar, and the gas composition ratio was 2:1 Ar to O$_2$.

\begin{figure}[H]
\centering \includegraphics[width=6.996cm,height=7.826cm]{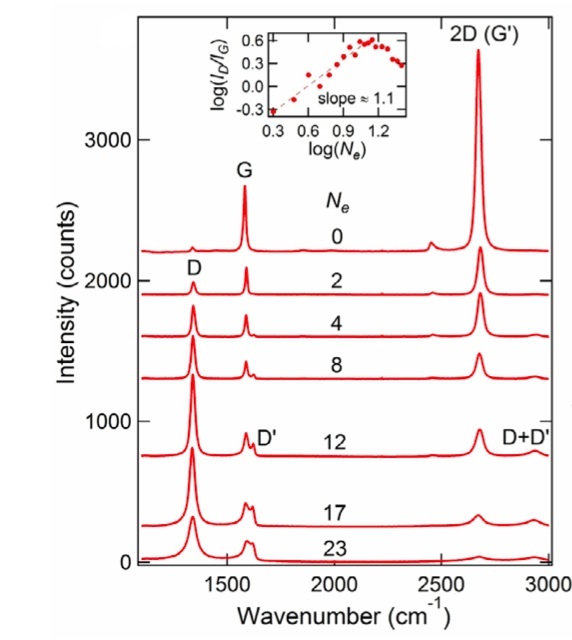}
\caption{\textbf{Raman spectra of graphene with increasing number of 1.5 sec plasma treatments} \cite{Childres2011}.}
\label{RFplasma}
\end{figure}

An important finding reported by \cite{Childres2011} is the temporal evolution of the Raman spectrum of graphene with increasing plasma treatment time. The spectra are reproduced in \figref{RFplasma}. The most noticeable change in the spectra is the
gradual reduction in intensity of the 2D and 2D' peaks, which are indicative of $sp^2$ hybridization. This indicates the disruption of the graphene lattice by introduction of oxygen groups and demonstrates oxidation. Further changes in the region of interest involve the development of a G peak that arises from the increased presence of ``disordered''
carbon. Similar findings were reported in other work using pulsed RF plasma treatment rather than continuous treatments.

\begin{figure}[H]
\centering \includegraphics[width=6.996cm,height=11.028cm]{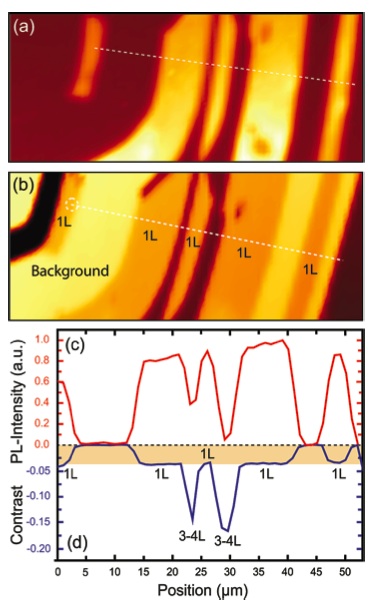}
\caption{\textbf{Photoluminescence in oxygen plasma treated GO.} (a) Dark regions are pristine graphene, while bright regions are single layer graphene oxide. (b) Few-layer graphene is bright while single-layer graphene oxide is dark. (c) Shaded zone shows the correlation between contrast and PL-intensity taken along the white dotted line in (a) \cite{Gokus2009}.}
\label{PL}
\end{figure}

In a sample with a heterogeneous surface, it was shown that only regions of single-layer graphene oxide displayed PL, while untreated graphene or multi-layer graphene did not. The first image shown in \figref{PL} is an image of the PL produced by laser fluorescence. The bright regions correspond to low intensity regions in an elastic scattering image (b), revealing that they are single-layer graphene, labeled 1L in the chart. (c) shows both a PL curve and a contrast curve taken along the white dotted line in (a). The middle-contrast regions in the blue curve correspond exactly to high PL.

Photoluminescence occurs after plasma treatment as a result of the introduction of defects in the graphene lattice. Such defects disrupt the electrical properties of pristine graphene and introduce a bandgap that is absent from native graphene monolayers. A bandgap is desirable for more than PL, and other work has reported plasma oxidation of single and few layer graphene for these purposes \cite{Nourbakhsh2010}. Specifically, a bandgap would allow for logic and optoelectronic applications as we shall see respectively in \sectionref{sec:transistors} and \sectionref{sec:optical}.\\

Once graphene has been produced, it is important to identify it and to characterize its structure, which is the topic of the next section.

\section{Characterization}
\label{Characterization}
A great many techniques are being used to characterize graphene. We discuss here some of the most important ones with a particular emphasis on the identification of graphene.  

\subsection{Raman spectroscopy}
Raman spectroscopy is an important characterization tool used to probe the phonon spectrum of graphene as discussed in section \sectionref{VibrationalProperties}. Raman spectroscopy of graphene can be used to determine the number of graphene layers and stacking order as well as density of defects and impurities. The three most prominent peaks in the Raman spectrum of graphene and other graphitic materials are the G band at $\sim$1580 cm$^{-1}$, the 2D band at $\sim$2680 cm$^{-1}$ and the disorder-induced D band at $\sim$1350 cm$^{-1}$.

The G band results from in-plane vibration of $sp^2$ carbon atoms and is the most prominent feature of most graphitic materials. This resonance corresponds to the in-plane optical phonons at the $\Gamma$ point. The 2D band arises as a result of a two phonon resonance process, involving phonons near the K point, and is very prominent in graphene as compared to bulk graphite \cite{ni08}.

The D band is induced by defects in the graphene lattice (corresponding to the in-plane optical phonons near the K point), and is not seen in highly ordered graphene layers. The intensity ratio of the G and D band can be used to characterize the number of defects in a graphene sample \cite{pim07}.

\begin{figure} [htbp]
\centering
\includegraphics[width=9.5cm]{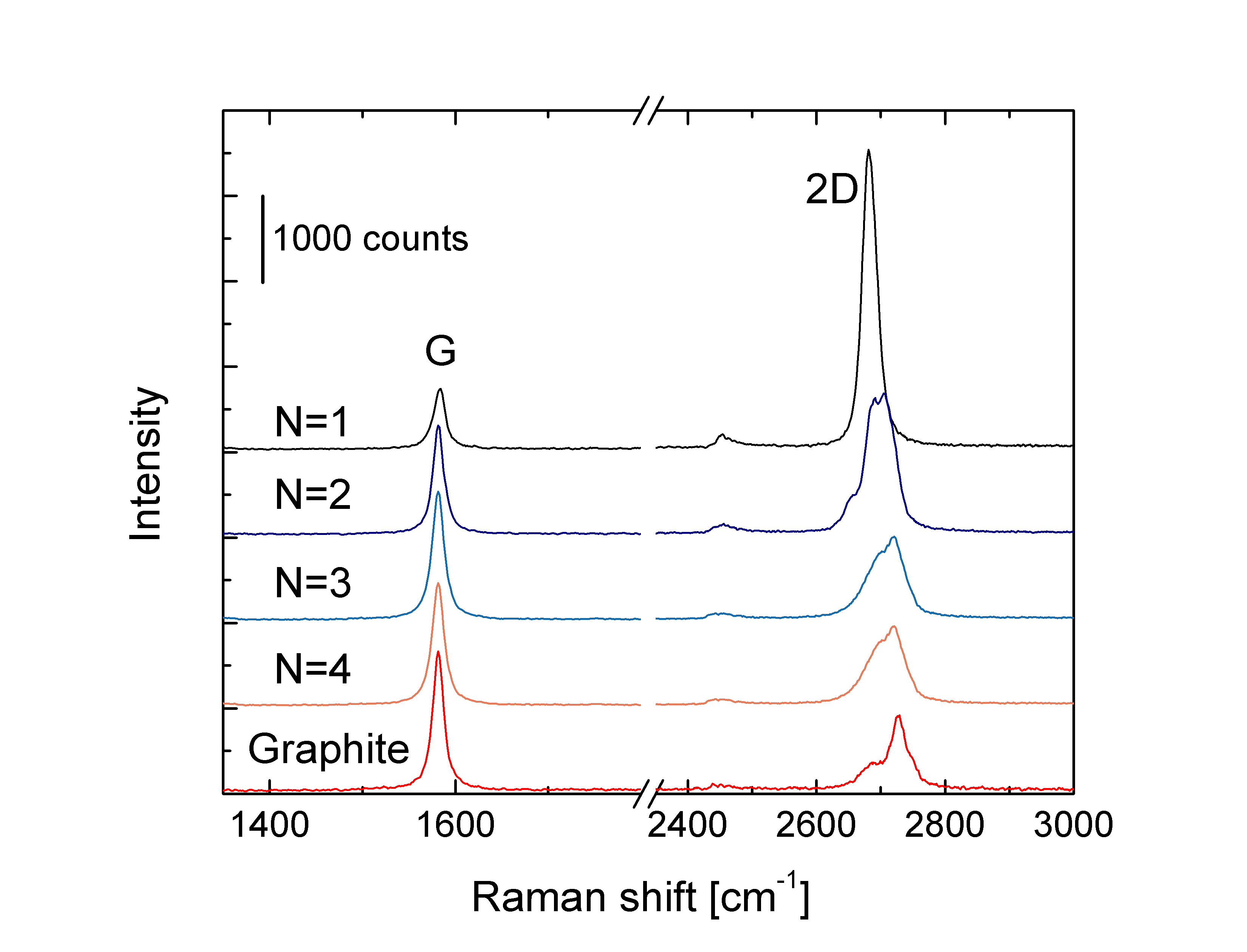}
\caption{\textbf{Layer dependence of graphene Raman spectrum.} Raman spectra of N = 1-4 layers of graphene on Si/SiO$_2$ and of bulk graphite. Figure adapted from \cite{yu10}.}
\label{layers}
\end{figure}

The line shape of the 2D peak, as well as its intensity relative to the G peak, can be used to characterize the number of layers of graphene present as illustrated in \figref{layers}. Single-layer graphene is characterized by a very sharp, symmetric, Lorentzian 2D peak with an intensity greater than twice the G peak. As the number of layers increases the 2D peak becomes broader, less symmetric and decreases in intensity \cite{wang08}.

\subsection{Optical microscopy}

Monolayer graphene becomes visible on SiO$_2$ using an optical microscope. The contrast depends on the thickness of SiO$_2$, the wavelength of light used \cite{blake07} and the angle of illumination \cite{yu09}.

\begin{figure} [htbp]
\centering
\includegraphics[width=\columnwidth]{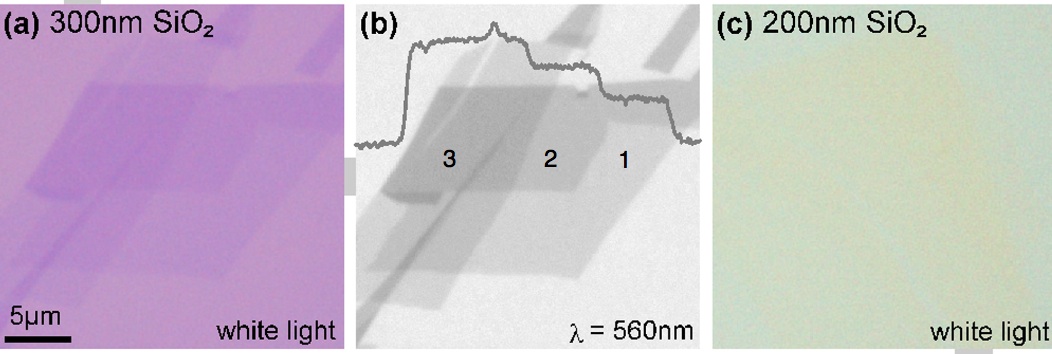}
\caption{\textbf{Optical microscope images of graphene.} Multilayer graphene sheet on Si/SiO$_2$ showing optical contrast at different wavelengths and thicknesses \cite{blake07}.}
\label{opt}
\end{figure}

This feature of graphene is useful for the quick identification of few- to single-layer graphene sheets, and is very important for mechanical exfoliation. \figref{opt} shows the optical contrast of one, two and three layers of exfoliated graphene under different wavelengths of illumination and different thicknesses of SiO$_2$.


\subsection{Electron microscopy}

Transmission electron microscopy has been used to image single-layer graphene suspended on a microfabricated scaffold. It was found that single-layer graphene displayed long range crystalline order despite the lack of a supporting substrate \cite{meyer07}. Suspended graphene was found to have considerable surface roughness with out-of-plane deformations of up to 1 nm.

\begin{figure} [htbp]
\centering
\includegraphics[width=0.75\columnwidth]{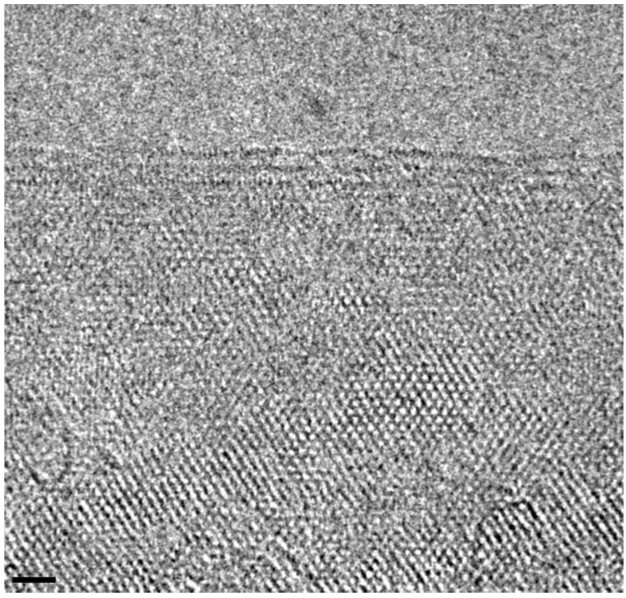}
\caption{\textbf{Atomic scale TEM image of suspended graphene.} Few- to single-layer graphene sheet showing long range crystalline order, scale bar 1 nm \cite{meyer07}.}
\label{tem1}
\end{figure}

\begin{figure} [htbp]
\centering
\includegraphics[width=\columnwidth]{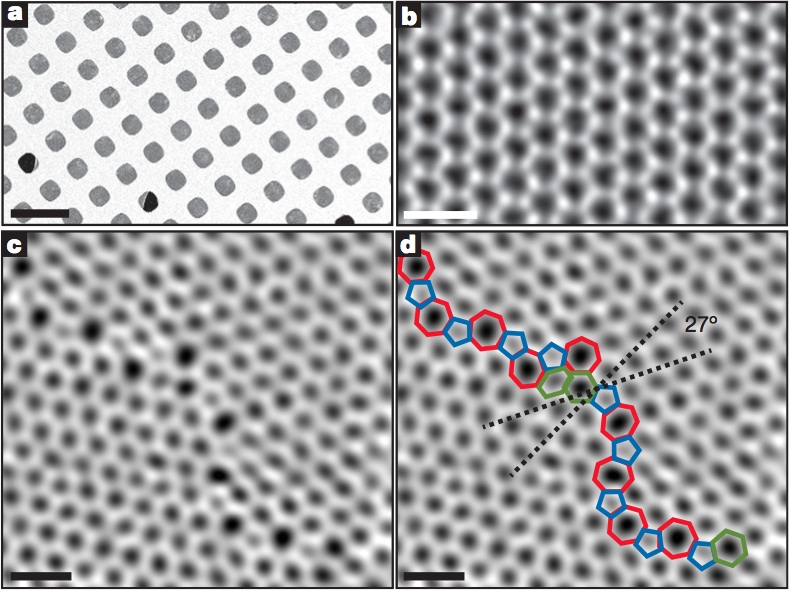}
\caption{\textbf{ADF-STEM imaging of graphene suspended on TEM grid.} (a) SEM image of graphene transferred to TEM grid, scale bar 5 $\mu$m. (b) Atomic scale ADF-STEM image showing the hexagonal lattice in the interior of a graphene grain. (c) ADF-STEM image showing intersection of two grains with a relative rotation of 27$^{\circ}$. (d) Same image with pentagons, heptagons and deformed hexagons formed along grain boundary highlighted (b, c and d have scale bars of 5 \AA) \cite{huang11}.}
\label{stem1}
\end{figure}

Aberration-corrected annular dark-field scanning transmission electron microscopy (ADF-STEM) was used in order to image CVD grown graphene suspended on a TEM grid \cite{huang11}. They found that along grain boundaries the hexagonal lattice structure breaks down and the grains are ``stitched together'' with pentagon-heptagon pairs as seen in \figref{stem1}.

\subsection{Measuring the electronic band structure}

A wide variety of experimental techniques exist for measuring the band structure of materials. Due to its particular characteristics, graphene places severe limitations on the techniques available. Most band structure measurement techniques are highly sensitive to the bulk of a material rather than the surface. Since graphene is so thin, we need techniques that are very sensitive to surface layers.

Angle-resolved photoemission spectroscopy (ARPES) is the most popular technique for measuring the band structure of graphene. Photons of sufficient energy (20-100 eV) are shot at the surface of the material being probed. Each photon is energetic enough that if it interacts with an electron, it has a significant chance of transferring enough energy to launch the electron out of the material completely. The electron must be given enough energy to overcome the work function of the material.

The electron, once free of the material, will have a chance of hitting the ARPES detector. The detector is oriented so that it can measure one specific angle of electron emission. Note that there are two degrees of freedom in angle, typically called $\phi$ and $\theta$. Using these two together, one can specify any direction. The detector is also able to accurately measure the energy $E$ of the outgoing electron. This means that ARPES will simultaneously measure the three variables $\phi$, $\theta$, and $E$. The three components (x,y,z) of the scattered electron's momentum prior to being struck by the photon can be found using the measured quantities. In this way the experimenter can map out the correspondence between energy and momentum within the material with high resolution.

\begin{figure}[htbp]
  	\centering
      	\includegraphics[width=0.45\textwidth]{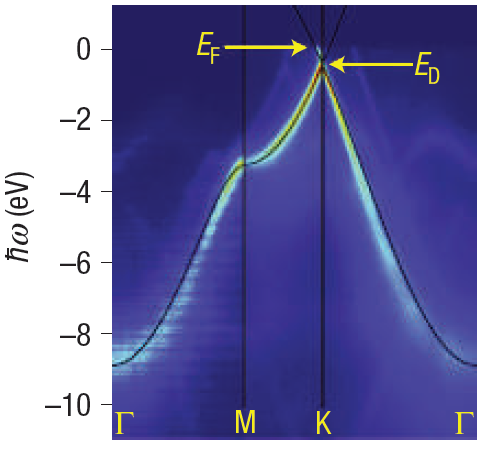}
  	\caption{\textbf{Band structure of graphene on top of SiC.} Vertical axis is the electron's energy, and horizontal axis is its momentum. Note the key locations in momentum space with reference to \figref{fig:BandStructure}, and also that $\Gamma$ is zone center, corresponding to zero momentum. The black line is a theoretical prediction based on the tight binding approximation. The fainter bands are believed to be due to interactions between the substrate and the graphene. Image adapted from \cite{Bostwick2007}.}
  	\label{fig:QuasiParticleDynamics}
\end{figure}

\begin{figure}[htbp]
  	\centering
      	\includegraphics[width=0.45\textwidth]{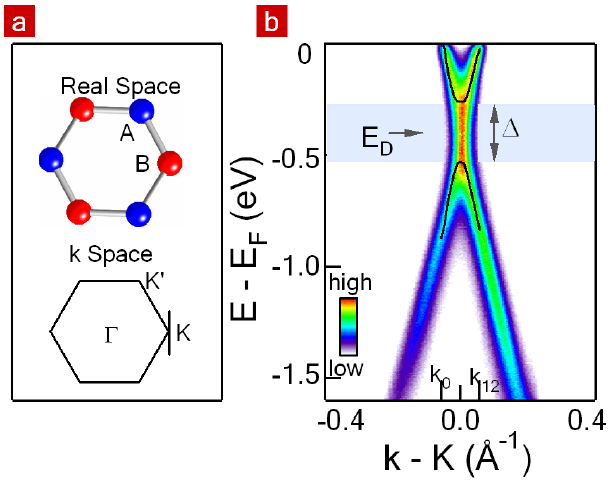}
  	\caption{\textbf{Substrate-induced band gap in single layer graphene on top of SiC.} (a) Real space and momentum space structure of graphene. (b) Band structure of graphene taken along vertical black line near the K point in panel (a). The black lines are dispersion relations estimated from energy distribution curves. Figure from \cite{Zhou2007}.}
  	\label{fig:SubstrateInducedBandGapARPES}
\end{figure}

ARPES is capable of scanning to within about 5 \AA{} of the surface when using electrons of 20-100 eV. This means that most of the signal will be from the first few atomic layers of the surface in question. This property makes ARPES particularly well-suited to measuring the band structure of incredibly thin materials such as graphene. On the other hand, this introduces some experimental difficulties because it means that the sample surface must be kept under ultra-high vacuum (UHV). Creating and measuring graphene without leaving UHV is a significant experimental challenge. Another commonly used method is to anneal the graphene by running a significant electrical current through it. The annealing process does a good job of cleaning the graphene so that it can be measured by techniques like ARPES even after it has been exposed to atmosphere.

ARPES measurements have been made on graphene in a wide variety of circumstances and with many different goals. For the purposes of this review, only a few studies will be mentioned out of this vast field. \figref{fig:QuasiParticleDynamics} shows the experimental band structure of graphene grown on top of SiC \cite{Bostwick2007}. The intent of this study was to better understand the dynamics of charge carriers in graphene. \figref{fig:SubstrateInducedBandGapARPES} is from a different group that was also probing the behavior of graphene grown on top of SiC \cite{Zhou2007}. This second experiment observed a notable band gap in their single-layer graphene samples. Additionally, they noticed that this gap shrank as the number of layers of graphene was increased from one to four. It is believed that the existence of the observed band gap is due to interactions with the substrate that cause the symmetry of graphene's $\pi$-bonds to be broken \cite{Zhou2008}.

There are a number of variations of ARPES that differ only in the wavelength of the probing photons. Angle-resolved ultraviolet photoemission spectroscopy (ARUPS) has also been used to study graphene's band structure \cite{Gierz2008}. The primary reason why this technique is employed is convenience. In ARUPS, a laboratory-based ultraviolet wave source can be used to produce the probing photons. This is a less expensive and simpler setup than ARPES, which typically uses X-rays produced from a synchrotron. It is also worth noting that information about the electronic structure of graphene can be inferred from the results of other material techniques such as optical spectroscopy \cite{Mak2010}.

In the previous sections we have discussed various ways to obtain and identify graphene. We now turn our attention to its physical properties, starting with electronic transport measurements.


\section{Electronic Transport and Field Effect}
\label{ElectronicProperties}
Owing to its unique band structure (see \sectionref{sec:BandStructure}), graphene exhibits novel transport effects such as ambipolar field effect and minimum conductivity which are absent in most conventional materials \cite{Wu10}. This unusual electronic behavior leads to exceptional transport properties in comparison to common semiconductors. This can be seen on \tblref{compareprop} which compares two of the main electronic properties (carrier mobility and saturated velocity) of graphene with those of common bulk semiconductors and 2DEGs. In what follows, we will first describe the experimental methods that are commonly used to measure the ambipolar field effect. We will then discuss the transport properties that can be extracted from this experimental data. The effect of different scattering mechanisms on the carrier mobility and minimum conductivity will then be discussed in detail. Finally, other electrical properties relevant to transistor technology will be reviewed.

\begin{table}
\begin{tabular}{ c*{6}{c}c}
\hline
Property & Si & Ge & GaAs & 2DEG & Graphene \\
\hline\hline
$E_{g}$ at 300 K & 1.1 & 0.67 & 1.43 & 3.3 & 0 \\
(eV) & & & & & \\
\hline
$m^*/m_e$ & 1.08 & 0.55 & 0.067 & 0.19 & 0 \\
 & & & & & \\
\hline
 $\mu_{e}$ at 300 K  & 1350 & 3900 & 4600 & 1500-2000 & $\sim$2$\times 10^5$ \\
(cm$^2$V$^{-1}$s$^{-1}$) & & & & & \\
\hline
$\nu_{sat}$ & 1 & 0.6 & 2 & 3 & $\sim$4 \\
($10^{7}$ cm/s ) & & & & & \\
\hline
\end{tabular}
\caption{\textbf{Comparison between the electronic properties of graphene and common bulk semiconductors}.   Energy band gap ($E_{g}$), electron effective mass ($m^*/m_e$), electron mobility ($\mu_{e}$) and electron saturation velocity ($\nu_{sat}$) of graphene is compared to those of conventional semiconductors and AlGaN/GaN 2DEG \cite{Giannazzo11}.}
\label{compareprop}
\end{table}

\subsection{Measurement of the ambipolar field effect}

\begin{figure}[htbp]
\centering
\includegraphics[scale=0.3]{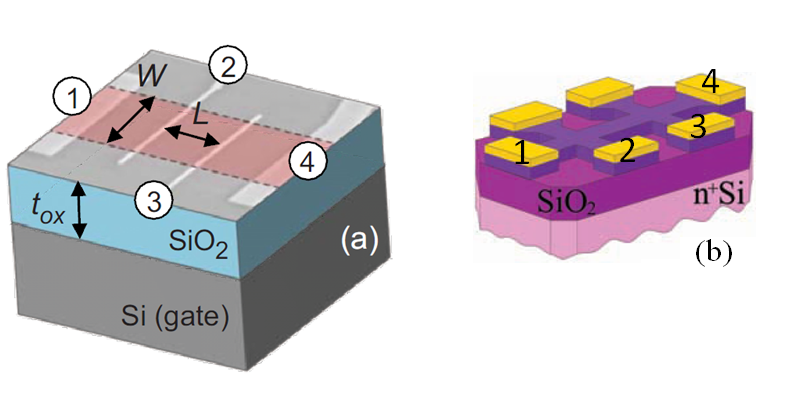}
\caption{\textbf{Schematic representation of common electronic devices.} (a) 4-lead \cite{Dorgan10} and (b) Hall bar \cite{nov04}.}
\label{graphenedevice}
\end{figure}

Transport properties are typically measured with a graphene device similar to those shown in \figref{graphenedevice}. To fabricate these devices, graphene (exfoliated, CVD, etc) is often deposited on an oxidized silicon wafer (SiO$_2$/Si). Later on we discuss other substrates that are sometimes used \cite{Ponomarenko09, Dean10}. Unless otherwise specified, all measurements reported here were made using exfoliated graphene flakes. Electrical contacts, usually made of gold, are then defined using a lithographic process or a stencil mask to avoid photoresist contamination. Electrodes are generally patterned in a 4-lead (\figref{graphenedevice}a) or Hall bar (\figref{graphenedevice}b) configuration. Lastly, the device can be cleaned by annealing at ultrahigh vacuum or in H$_2$/Ar gas, or by applying a large current density ($\sim 10^8$ A/cm$^2$) through it to remove adsorbed contamination \cite{Moser07}.

\begin{figure}[H]
\centering
\includegraphics[scale=0.2]{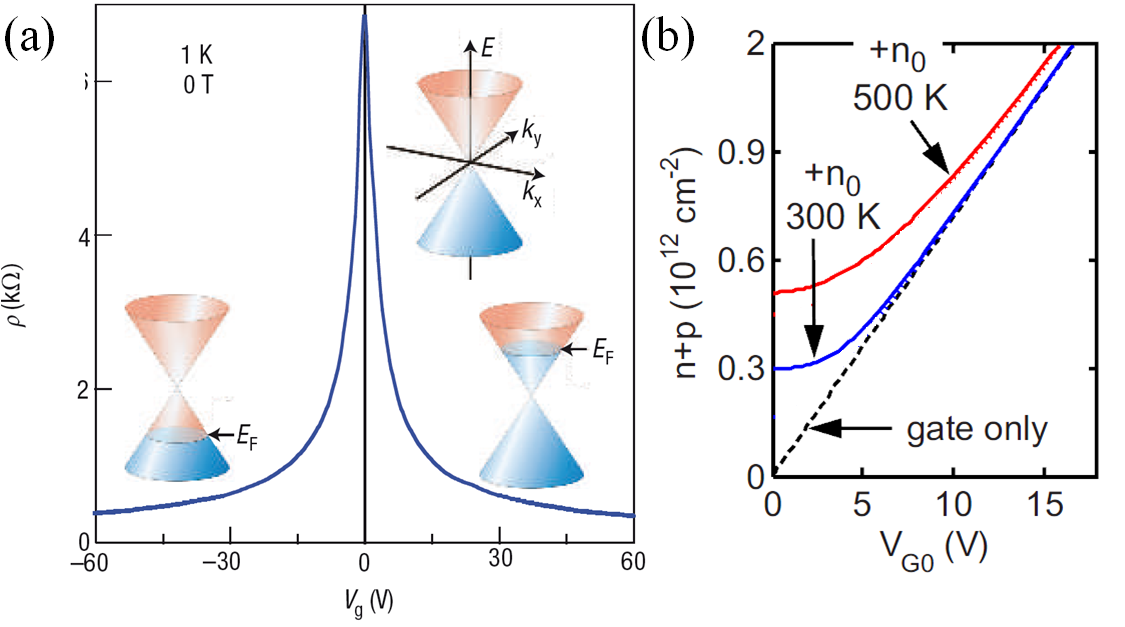}
\caption{\textbf{Ambipolar electric field effect in graphene.} The insets of (a) show the changes in the position of the Fermi level $E_f$ as a function of gate voltage \cite{geim07}. (b) Calculated charge density vs. gate voltage at 300 K and 500 K \cite{Dorgan10}. Solid lines include contribution from $n^*$, $n_th$ and $ n_g$. Dashed line shows only the contribution from the gating ($n_g$).}
\label{ambipolar}
\end{figure}

With this graphene device in hand, one can tune the charge carrier density between holes and electrons by applying a gate voltage ($V_g$) between the (doped) silicon substrate and the graphene flake. The gate voltage induces a surface charge density $n = \epsilon_0 \epsilon V_g/te$ where $\epsilon_0\epsilon$ is the the permittivity of SiO$_2$, $e$ is the electron charge and $t$ is the thickness of the SiO$_2$ layer. This charge density change shifts accordingly the Fermi level position ($E_f$) in the band structure (see the insets of \figref{ambipolar}a). At the Dirac point, $n$ should theoretically vanish \cite{geim07}, but as will be explained further on, thermally generated carriers ($n_th$) and electrostatic spatial inhomogeneity ($n^*$) limit the minimum charge density \cite{Dorgan10}. \figref{ambipolar}b, which shows the calculated carrier density as a function of gate voltage, clearly illustrates the fact that charge density is well controlled by the gate away from the Dirac point. The linear relation between $V_g$ and $n$ was verified experimentally \cite{nov04} in that region by measuring the Hall coefficient $R_H=1/ne$ as a function of $V_g$ (see \sectionref{sec:CarrierDensityTuning}). Typically, charge density can be tuned from $10^{11}$ to $10^{13}$ cm$^{-2}$ by applying a gate voltage that moves $E_f$ 10 to 400 meV away from the Dirac point \cite{Giannazzo11}.

In 2004, the ambipolar field effect corresponding to the change in resistivity $\rho$ (or conductivity $\sigma = 1/\rho$) that occurs when the charge density is modified by the gate voltage was observed and analyzed \cite{nov04}. Experimentally, $\rho$ is measured using a standard 4-probe technique \cite{VanDerPauw09} and is given by $\rho = (W/L)(V_{23}/I_{14})$ where $W$ and $L$ are respectively the width and the length of graphene, $V_{23}$ is the voltage across electrodes 2 and 3 (see \figref{graphenedevice}) and $I_{14}$ is the current between contact 1 and 4. Note that because of the uncertainty on the aspect ratio $L/W$, the error on the absolute magnitude of $\rho$ is usually around 10\% \cite{Chen08, Chen208}. As \figref{ambipolar}b shows, resistivity rapidly increases as we remove charge carriers, reaching its maximum value at the Dirac point. From this curve one can extract the carrier mobility $\mu = 1/en\rho$ and the minimum conductivity $\sigma_{min}$. Other definitions for mobility are sometimes used such as the field effect mobility $\mu_{FE} = (1/C)d\sigma/dV_g$ (where $C$ is the gate capacitance) \cite{Dean10} and the Hall mobility $\mu_{Hall}= R_H/\rho$ \cite{nov04}. Note that in practice, the carrier mobility is only meaningful away from the Dirac point, where $n$ is accurately tuned by the gate voltage.

\subsection{Transport and scattering mechanisms}

In contrast with the ideal, theoretical graphene, experimental graphene contains defects \cite{Chen09} and impurities \cite{Chen208, Zhang09}, interacts with the substrate \cite{Chen08}, has edges and ripples \cite{Katsnelson08} and is affected by phonons \cite{Bolotin08}. These perturbations alter the electronic properties of a perfect graphene sheet first by introducing spatial inhomogeneities in the carrier density and, second, by acting as scattering sources which reduce the electron mean free path \cite{Giannazzo11}. The former effect dominates when the Fermi level is close to the Dirac point and alters the minimum conductivity of graphene whereas the latter effect prevails away from the Dirac point and affects the carrier mobility.
The impact of these perturbations has been subjected to intensive and ongoing investigation, on both the theoretical and experimental side, in order to determine the mechanisms that limit the mobility and the minimum conductivity. From a theoretical point of view, two transport regimes are often considered depending on the mean free path length $l$ and the graphene length $L$. When $l > L$, transport is said to be ballistic since carriers can travel at Fermi velocity ($\nu_f$) from one electrode to the other without scattering. In this regime, transport is described by the Landauer formalism \cite{Peres2009} and the conductivity is expressed as:

\begin{align}
	\sigma_{ball}=\frac{L}{W}\frac{4e^2}{h}\sum^{\infty}_{n=1}T_n
 \label{landauer}
\end{align}

where the sum is over all available transport modes of transmission probability $T_n$. For ballistic transport mediated by evanescent modes, this theory predicts that at the Dirac point the minimum conductivity is:

\begin{align}
	\sigma_{min}=\frac{4e^2}{\pi h}=4.92\times10^{-5}~\Omega^{-1}
 \label{ballistic} 
\end{align}	

On the other hand, when $l < L$, carriers undergo elastic and inelastic collisions and transport enters the diffusive regime. This regime prevails when the carrier density $n$ is much larger than the impurity density $n_i$. In that case, transport is often described by the semiclassical Boltzmann transport theory \cite{Sarma2010} and at very low temperature carrier mobility can be expressed in terms of the total relaxation time $\tau$ as:

\begin{align}
	\sigma_{sc}=\frac{e^2\nu_f\tau}{\hbar}\sqrt{\frac{n}{\pi}}
 \label{boltzmann}
\end{align}

This equation describes the diffusive motion of carriers scattering independently off various impurities. The relaxation time depends on the scattering mechanism dominating the carrier transport or a combination thereof. The scattering mechanisms mostly discussed in the literature include Coulomb scattering by charged impurities (long range scattering), short-range scattering (defects, adsorbates) and electron-phonon scattering. In the following, we provide a brief theoretical introduction of these scattering processes and relevant transport measurements. 

\subsubsection{Phonon scattering}

Phonons can be considered an intrinsic scattering source since they limit the mobility at finite temperature even when there is no extrinsic scatterer. As explained in section \sectionref{VibrationalProperties}, the dispersion relation of graphene comprises six branches. Longitudinal acoustic (LA) phonons are known to have a higher electron-phonon scattering cross-section than those in the other branches \cite{hwang}. The scattering of electrons by LA phonons can be considered quasi-elastic since the phonon energies $\hbar\omega_q$ are negligible in comparison with $E_F$, the Fermi energy of electrons. 

In order to determine the effect of electron-phonon scattering on resistivity, one must consider two distinct transport regimes separated by a characteristic temperature $T_{BG}$ called the {\it Bloch-Gr\"{u}neissen} temperature, defined as \cite{hwang}:

\begin{align}
k_{B}T_{BG}=2k_{F}v_{ph}
\end{align}

where $k_B$ is the Boltzmann constant, $v_{ph}$ is the sound velocity and $k_F$ is the Fermi wave vector with reference to the K point in the BZ.

\begin{align}
k_{F}=\sqrt{n\pi}
\end{align}

where $n$ is the electron density in the conduction band \cite{pisana}. If one measures $n$ in units of $n=10^{12}$ cm$^{-2}$ we get $T_{BG}\approx 54\sqrt{n}~$ K.\\

\begin{figure}[h!]
\begin{center}
\includegraphics*[width=3.0in]{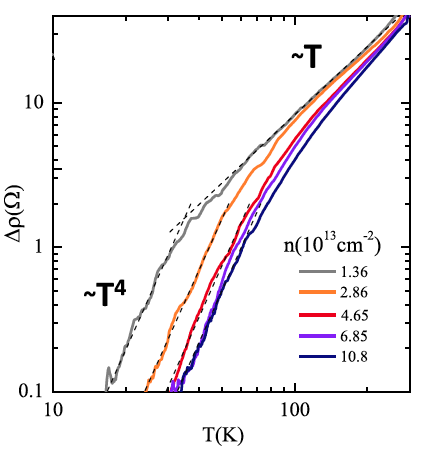}
\caption{\textbf{Electric resistivity of graphene at ultrahigh carrier densities}. Resistivity over wide range of $T$, showing the cross-over from the low $T \sim (BG)$ regime to the high $T \sim (EP)$ one \cite{efetov}.}
\label{figresist}
\end{center}
\end{figure}

Consider first $T \gg T_{BG}$, the equipartition $(EP)$ limit. In this case the Bose-Einstein distribution function for the phonons is $N(\omega_q)\approx k_{B}T/\hbar\omega_q $, which leads to a linear $T$-dependence of the scattering rate and hence the resistivity $\rho\sim T$. In the $BG$ or degenerate regime, on the other hand, where $T \ll T_{BG}$, one obtains at very low $T$, $\rho\sim T^4$ \cite{hwang} as shown in \figref{figresist}.

\subsubsection{Coulomb scattering}

Coulomb scattering stems from long-ranged variations in the electrostatic potential caused by the presence of charged impurities close to the graphene sheet. These impurities are often thought of as trapped ions in the underlying substrate and screened by the conduction electrons of graphene. Assuming random distribution of charged impurities with density $n_i$ and employing a semiclassical approach, it was predicted \cite{Adam2007} that the charged-impurity scattering is proportional to $\frac{\sqrt{n}}{n_i}$. With \eqnref{boltzmann}, the conductivity at high carrier density ($n \gg n_i$) is given by:

\begin{align}
\sigma_i=\frac{Ce^2}{h}\frac{n}{n_i}
\label{coulomb}
\end{align}

where $C$ is a dimensionless parameter related to the scattering strength. Considering the random phase approximation and the dielectric screening from the SiO$_2$ substrate, it was predicted that $C \approx 20$. \\

\begin{figure}[htbp]
	\centering
		\includegraphics[width=0.50\textwidth]{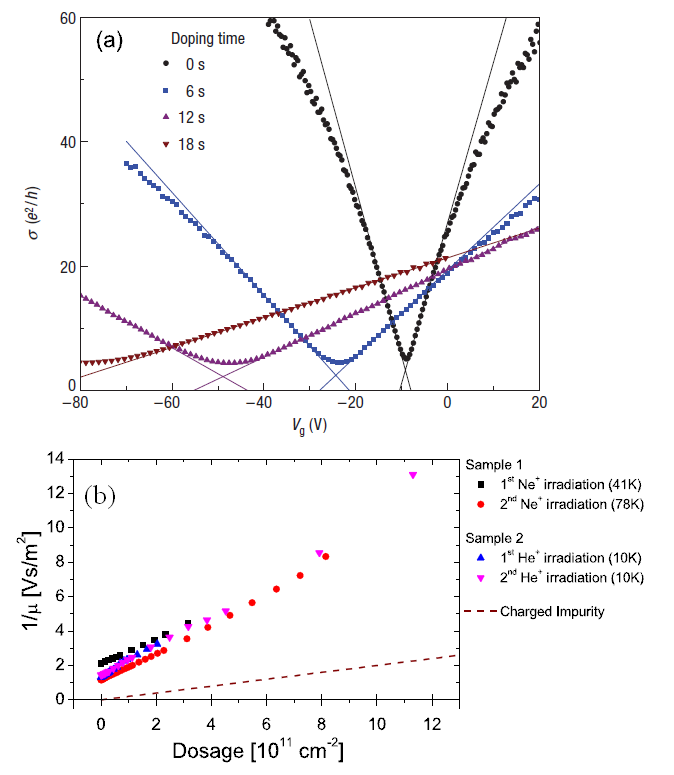}
	\caption{\textbf{Effect of charge impurities and defects on transport properties of graphene}. (a) The conductivity ($\sigma$) vs. gate voltage for a pristine sample (black curve) and three different potassium doping concentrations at 20 K in UHV. Lines represent empirical fits \cite{Chen208}. (b) Inverse of mobility ($1/\mu$) as a function of ion dosage for different samples and irradiated ions (Ne$^+$ and He$^+$). The dashed line corresponds to the behavior for the same concentration of potassium doping \cite{Chen09}.}
	\label{fig:charge+defect}
\end{figure}

Chen \textit{et al.} investigated experimentally the effect of charged impurities on the carrier mobility and conductivity by doping a graphene flake with a controlled potassium flux in UHV \cite{Chen208}. \figref{fig:charge+defect}a shows the conductivity as a function of gate voltage for a pristine sample and three different doping concentrations. It can be clearly seen that the gate voltage of minimum conductivity becomes more negative with increasing doping. As it was previously shown \cite{Schedin07}, this is because K atoms dope graphene with electrons (n-doping), which in effect moves the Fermi level up with respect to the Dirac point. From \figref{fig:charge+defect}a, one can also see that $\sigma(V_g)$ becomes more linear and mobility decreases as the doping concentration $n_i$ increases which is in good agreement with \eqnref{coulomb}. The dashed line in \figref {fig:charge+defect}b shows that mobility scales linearly with $1/n_i$ when transport is limited by charged-impurity scattering. In these measurements $C \approx 20$ in \eqnref{coulomb} was obtained \cite{Chen208, Tan2007}.

\subsubsection{Short-range scattering}

Finally, short-range defects such as vacancies and cracks in graphene flakes are predicted to produce midgap states in graphene \cite{Stauber2007}. Vacancies can be modeled as a deep circular potential well of radius $R$ and this strong disorder gives rise to a conductivity which is roughly linear in $n$:

\begin{align}
\sigma_d=\frac{2e^2}{\pi h}\frac{n}{n_d}ln^2(\sqrt{\pi n}R)
\label{defect}
\end{align}

where $n_d$ is the defect density. This equation mimics the one for charged impurities (\eqnref{coulomb}), with a slightly logarithmic dependence of the conductivity on the charge carrier density. Defect scattering was experimentally studied by irradiating a graphene flake with 500 eV He and Ne ions in UHV \cite{Chen09}. The resulting conductivity was also demonstrated to be approximately linear with charge density, with mobility inversely proportional to the ion (or defect) dose $n_d$. As shown in \figref{fig:charge+defect}b, the mobility decrease was found to be 4 times larger than the same concentration of charged impurities. From the linear fits of \figref{fig:charge+defect} with \eqnref{defect}, the impurity radius was found to be $R \approx 2.9$ \AA{} which is a reasonable value for a single-carbon vacancy.

\subsection{Mobility}
\subsubsection{Graphene on SiO$_2$}

\begin{figure} [htbp]
	\centering
		\includegraphics[width=.50\textwidth]{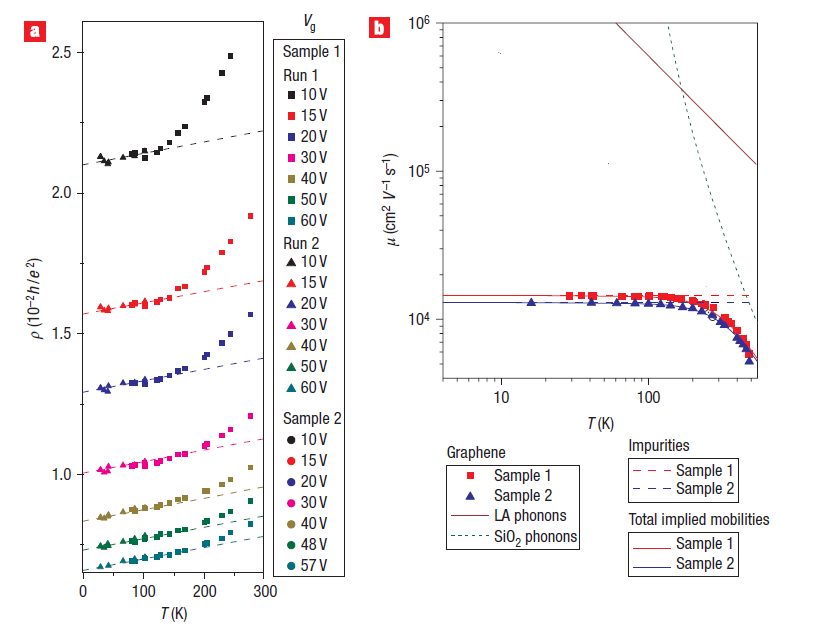}
	\caption{{\textbf{Temperature dependence of resistivity and mobility in graphene on SiO$_2$. } (a) Temperature-dependent resistivity of graphene on SiO$_2$} for different gate voltages, samples and runs. Dashed lines are linear fits at low temperature. (b)Temperature-dependent mobility in graphene on SiO$_2$ for two samples at $n =1\times10^{12}$ cm$^{-2}$. Mobility limits due to LA phonons (dark red solid line), substrate surface phonons (green dashed line), and impurities/defects (red and blue dashed lines) are presented. Matthiessen's rule was used to obtain the net mobility for each sample (red and blue solid lines) \cite{Chen08}.}
	\label{fig:all+ro}
\end{figure}

In a graphene electronic device, all of the scattering mechanisms mentioned above come into play. From a technological point of view, determining the exact nature of the scattering that limits the mobility is essential in order to develop high-speed electronic devices. To do so, one must also take into account the effect of the underlying substrate on the electronic transport. Chen \textit{et al.} performed a general study of scattering mechanisms in graphene on SiO$_2$ by measuring the temperature dependence of the ambipolar effect \cite{Chen08}. The measurements were fitted using three terms: $\rho(n,T) = \rho_0(n) + \rho_{LA}(T) + \rho_{SPP}(n,T)$. Each term was associated with a certain scattering mechanism. As shown in \figref{fig:all+ro}a, the first two terms were determined with a linear fit at low $T$. The y-intercepts correspond to $\rho_0(n)$ and were found to scale linearly with $n^{-1}$. According to \eqnref{coulomb} and \eqnref{defect}, this behavior can be associated with charged impurities and defects. $\rho_{LA}(T)$ was determined by the slope of the linear fit in \figref{fig:all+ro} and appears to be independent of the charge density. These results agree very well with predictions for longitudinal acoustic (LA) phonon scattering in the regime where $T \gg T_{BG}$. Finally, the third term $\rho_{SPP}$, was found to have a strong temperature dependence. Such behavior was explained by considering the surface polar phonons (SPP) of the SiO$_2$ substrate which produce an electrical field that couples to electrons in graphene. Theoretical expressions for SPP-limited resistivity showed good agreement with the data of \figref{fig:all+ro}a at high temperature. \\

 \figref{fig:all+ro}b presents the temperature-dependent mobility and the theoretical limits of the three scattering mechanisms (LA phonons, SPP and impurities/defects). One can see that at room temperature, SPP and impurities/defects are by far the two dominant scattering mechanisms for graphene on SiO$_2$. At low and room temperature, mobility is mainly limited by the former type of scattering. Typically, mobility of graphene on SiO$_2$ ranges from 10000 to 15000 cm$^2$V$^{-1}$s$^{-1}$. It was suggested that the scattering impurities are trapped charges in the underlying SiO$_2$ substrate, but this remains under debate \cite{Ponomarenko09}.

\subsubsection{Suspended graphene}
\label{sec:SuspendedGraphene}

\begin{figure}
	\centering
		\includegraphics[width=0.40\textwidth]{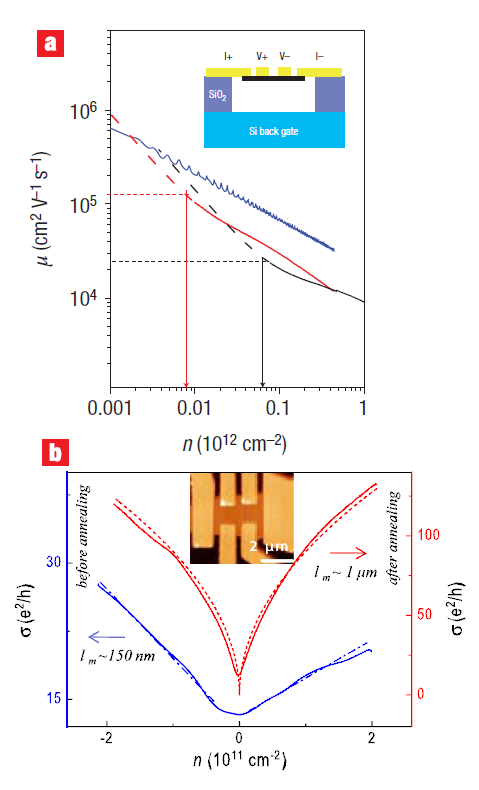}
						\caption{ \textbf{Conductance and mobility in suspended graphene as a function of charge density.} (a) Mobility vs. charge density for suspended (red line) and non-suspended (black line) graphene at $T$ = 100 K. The blue line represents the ballistic model prediction. Inset: schematic representation of the suspended graphene device \cite{Du08}. (b) Conductance vs. gate voltage (at $T$ = 40 K) for a suspended graphene device before (blue line) and after (red line) annealing and current-induced cleaning. The red dotted line was calculated using a ballistic model. Inset: AFM image of the suspended graphene device \cite{Bolotin08}.}
\label{fig:suspended}
\end{figure}

Removing the substrate or using one which is free of trapped charge are two possible ways to improve the carrier mobility. The former approach was used by \cite{Du08}, in which the authors fabricated a suspended graphene device by chemically etching the underlying SiO$_2$ (see the inset of \figref{fig:suspended}a). From this device, they obtained mobilities as high as 200,000 cm$^2$V$^{-1}$s$^{-1}$ for charge density below 5$\times10^9$ cm$^{-2}$. \figref{fig:suspended}a shows the charge density dependence of mobility for suspended and non-suspended (on SiO$_2$) graphene devices at 100 K. The arrows indicate the minimum charge density that is accurately controlled by the gate voltage. From this figure, one can see that mobility in suspended graphene approaches the ballistic model prediction. Furthermore, the mobility deterioration caused by adsorbed impurities on the surface of suspended graphene has been demonstrated \cite{Bolotin08, Bolotin208}. As \figref{fig:suspended}b shows, mobility and mean free path $l$ clearly increase after current-induced cleaning and high-temperature annealing of the device. At large carrier density ($n > 0.5\times10^{11}$ cm$^{-2}$) and temperatures above 50 K, resistivity of clean suspended graphene was found to be linear with temperature, suggesting LA phonon scattering. At low temperature ($\sim$ 5 K), mobility as high as 170,000 cm$^2$V$^{-1}$s$^{-1}$ was obtained and the mean free path reached the device dimension. In these conditions, conductivity is well described by the ballistic model as shown in \figref{fig:suspended}b.

\subsubsection{Other substrates}

Although suspended graphene shows impressive transport properties, this geometry imposes evident constraints on the device architecture. To overcome this problem, boron nitride (BN) was proposed as a substrate \cite{Dean10}. Compared to SiO$_2$, BN has an atomically smooth surface, is relatively free of charged impurities, has a lattice constant similar to that of graphene and high surface phonon frequencies. All these advantages result in a mobility about three times higher than that of graphene on SiO$_2$. However, graphene/BN devices are troublesome to fabricate and are thus not ideal for industrial applications. Currently, large area graphene synthesized by CVD or thermal segregation is the most promising material for technological applications since it can be produced on a wafer scale \cite{li09, Kim09}. Depending on the technique, large scale graphene (on SiO$_2$) typically shows mobilities lower than 5000 cm$^2$V$^{-1}$s$^{-1}$ \cite{Avouris10}. The scattering mechanisms responsible for this reduced mobility are still under investigation. Preliminary studies suggested that grain boundaries \cite{Yazyev10}, doping defects \cite{Wei09} and ineffective gate voltage control \cite{Cao10} might limit the mobility. In addition to mobility, a common way to characterize the quality of large scale graphene is to measure its sheet resistance $R_s$ (resistivity at $V_g = 0$). This property varies with the synthesizing technique used and can be as low as 125 $\Omega/\Box$ \cite{bae10}. It was also shown that sheet resistance increases with decreasing temperature \cite{Park10}. This observation contrasts with the metallic behavior ($d\sigma/dT<0$) of pristine graphene flakes but agrees with the insulating behavior of irradiated graphene (with defects) \cite{Chen09}.

\begin{table}[h!]
  	\centering
\begin{tabular}{ c c c c c }
\hline
Substrate & Production & $\mu$ & $\sigma_{min}$ & Ref.\\
   	& technique & ($\times10^3$ cm$^2$V$^{-1}$s$^{-1}$) & ($e^2/h$) &\\
\hline \hline
SiO$_2$/Si & Exfoliation & 10-15 & 4 & a \\
\hline
Boron nitride & Exfoliation & 25-140 & 6 & b \\
\hline
Suspended & Exfoliation & 120-200 & 1.7/$\pi$ & c \\
\hline
SiC & Thermal-SiC & 1-5 & - & d \\
\hline
SiO$_2$/Si & Ni-CVD & 2-5 & - & e \\
\hline
SiO$_2$/Si & Cu-CVD & 1-16 & - & f \\
\hline
\end{tabular}
\caption{ \textbf{Mobility range ($\mu$) and minimum conductivity ($\sigma_{min}$) of graphene produced by different techniques and deposited on different substrates}. a: \cite{nov04}, b: \cite{Dean10}, c: \cite{Bolotin208}, d: \cite{emts09}, e: \cite{Kim09}, f: \cite{li10} }
\label{mobcond}
\end{table}

\subsection{Minimum conductivity}

The presence of disorder in graphene (ripples, defects, impurities, etc.) produces fluctuations in its electrostatic potential. We mentioned above that these perturbations are significant at the Dirac point where the screening of the potential fluctuations is weak due to the low charge density. These fluctuations in the charge density can be thought of as electron-hole puddles and have been observed experimentally using scanning probe methods \cite{Zhang09, Martin08} on graphene/SiO$_2$ samples. Experimentally, these disordered graphene samples have a minimum conductivity about $\pi$ times larger than that predicted for ballistic transport given in \eqnref{ballistic}. This discrepancy between theory and experiments was observed in 2004 \cite{nov04} and was known as ``the mystery of the missing pi.'' The effect of doping on the minimum conductivity was also investigated in a study of charged-impurity scattering \cite{Chen208}. It was found that $\sigma_{min}$ decreases on initial doping and reaches a minimum near $4e^2/h$ only for non-zero charged impurities. This suggests that charged impurities trapped in the SiO$_2$ substrate, located at the graphene/SiO$_2$ interface or on the graphene surface are responsible for the minimum conductivity obtained experimentally.

\begin{figure}
  	\centering
      	\includegraphics[width=.40\textwidth]{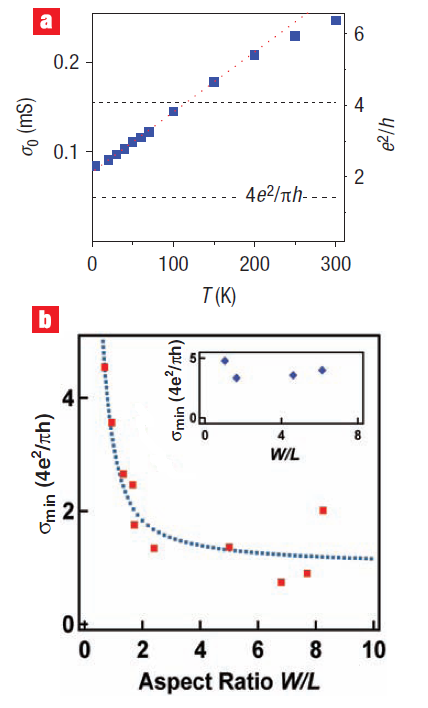}
  	\caption{\textbf{Minimum conductivity of graphene.} (a) Minimum conductivity vs. temperature for a suspended graphene device. The upper dashed line represents the $4e^2/h$ limit for graphene with charged impurities. The lower dashed line corresponds to the theoretical value in the ballistic regime \cite{Du08}. (b) Minimum conductivity vs. aspect ratio for small and large (inset) graphene/SiO$_2$ devices \cite{Miao07}.}
  	\label{fig:minconduct}
\end{figure}

To verify this conclusion, the minimum conductivity of clean suspended graphene was measured as a function of temperature \cite{Du08}. As \figref{fig:minconduct}a shows, minimum conductivity decreases with temperature and approaches the ballistic prediction down to a factor of 1.7. However, the linear temperature dependence was not expected. Another group also investigated the near-ballistic regime by measuring the minimum conductivity on samples with different aspect ratios ($W/L$) and surface areas at low temperature ($\sim$1.5 K) \cite{Miao07}. For large-area samples ($\sim$3 $\mu$m$^2$), the minimum conductivity is mostly independent of the aspect ratio (inset of \figref{fig:minconduct}b). On the other hand, small-area devices ($<$0.2 $\mu$m$^2$) yield results that are clearly dependent on the aspect ratio (\figref{fig:minconduct}b). Devices with wide electrodes and short channels (large $W/L$) approach the theoretical minimum conductivity in the ballistic regime. \tblref{mobcond} compares the carrier mobility and minimum conductivity of graphene produced by different techniques and on different substrates.

\subsection{Other transport properties}

Concerning electronic applications, graphene has attracted considerable attention due to its high mobility. However, other transport properties must be taken into account in order to develop graphene-based technologies successfully. Among those properties, saturation velocity is particularly relevant for field-effect transistor (FET) applications. In modern FETs, short channel lengths result in high electrical fields \cite{Schwierz:2010ix} of around $\sim$100 kV/cm. In such conditions, the carriers acquire enough kinetic energy to excite the optical phonon modes of graphene \cite{Avouris10}. As a consequence, carrier velocity saturates, reducing the relevance of mobility to device performance. The optical phonons of graphene have higher energy ($\sim$160 meV) than those of common semiconductors such as Si (55 meV) \cite{Dorgan10}. Consequently, the intrinsic saturation velocity is higher in graphene than in conventional semiconductors. 

\begin{figure}
  	\centering
      	\includegraphics[width=0.50\textwidth]{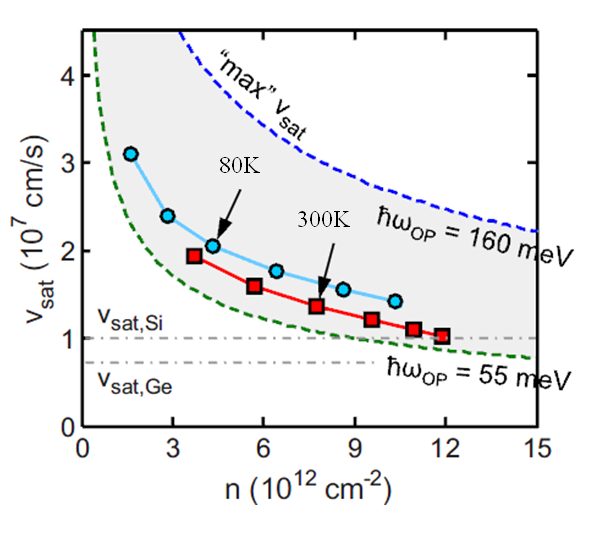}
          	\caption{\textbf{Electron saturation velocity of graphene on SiO$_2$.} Electron saturation velocity vs. charge density for an electric field of 2 V/$\mu$m at $T$ = 80 K and 300 K. The upper and lower dashed curves correspond to theoretical saturation velocities limited by optical phonons of graphene and SiO$_2$, respectively \cite{Dorgan10}.}
  	\label{fig:sat v}
\end{figure}

Numerical values of saturation velocity are presented in \tblref{compareprop}, and \figref{fig:sat v} shows the experimental and theoretical dependence of saturation velocity on charge density for graphene/SiO$_2$. A simple phonon emission model \cite{Meric11} predicts that saturation velocity is proportional to $1/n^{1/2}$. In \figref{fig:sat v}, the upper and lower theoretical curves correspond to saturation velocities limited by optical phonons of graphene and SiO$_2$, respectively. The experimental results suggest that both kinds of phonons play a role in limiting $\nu_{sat}$ but that substrate phonons of SiO$_2$ are dominant for this device \cite{Dorgan10}. Some studies also point out that current never reaches a complete saturation in some graphene devices. This incomplete saturation might be due to a competition between disorder and optical phonon scattering \cite{Barreiro09}, or the formation of a ``pinch-off'' region \cite{Meric11}. Mechanisms limiting the saturation velocity are under active investigation.

Graphene has excellent transport properties, but to take advantage of them, carriers must be injected and collected through metal contacts. These electrical contacts produce energy barriers that limit the charge transfer at the graphene/metal junction \cite{Avouris10}. This limitation results in a contact resistance which can be measured and should be reduced to create high-performance graphene devices. One way to determine the contact resistance is to fabricate variable channel length devices and to extrapolate the resistance to zero channel length. It was shown that contact resistance is temperature and gate voltage dependent, and that it varies from about 100 $\Omega~\mu$m to a few k$\Omega~\mu$m \cite{Xia11}. Finally, it is worth pointing out that graphene can sustain current densities greater than $10^8$ A/cm$^{2}$, which is 100 times higher than those supported by copper \cite{Moser07}. Graphene can thus be used as interconnects in integrated circuits.

Since Geim's and Novoselov's seminal work \cite{nov04}, graphene's unique electronic properties have attracted massive interest and created an explosion of scientific activity. Much ink has been spilled about the ambipolar field effect, the ultra-high mobility of graphene and the limiting scattering mechanisms, as well as the minimum conductivity. In this section, we reported some of the main experimental studies on these ever growing subjects. In the next section, we turn our attention to the related but more specific topics of magnetotransport and quantum Hall effect in graphene.


\section{Magnetoresistance and Quantum Hall Effect}


The first demonstration, in the early 1980s, of the existence of a so-called quantum Hall effect (QHE) of the two-dimensional electron gas (2DEG) at the interfaces of semiconductor heterostructures sparked great interest in the experimental study of the properties of low-dimensional systems \cite{VonKlitzing1980}. Thus it is perhaps no surprise that when the one atom thick, two-dimensional lattice of graphene could finally be isolated in 2004, physicists immediately tried to characterize its magnetotransport properties, which are still an important field of research today.
The purpose of this section is to recapitulate the present state of experimental knowledge about magnetotransport measurements in graphene. Firstly, we will describe the general experimental procedures that are necessary to carry such measurements. Secondly, we will briefly review the fundamental physics relevant to localization phenomena and to the QHE. Finally, we will show how these techniques provide information on the electronic properties of graphene through weak (anti-)localization, the integer QHE and the fractional quantum Hall effect (FQHE).

\subsection{Experimental procedures}

We first describe the basic experimental procedures required to perform low temperature magnetotransport measurements in graphene.

\subsubsection{Setup}

\begin{figure}[!ht]
\centering
\includegraphics[width=0.6\columnwidth]{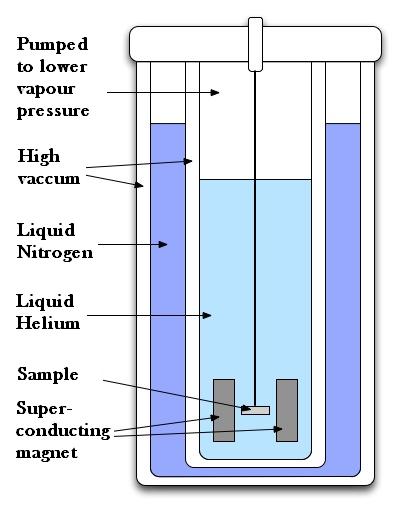}
\caption{\textbf{Setup for magnetotransport measurements.} The cryostat is cooled to $T$ = 4 K in two steps using liquid nitrogen and liquid helium. A solenoidal superconducting magnet creates a magnetic field perpendicular to the plane of the graphene. The helium bath can be pumped on to yield even lower temperatures.}
\label{QHEExpSetup}
\end{figure}

The setup consists of a superconducting electromagnet (typically Nb based type-II superconductor, e.g. NbTi or NbSn$_3$) immersed in a liquid helium cryostat. It is represented in \figref{QHEExpSetup}. Thus the sample can be refrigerated to temperatures of 4 K and below. Lower temperatures are achieved by reducing the vapor pressure of helium by pumping. Furthermore, in liquid helium the fields generated in the electromagnet can reach high maximal values typically ranging between 6 and 14 T.

\begin{figure}[!ht]
\centering
\includegraphics[width=0.6\columnwidth]{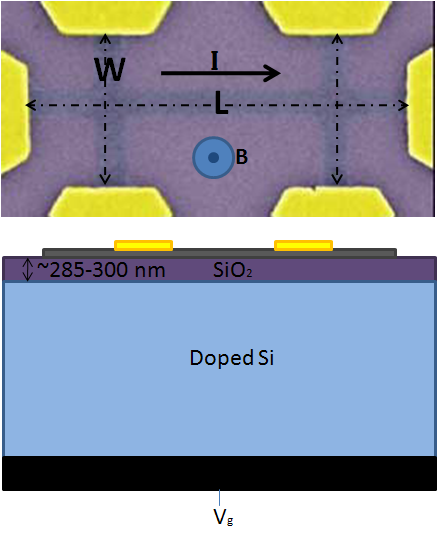}
\caption{\textbf{Magnetotransport measurements.} Top: Graphene is etched into a 4-terminal Hall bar. Gold electrodes are used to perform electrical measurements. The width of the central wire is 0.2 $\mu$m. Bottom: Graphene is placed on a thin SiO$_2$ layer (285-300 nm), which is itself on a doped Si wafer. A gate voltage $V_g$ allows the carrier density to be tuned. Figure from \cite{nov05}.}
\label{FourTerminalDevice}
\end{figure}

The sample is placed at the bottom of the cryostat in such a way that the field is perpendicular to the plane of the graphene. The device is usually conceived for 4-terminal measurements, as illustrated in \figref{FourTerminalDevice}. A small current is driven through the end terminals and voltage measurements are carried out between different combinations of the lateral terminals. This avoids the measurement of the contact resistance at the current injection points. The longitudinal resistivity $\rho_{xx}$ is the ratio of the longitudinal voltage to the current normalized by the aspect ratio. The \emph{Hall resistivity} $\rho_{xy}$ and the \emph{Hall Resistance} $R_{xy}$ are identical and are defined as the ratio of the transverse voltage to the current. The \emph{Hall conductivity} $\sigma_{xy}$ is related to the resistivity by the inverse tensor relation $\sigma_{xy} = \frac{\rho_{xy}}{\rho_{xy}^2+\rho_{xx}^2}$.

\subsubsection{Carrier density tuning}
\label{sec:CarrierDensityTuning}

In graphene, the carrier density $n$ can be tuned using the electric field effect, as discussed
 in \sectionref{ElectronicProperties}. A graphene sheet is placed on a SiO$_2$ substrate about 330 nm thick, which in turn lies on a doped Si wafer. A gate voltage is applied on the Si to inject electron carriers in the graphene or withdraw them. In the latter case, \emph{holes} carry the current. Thus graphene can be studied above and below the \emph{neutral point} corresponding to the Fermi energy and for which in principle $n=0$.

\begin{figure}[!ht]
\centering
\includegraphics[width=0.6\columnwidth]{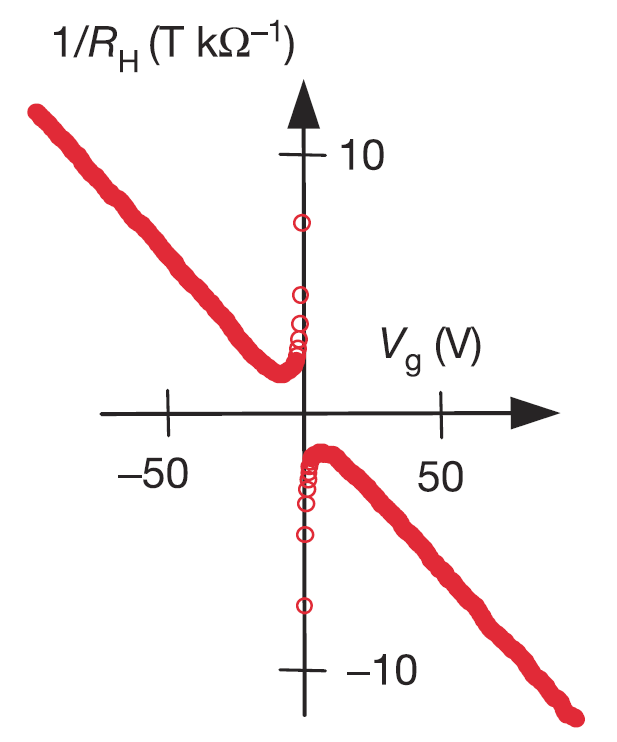}
\caption{\textbf{Measurement of the conventional Hall effect} for different gate voltages $V_g$ gives $V_g\propto n$. The Hall constant $R_H$ is related to the carrier density by $R_H=-\frac{1}{ne}$. Figure from \cite{nov05}.}
\label{Calibration}
\end{figure}

The calibration is done using the conventional Hall effect. In a (low) applied magnetic field, the Hall resistance is related to the current by $R_{xy}=R_H B I$, where $R_H=\frac{1}{n q}$ is the Hall constant and $q$ is the charge of the carriers. A measurement of $\frac{1}{R_H}$ for different gate voltages $V_g$ allows one to establish a linear map between the carrier density $n$ and $V_g$ for both holes and electrons. An example of calibration is shown on \figref{Calibration}.

\subsubsection{Observation of the quantum Hall effect}

The observation of the QHE is conditional on many factors.

\begin{enumerate}
  \item High quality samples must be used in order to maximize the momentum scattering time $\tau_q$ of the electrons.
  \item High magnetic fields $B$ are required in order for the cyclotron period of the carriers to be much shorter than the scattering time $\tau_q$. Equivalently, this means that the cyclotron frequency $\omega_c$ of an electron must be much higher than the broadening of the carrier energy levels (the \emph{Landau levels}, see \sectionref{Theory}): $\omega_c\tau_q \gg 1$.
  \item Low temperatures are needed in order for the thermal energy $k_BT$ to be much less than the spacing $\hbar\omega_c$ of the Landau levels: $k_BT<<\hbar\omega_c$.
\end{enumerate}

In graphene, the integer QHE can be observed at liquid helium temperature for fields ranging from about 1 T and higher. For high magnetic fields the effect can be observed for much higher temperatures \cite{nov05}. As an extreme example it has been shown that for a gigantic field of $B$ = 45 T, the QHE remains detectable at room temperature \cite{Novoselov2007}. The observation of the FQHE requires much more extreme conditions. The FQHE can be observed in graphene for fields above 2 T for temperatures lower than 1 K and it can subsist up to $T$ = 20 K at B = 12 T, provided that the sample is ultraclean and that the graphene is suspended to suppress scattering. This yields very low charge inhomogeneity of order $n=10^9$ cm$^{-2}$ \cite{Du2009}. In such small devices the contacts of the 4-terminal measurement prevent the appearance of the QHE and one must perform a two-terminal measurement instead \cite{Du2009}. More recently, FQHE measurements in graphene have been performed on graphene lying on a hexagonal boron-nitride substrate at fields of the order of B = 35 T and temperatures of the order of $T$ = 0.3 K \cite{Dean2010}. Such fields cannot be reached with the simple setup illustrated in \figref{QHEExpSetup}. Instead, these measurements are performed in High Magnetic Field Laboratories such as the National High Magnetic Field Laboratory at Florida State University or Grenoble High Magnetic Field Laboratory in France.


\subsection{Theoretical background}
\label{Theory}

\subsubsection{Chiral electrons and pseudospin}

Imagine a gas of electrons in the x-y plane confined by some potential well in the z-direction. We assume that this well is deep and narrow enough, and hence that the energy levels of the well are separated enough, so that the electrons are forbidden to access its excited states through any excitations. Then we say that the electron motion is two-dimensional.

At low carrier density $n$, the dispersion of graphene is linear around a Dirac point (call it K) so that we cannot define an effective mass such that $E=\frac{\hbar^2\vec{k}^2}{2m^*}$ as in semiconductor samples. Instead it can be shown that in the continuum limit (i.e. taking the position of the electron on the lattice to be a continuous variable), the motion of the electron around \emph{a single Dirac point} should be described by the Dirac-Weyl equations for massless fermions discussed in \sectionref{sec:BandStructure} using the Hamiltonian of \eqnref{HK} \cite{CastroNeto2009}:

\begin{align}
 \left( \hbar v_F \vec{\sigma}\cdot\vec{k} \right)\Psi_K = E\Psi_K, \label{DiracWeyl}
\end{align}

where $\Psi_K$ is a spinor. It can be shown that the state $\Psi_K$ acquires an extra phase (a \emph{Berry} phase) of $\pi$ on a closed trajectory \cite{CastroNeto2009}. This is key to understanding the weak localization measurements in graphene discussed below.

Since in graphene there are two inequivalent Dirac valleys K and K' with the same dispersion, there is an equation similar to \eqnref{DiracWeyl} for K'. This leads to an extra \emph{valley degeneracy} for the state of the electron and to chiral electrons: the state $\Psi_K$ in valley K has a different helicity than a state $\Psi_{K'}$ in valley K'. We call the doublet (K, K') the valley pseudospin. We conclude that the energy levels should be 4 times degenerate: two times for the pseudospin and two times for the \emph{actual} spin of the electron.

In a graphene \emph{bilayer}, two sheets of graphene are stacked onto each other (in the natural \emph{Bernal stacking} of graphite) \cite{CastroNeto2009}. In such a system, the carriers can hop between the layers and the dispersion relation is not linear. Instead, the valence and conduction bands consist of two parabolic bands of the same curvature touching at the neutral point \cite{CastroNeto2009}. The theory of \eqnref{DiracWeyl} can then be extended to show that the carriers should be \emph{massive} Dirac fermions with a Berry phase of $2\pi$ \cite{Novoselov2006}.

\subsubsection{Landau levels}

\eqnref{DiracWeyl} can be solved in the presence of a magnetic field and the result predicts a sequence of energy levels called the \emph{Landau levels} (LL) \cite{Ezawa2008,CastroNeto2009,Gusynin2005,Apalkov2006}:

\begin{align}
  E_N = sgn(N)\sqrt{2e\hbar v_F^2 \left|N\right|} \ \ N=0,\pm1,\pm2... \label{LLGraphene}
\end{align}

\begin{figure}[!ht]
\centering
\includegraphics[width=0.8\columnwidth]{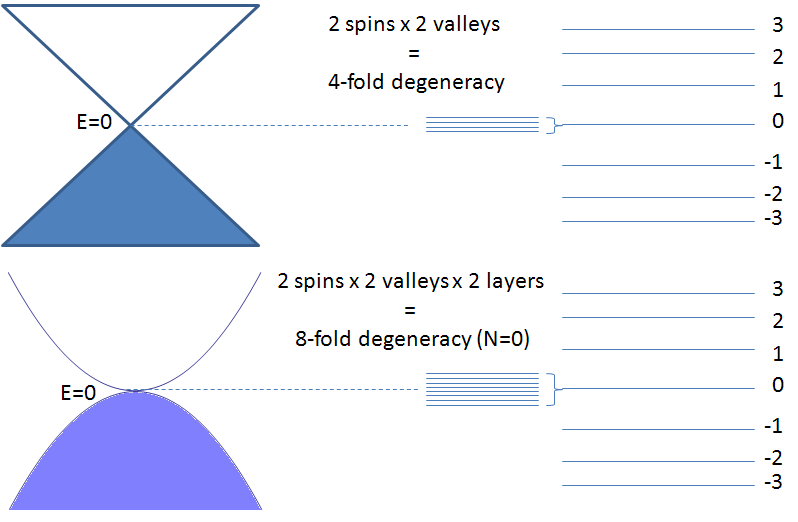}
\caption{\textbf{Schematic representation of the formation of Landau levels} in graphene (top) and bilayer graphene (bottom). In both cases the levels are not equally spaced and the neutral point has an associated LL. In graphene each level is 4 times degenerate because of the spin and valley pseudospin. In bilayer graphene the lowest LL $N = 0$ is 8 times degenerate because of the extra layer degree of freedom.}
\label{BandSplitting}
\end{figure}

The energy is measured with respect to the Dirac point (i.e. the energy at the Dirac point is zero). Each such level should also have the 4-fold degeneracy discussed above. Note that in a semiconductor 2DEG, the LLs would be equally separated by an amount $\hbar\omega_c$ where $\omega_c=\frac{e \left|B\right|}{m_c}$ is the cyclotron frequency and $m_c$ the cyclotron mass \cite{Datta1997}, which from \eqnref{LLGraphene} is clearly not the case for graphene. We sketch a figure of the splitting of the bands in LLs for graphene in \figref{BandSplitting}.

For bilayer graphene, we find a different expression:

\begin{align}
  E_N = sgn(N)\hbar\omega_c\sqrt{\left|N\right|\left(\left|N\right|+1\right)} \ \ N=0,\pm1,\pm2... \label{LLBGraphene}
\end{align}

It can be shown that the extra layer degree of freedom should give rise to an extra degeneracy for the level $N = 0$ and \emph{only} for that level. Thus, the $N = 0$ level is 8 times degenerate while the others keep the 4-fold degeneracy of graphene. The splitting of the bands in LLs for bilayer graphene is illustrated schematically in \figref{BandSplitting}.


\subsection{Measurements of weak localization and antilocalization in graphene}

\subsubsection{Quantum interferences}

Quantum interferences have long been known to affect transport measurements at low temperatures. In particular, two-dimensional electron systems in the presence of a (low) magnetic field show variations of conductance with respect to the value in the absence of a field. This effect is known as \emph{weak (anti-)localization}.

Consider an electron hitting an impurity and going back in the direction from which it came. In a classical picture, this could happen in a number of ways. In particular, the electron could go around the impurity clockwise or counter-clockwise (see \figref{TransitionWL}). However, since electrons are quantum mechanical waves, we must add each of the possible paths and make them interfere together to get the net probability of the electron to have backscattered. Electrons in conventional 2DEGs, in the absence of spin-orbit interaction and scattering by magnetic impurities, gain the same phase on both trajectories and interfere constructively, leading to an increased probability of backscattering compared to the value expected from the Drude model. We say that the electrons exhibit \emph{weak localization} (WL). If a magnetic field is applied, an additional (Aharonov-Bohm) phase is added between the two paths, destructive interference occurs and the conductance \emph{increases} (positive magnetoconductance) \cite{Tikhonenko2009}.

In graphene, however, it can be shown that because of the additional Berry phase of $\pi$ of the wave function, the two paths end up in opposite phase in the absence of spin-orbit interaction and scattering within or between Dirac valleys. Therefore application of a magnetic field can only restore the constructive interference and \emph{decrease} the conductance (negative magnetoconductance) \cite{Morozov2006,Tikhonenko2009}. We say that the electrons exhibit \emph{weak anti-localization} (WAL). Note that conventional electrons which couple strongly through spin-orbit interactions can also exhibit WAL. However, this effect is not expected in graphene because of the small mass of the carbon atom \cite{Tikhonenko2009}.

\subsubsection{Measurements}

\begin{figure}[!ht]
\centering
\includegraphics[width=\columnwidth]{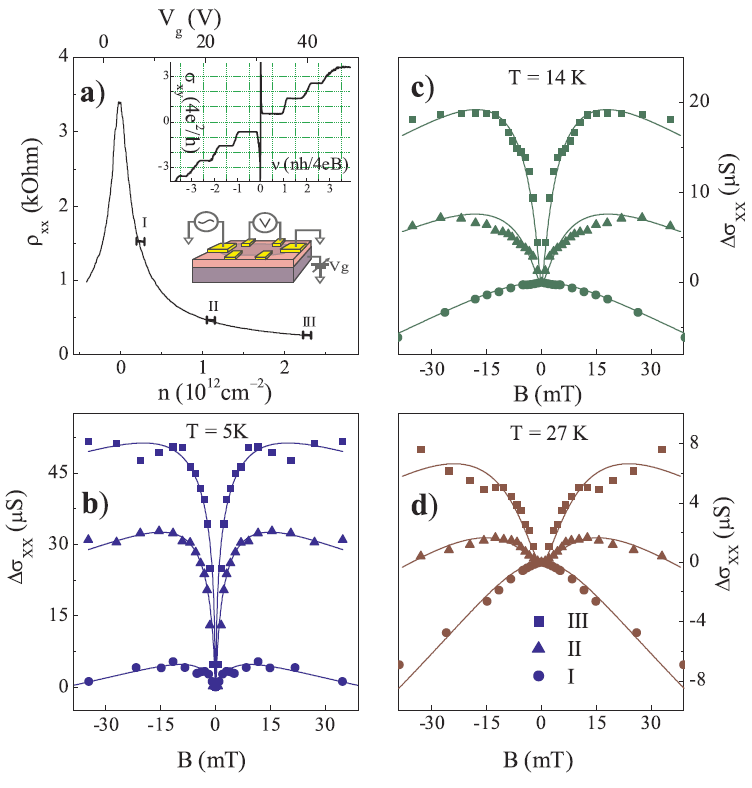}
\caption{\textbf{Weak localization and anti-localization in graphene.} (a) The sample is studied for various carrier densities (here I, II and III). The QHE shows that the sample is indeed two-dimensional (see \sectionref{QHE}). The sample is annealed at $T\approx$ 400 K for 2 hours to increase its homogeneity. (b), (c) and (d) Variation of the magnetoresistance behavior as $n$ and $T$ are changed. A transition occurs between WL and WAL and negative magnetoconductance (WAL) is observed for low $n$ and high $T$ due to the competition between valley scattering and dephasing processes (see text). Figure from \cite{Tikhonenko2009}.}
\label{WeakLocalization}
\end{figure}

The early quantum interference measurements on graphene showed no WAL and a strongly suppressed WL. This was attributed to suppression of interference within one Dirac valley due to ripples and large defects \cite{Morozov2006}. More recently, it was demonstrated that both WAL and WL could be observed in graphene under the proper conditions \cite{Tikhonenko2009}. The behavior of the magnetoconductance is shown on \figref{WeakLocalization}.

\begin{figure}[!ht]
\centering
\includegraphics[width=\columnwidth]{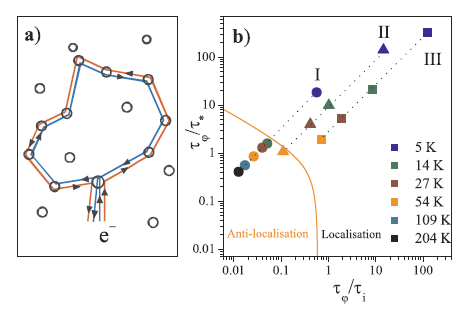}
\caption{\textbf{Phase diagram for weak (anti-)localization in graphene.} (a) A schematic representation of interference between electron paths. For graphene it is expected to be naturally destructive and a magnetic field is expected to partly restore constructive interference (see text). (b) A phase diagram for weak localization behavior in graphene. It is seen that the ratios of the dephasing time to the two valley scattering times determine the magnetoresistance behavior. Figure from \cite{Tikhonenko2009}.}
\label{TransitionWL}
\end{figure}

The data shows WAL (negative magnetoconductance) for \emph{both} decreasing carrier density and increasing temperature. The increasing temperature reduces the dephasing time $\tau_{\phi}$ of the phase of the electron due to thermal fluctuations, while decreasing carrier density increases the intervalley and intravalley elastic scattering times $\tau_i$ and $\tau_*$ defined in the theory of reference \cite{McCann2006}. The latter are roughly temperature independent. Thus it is observed that the dephasing time is not the only parameter controlling weak localization behavior in graphene. Instead, one must consider the ratios $r_i=\frac{\tau_{\phi}}{\tau_i}$ and $r_*=\frac{\tau_{\phi}}{\tau_*}$. When they are small, anti-localization occurs while when they are big, localization occurs. Note that there exist combinations of these parameters for which the correction to the magnetoconductance is suppressed. This is shown in \figref{TransitionWL}.

In graphene, the quantum corrections to the conductance survive at much higher temperatures than for 2DEG semiconductor structures because the electron-phonon scattering is expected to be weak in the system. Indeed, weak (anti-)localization can be observed up to much high temperatures in graphene, in particular in large scale CVD grown graphene \cite{Whiteway2010}. It turns out that electron-electron scattering could be responsible for the disappearance of magnetoconductance at high temperatures \cite{Tikhonenko2009}.

\subsection{Measurements of the quantum Hall effect in graphene}
\label{QHE}

Here we review the principal features of the QHE in graphene and compare them to the conventional results. We first discuss the general features of the QHE.

\subsubsection{Shubnikov-de Haas oscillations}

\begin{figure}[!ht]
\centering
\includegraphics[width=\columnwidth]{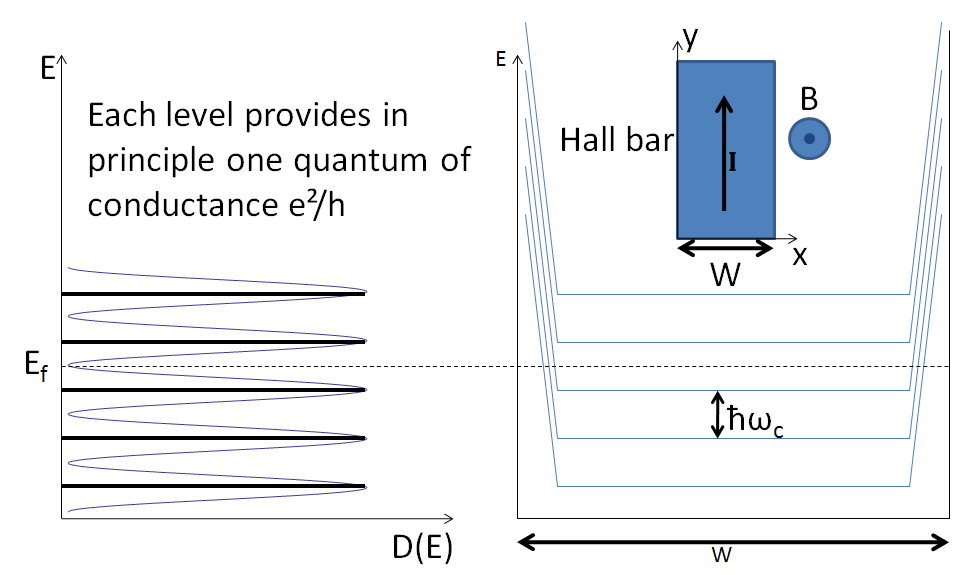}
\caption{\textbf{Quantization of Hall conductance in graphene.} (a) The density of states is peaked at the LLs and broadened by the presence of disorder. (b) Edge states topologically separated by the magnetic field conduct the current between the leads at fixed potential. In the absence of scattering, the quantum of conductance $e^2/h$ carried by each channel translates into the quantization of the Hall conductivity $\sigma_{xy}$ (see text).}
\label{HallConductance}
\end{figure}

The existence of LLs implies the appearance of the so-called Shubnikov-de Haas oscillations (SdHO) in the QH system. The SdHO consist of a strong modulation of the longitudinal resistivity $\rho_{xx}$ as a function of the carrier density $n$. The minima of these peaks correspond to a completely vanishing longitudinal resistivity. This occurs when the chemical potentials of both leads sit between the LLs, as illustrated in \figref{HallConductance}. Then all the $\vec{k}$ carrying states of the LLs are completely filled so that the electrons cannot carry current in a LL. The only states that carry current are the \emph{edge states} at the Fermi level (see \figref{HallConductance}). Furthermore, the magnetic field is such that each edge carries current in different directions. Hence, the only way for backscattering to occur is for an electron to go from one edge to another, which is not feasible. Thus scattering is suppressed and no voltage can build up, hence $\rho_{xx}$ vanishes. As the Fermi level is raised, a LL may sit at the chemical potential of the leads, which allows for some backscattering, hence producing a maximum in resistivity.

\subsubsection{Conductance quantization}

Associated with the SdHO is the celebrated phenomenon of Hall conductance quantization. It is observed that at the values of $n$ for which the zeros in $\rho_{xx}$ occur, the Hall resistivity $\rho_{xy}$ is constant and forms a plateau that is a multiple $\nu^{-1}$ of $h/e^2$. The integer $\nu$ is said to be the filling factor of the QH state and represents the total number of levels below the Fermi energy. In fact, there are as many conducting edge channels as there are filled levels, and one can show that each such channel should contribute one \emph{quantum of conductance} $\frac{e^2}{h}$ \cite{Datta1997}. In the absence of backscattering, one expects each edge to be at the same potential as the lead that provides its carriers: this implies that the transverse Hall voltage is the same as the potential difference between the leads. Therefore the Hall conductance should be the same as the sum of the conductance of each of these $\nu$ channels. Thus:

\begin{align}
  \nu = \frac{G_{xy}}{\left(e^2/h\right)}
\end{align}

This phenomenon is known as the quantum Hall effect. The reader is referred to \cite{Datta1997} for a thorough theoretical treatment of ballistic transport in micron scale samples.

\subsubsection{Integer quantum Hall effect}

\paragraph{Monolayer graphene}
For the plateaus of the conventional QHE in semiconductor heterostructures, $\nu$ is a positive integer and a finite number of charge carriers is required to occupy the lowest LL. At low magnetic field, only the even filling factors appear because the LLs can be filled with both spin up and spin down electrons. We have the sequence:

\begin{align}
  \nu = 2,4,6,8,10...
\end{align}

Therefore the sequence of $\nu$'s provide information on the degeneracies of the electron states or equivalently, on their symmetries. If the magnetic field is increased, the degeneracies can be completely lifted via Zeeman interaction of the spin with the magnetic field, revealing the entire sequence of filling factors.

\begin{figure}[!ht]
\centering
\includegraphics[width=0.8\columnwidth]{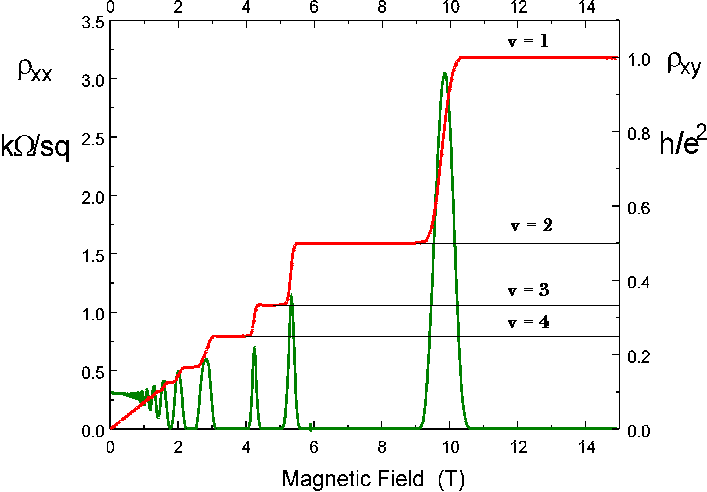}
\caption{\textbf{Typical measurement of the integer quantum Hall effect.} The Hall resistivity exhibits plateaus quantized in exact multiples of $\frac{h}{e^2}$.  Image created by D.R. Leadley, Warwick University (1997).}
\label{IntegerQHE}
\end{figure}

\begin{figure}[!ht]
\centering
\includegraphics[width=0.8\columnwidth]{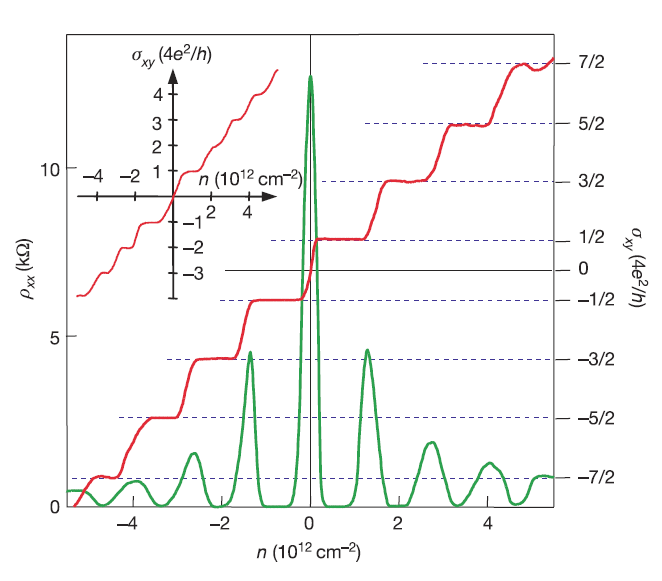}
\caption{\textbf{Integer quantum Hall effect in graphene.} The carrier density $n$ can be tuned across the neutral point $n=0$. The absence of a plateau at $n=0$ indicates the presence of a LL at the neutral point. The inset shows bilayer graphene, discussed in \sectionref{QHEBilayer}. Figure from \cite{nov05}.}
\label{IntegerQHEGraphene}
\end{figure}

As a two-dimensional system, graphene is expected to exhibit the QHE. It was observed in 2005 by two research groups \cite{nov05,Zhang2005} and showed features that were characteristic of the linear dispersion of graphene. A measurement of the conventional QHE in graphene is shown in \figref{IntegerQHEGraphene}.

The sequence of observed plateaus is very different from that of the conventional QHE. The precise sequence is:

\begin{align}
  \nu = \pm2,\pm6,\pm10...
\end{align}

We notice that the carrier density (i.e. the Fermi level) of graphene can be tuned through both the valence and conduction band of graphene, resulting in negative $\nu$ for holes and positive $\nu$ for electrons. From the sequence we immediately see that each plateau contributes $4e^2/h$ to the conductance, instead of $2e^2/h$. This is indicative of the expected extra valley degeneracy of the states, in addition to the spin degeneracy. Another interesting feature of this sequence is that there is no plateau at zero carrier density. Thus there is a LL $N=0$ at the neutral point with degenerate holes and electrons as seen from \eqnref{LLGraphene}. The structure of electron and hole states is also completely symmetric.

\begin{figure}[!ht]
\centering
\includegraphics[width=0.8\columnwidth]{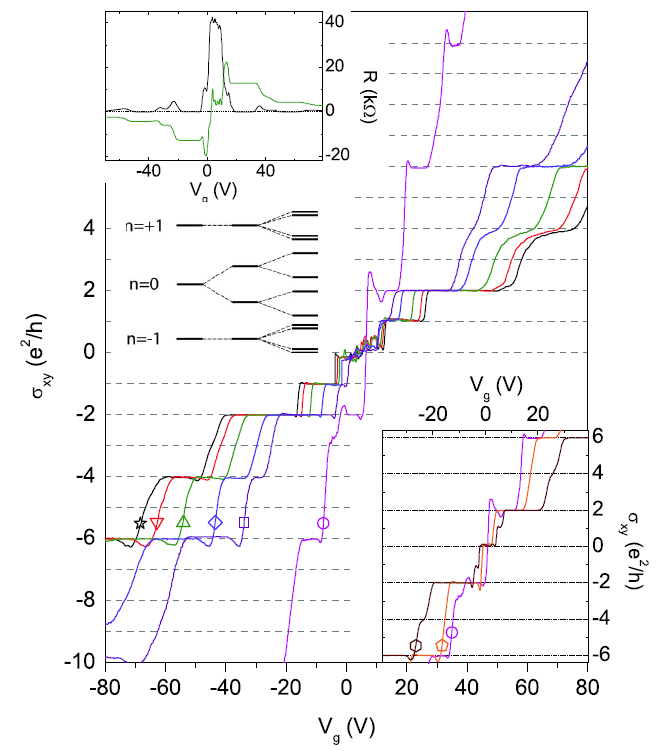}
\caption{\textbf{Complete lifting of the degeneracy of the lowest LL} with fields from 9 T (circle) up to 45 T (star). The expected 4-fold degeneracy is shown explicitly by the presence of plateaus at $\nu=0$ and $\nu=\pm 1$. The plateaus at $\nu=\pm4$ show partial degeneracy lifting of the levels $N=\pm1$. All degeneracies can be lifted in clean samples, see \figref{FQHEGraphene}. Figure from \cite{Zhang2006}.}
\label{LLSplittingZhang}
\end{figure}

\figref{LLSplittingZhang} shows a complete lift of the degeneracy of the $N=0$ LL at ultrahigh magnetic fields, confirming explicitly the 4-fold degeneracy of the level \cite{Zhang2006}. The Zeeman coupling in graphene is too weak to cause the lifting of the spin degeneracy and does not explain the breaking of the valley pseudospin symmetry \cite{Zhang2005,Zhang2006}. Instead, the lifting of the degeneracies is suspected to be caused by enhancement of Coulomb and exchange interactions in the sample. This is due to the fact that as the field is increased the cyclotron orbits become smaller and the carriers come closer to each other \cite{Dean2010,Ezawa2008}. All integer filling fractions can be observed for the cleanest samples \cite{Dean2010}.

Just like the conductance quantization, the SdHO oscillations show unusual behavior. We still observe their minima at the positions of the plateaus, as seen in \figref{IntegerQHEGraphene}.

\begin{figure}[!ht]
\centering
\includegraphics[width=\columnwidth]{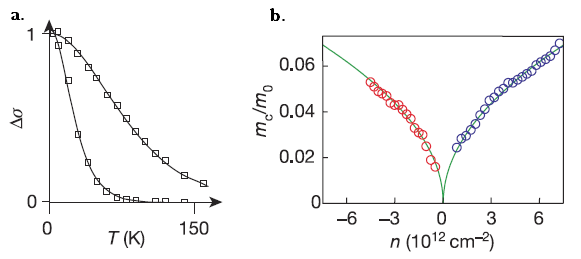}
\caption{\textbf{Determination of the dispersion of graphene through SdHO oscillations.} (a) The SdHO decay more rapidly with temperature as the carrier density is increased. The data can be fitted to yield the cyclotron mass $m_c$. (b) The cyclotron mass varies with carrier density as $\sqrt{n}$, implying the linear dispersion of graphene. Figure from \cite{nov05}.}
\label{CyclotronVsCarrier}
\end{figure}

However, their amplitude decays more rapidly with temperature as the density $n$ is increased. Theoretical studies have shown that for a given $n$, the amplitude $A$ of the SdHO should follow \cite{CastroNeto2009}:

\begin{align}
  A \propto \frac{T}{\sinh\frac{2\pi^2 T k_B m_c}{\hbar e B}}  
\end{align}

where $m_c$ is the cyclotron mass of the electrons at the Fermi level \cite{nov05}.

A fit to the data shows that the rapid decay of the SdHO implies that $m_c \propto \sqrt{n}$, as shown in \figref{CyclotronVsCarrier}. Unlike in semiconductor devices, the cyclotron mass in graphene varies with density. It is possible to show that this dependence of $m_c$ on $n$ implies the linear dispersion of graphene $E(k) = \hbar v_F k$ and thus the form of the LLs given in \eqnref{LLGraphene} \cite{nov05}.

\paragraph{Bilayer graphene}
\label{QHEBilayer}

\begin{figure}[!ht]
\centering
\includegraphics[width=\columnwidth]{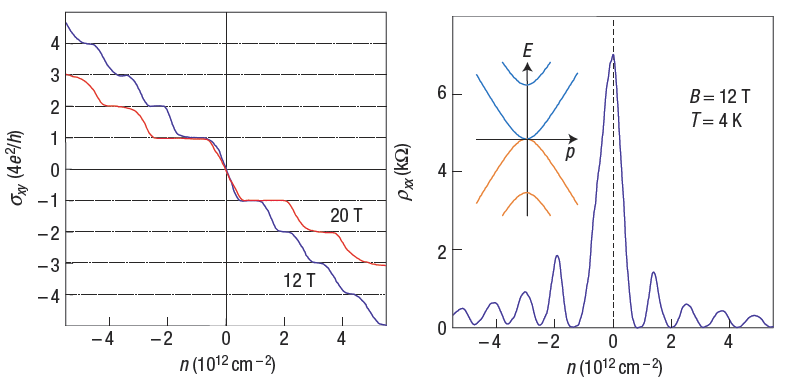}
\caption{\textbf{Quantum Hall effect in bilayer graphene.} As in the case of the monolayer, the quantum Hall effect in graphene bilayer shows no plateau at $\nu=0$, indicating the presence of a LL at zero carrier density and the relativistic behavior of the bilayer. However, contrarily to the monolayer, the $N = 0$ level is 8-fold degenerate as the height of the step indicates. This is consistent with the theoretical prediction of a ``parabolic Dirac point'' at $\nu = 0$, i.e. \emph{massive} Dirac fermions. Figure from \cite{Novoselov2006}.}
\label{bilayerQHE}
\end{figure}

The quantum Hall effect can also be observed in multilayer graphene. The inset of \figref{IntegerQHEGraphene} as well as \figref{bilayerQHE} shows measurements carried out on the graphene bilayer \cite{Novoselov2006}. The observed sequence is as follows:

\begin{align}
  \nu = \pm4,\pm8,\pm12,\pm16...
\end{align}

\begin{figure}[!ht]
\centering
\includegraphics[width=\columnwidth]{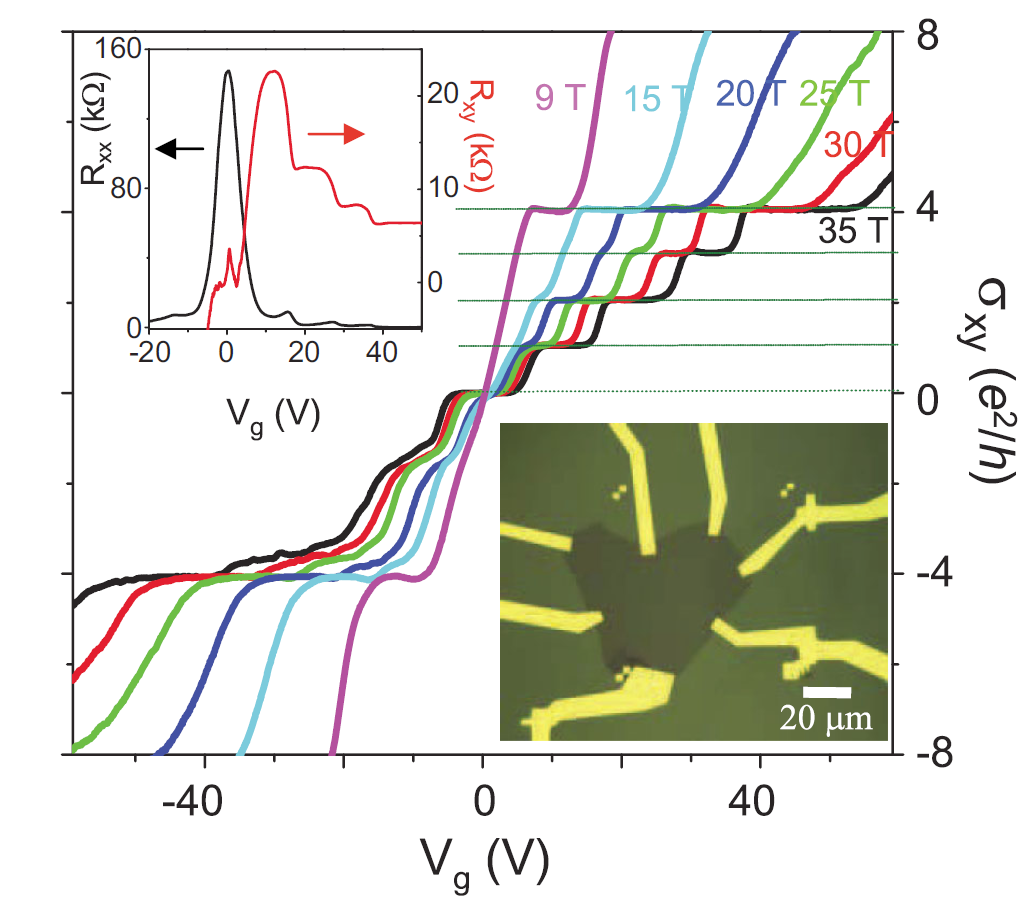}
\caption{\textbf{The degeneracy of the lowest LL of bilayer graphene is lifted by magnetic fields ranging from 9 T to 35 T.}  We clearly distinguish for levels above the neutral point and 4 others start to be visible below the neutral points, accounting for the expected 8-fold degeneracy of this level. Figure from \cite{Zhao2010}.}
\label{LLSplittingZhao}
\end{figure}

The absence of a plateau at $\nu = 0$ is a common characteristic of the QHE in graphene and bilayer graphene. Once again it indicates the presence of a LL at zero energy. However, this level is eight times degenerate as can be seen from the $8e^2/h$ jump in the Hall conductivity. The other levels keep the 4-fold degeneracy of graphene. This is what we expect from the additional \emph{layer} degree of freedom of the electron for a given spin and valley \cite{Novoselov2006, Ezawa2008}. The degeneracy of the lowest LL can be completely lifted by many-body interactions in ultra-high magnetic fields \cite{Zhao2010}, as shown in \figref{LLSplittingZhao}.

\subsubsection{Fractional quantum Hall effect}

The extreme two-dimensional confinement in graphene is expected to enhance many-body interactions between electrons. Since 1982, we know that these interactions can manifest through the so called fractional quantum Hall effect (FQHE), consisting of the appearance of Hall plateaus at fractional (rational) values of the filling factor $\nu$ \cite{Tsui1982}. The efforts to observe the FQHE in graphene were without success for a long time because of the existence of a competing insulating state induced by disorder near the neutral point $\nu = 0$. The improvements in the sample fabrication made its observation possible in recent years.

\begin{figure}[!ht]
\centering
\includegraphics[width=\columnwidth]{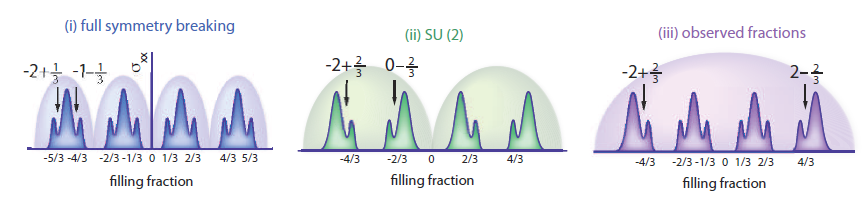}
\caption{\textbf{Schematic SdHO structures corresponding to different possible symmetries of the lowest LL in graphene.} (a) Total symmetry breaking. (b) Either the spin or the valley symmetry is broken. (c) Full symmetry of the spin and valley degrees of freedom. The latter is the best fit to the observations. Figure from \cite{Dean2010}.}
\label{Symmetries}
\end{figure}

FQH states can be understood as the realization of the integer QHE for weakly interacting quasiparticles called composite fermions. In a heuristic picture, an even number of magnetic flux vortices bind with an electron to form an object with reduced effective charge that is a fraction of the elementary charge $e$ \cite{Jain1989, Dean2010}. In graphene, the multiple components of the electron wave function are expected to give rise to new interacting ground states with various spatial arrangements of pseudospin (textures) that depend on the symmetries of the states. These special states are theoretically predicted to occur at filling factors that are multiples of $\frac{1}{m}$, where $m$ is an odd integer \cite{Ezawa2008}. The FQHE allows for a characterization of these symmetries and puts hard constraints on theoretical models. Different symmetry scenarios are presented in \figref{Symmetries}.

\begin{figure}[!ht]
\centering
\includegraphics[width=\columnwidth]{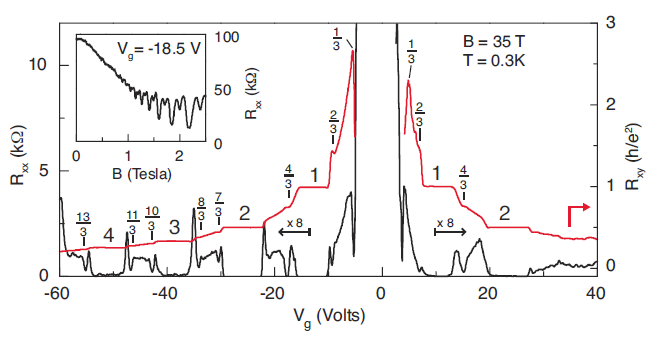}
\caption{\textbf{The FQHE in graphene.} Multiple plateaus at rational filling factors show that electron interactions give rise to new states. Figure from \cite{Dean2010}.}
\label{FQHEGraphene}
\end{figure}

\figref{FQHEGraphene} shows a clean and recent 6-terminal measurement of the FQHE in graphene made on a hexagonal boron-nitride substrate at ultrahigh fields \cite{Dean2010}. The filling factors appear in the following sequence:

\begin{align}
  \nu = \pm\frac{1}{3}, \pm\frac{2}{3}, \pm1, \pm\frac{4}{3}, \pm2, \frac{7}{3}, \frac{8}{3},3,\frac{10}{3},\frac{11}{3},4,\frac{13}{3} ...
\end{align}

The measurements for hole carriers were not performed for $\nu < -2$. The filling factors are all multiples of $\frac{1}{3}$, although $\frac{5}{3}$ is missing. Furthermore, a peak in the SdHO may indicate the emergence of a plateau at $\nu=\frac{8}{5}$. The effect is much more robust under increase in temperature than its counterpart in semiconducting devices, revealing the suspected enhancement of electron-electron interactions in graphene compared to those occurring in 2DEGs \cite{Dean2010,Du2009}. The absence $\nu=\pm\frac{5}{3}$ and the presence of $\nu=\pm\frac{1}{3}$ is consistent with a global $SU(4)$ symmetry of the FQH state in the lowest LL \cite{Dean2010}.

\begin{figure}[!ht]
\centering
\includegraphics[width=\columnwidth]{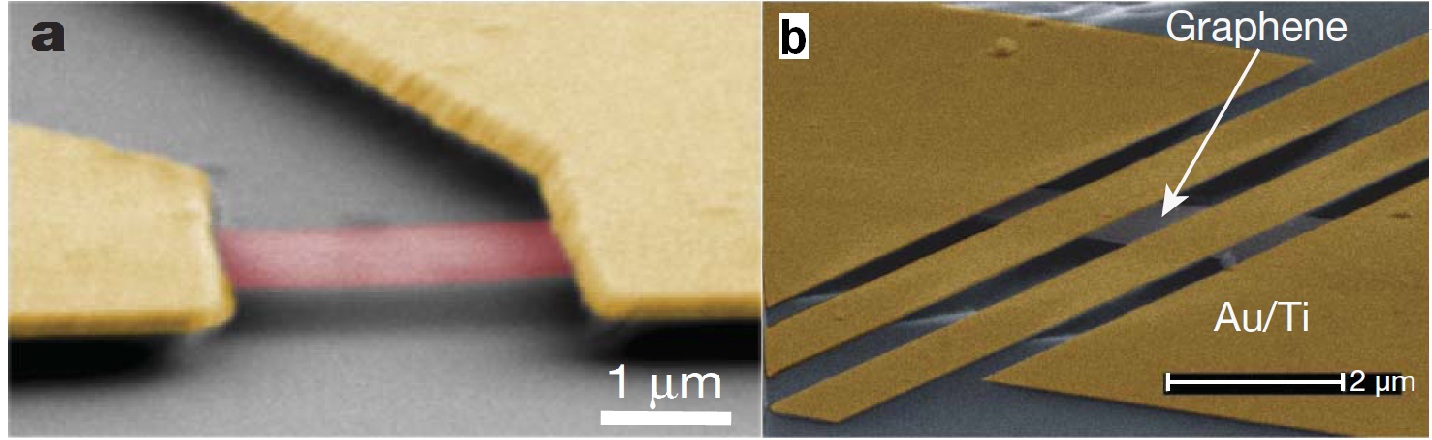}
\caption{\textbf{High quality suspended graphene samples} used by Bolotin \textit{et al.} (left) and Du \textit{et al.} (right). The contacts are less than 2 $\mu$m apart in both cases, making a 4-terminal measurement impossible. However, the FQHE can be observed clearly at relatively low fields ranging from 5 T to 12 T. Figures from \cite{Bolotin2009,Du2009}.}
\label{Suspended}
\end{figure}

Previous measurements have been performed on very small ($\leq$2 $\mu$m) 2-terminal ultraclean samples of suspended graphene \cite{Du2009, Bolotin2009}. For a discussion of the properties of suspended graphene see \sectionref{sec:SuspendedGraphene}. Their results are presented in \figref{Suspended}. They also obtain the FQHE in the lowest LL at $\nu=\frac{1}{3}$ and $\nu=\frac{2}{3}$. While 2-terminal measurements make the results less precise and harder to interpret (for example, the contact resistance makes it very hard to extract information from the longitudinal resistance data and may produce unusual features at $\nu=\frac{1}{2}$), they possess the advantage of being realizable with lower field magnets in standard physics laboratories.

The observation of the anomalous quantum Hall effect, including its fractional daughter are one of the brightest highlights, which propelled graphene as one of the most studied materials in recent years. We now move away from magneto-transport to discuss the mechanical properties in the next section, which are very important for future applications in nanodevices.




\section{Mechanical Properties}
\label{MechanicalProperties}

Microelectromechanical systems (MEMS) have been deployed to perform common tasks such as opening and closing valves, regulating electric current, or turning mirrors. This is realized by employing microscopic machines such as beams, cantilevers, gears and membranes. These MEMS are found in many commercially available products from accelerometers found in air bag deployment systems, gyroscopes in car electronics for stability control and ink jet printer nozzles \cite{craighead00}. The nanoscopic version of MEMS, nanoelectromechanical systems (NEMS), demonstrate their own advantages in engineering and fundamental science in areas such as mass, force and charge detection \cite{ekinci05}. All these electromechanical devices are functional only in response to an external applied force. The utmost limit would be a one atom thick resonator. Robustness, stiffness and stability is thus important when reaching this limit \cite{bunch07}.

The fabrication of graphene-based mechanical resonators is still under active development. Presented here is an overview of experimental results for single and multilayer graphene sheets placed over predefined trenches. A simple drive and detection system is used to probe the mechanical properties of these graphene resonators in order to extract the fundamental resonance frequency. This allows us to characterize the quality factor, Young's modulus and built-in tension \cite{bunch07}. One can also extract this information from a separate experiment using an atomic force microscope (AFM) tip \cite{frank07}. In addition to their use as mechanical resonators, graphene sheets are robust (impermeable) enough to act as a thin membrane between two dissimilar environments \cite{bunch08}.

\subsection{Overview of the harmonic oscillator}
In order to study the mechanical properties of graphene and graphite sheets, one needs to comprehend the basics of a harmonic oscillator. A classic example is a mass attached to a spring. In an ideal situation, the spring obeys Hooke's law, but for large displacements, the spring does not follow Hooke's law as the system is subject to damping, denoted by $\gamma$, which dissipates the vibrational energy of the system. In addition, a driving force $F$ with frequency $\omega$ is necessary to drive oscillations to the system. Therefore, the equation of motion for a damped harmonic oscillator is given by:

\begin{equation}
m\frac{d^2x}{dt^2} + \gamma\frac{dx}{dt} + m\omega_0^2x = F\cos(\omega t).
\label{eq:driving_f}
\end{equation}

The general solution is given by:

\begin{equation}
x = A\cos(\omega t - \theta)
\end{equation}

where $\theta$ is the phase shift, the amplitude

\begin{equation}
A = \frac{F}{m\omega_0^2}\frac{Q}{\sqrt{Q^2(1-\frac{\omega^2}{\omega_0^2})^2 + (\frac{\omega}{\omega_0})^2}},
\label{eq:amplitude}
\end{equation}


and the quality factor

\begin{equation}
Q = \frac{m\omega_0}{\gamma}.
\end{equation}

For a given peak with resonance frequency $\omega_0 = 2 \pi f_0$, the peak amplitude is $QF/m\omega_0^2$ and its full-width half maximum (FWHM) determines the Q factor, given by {\it f$_0$/Q}.

\subsection{Tuning resonance frequency by electrical actuation}

\begin{figure}
  	\centering
      	\includegraphics[scale=0.5]{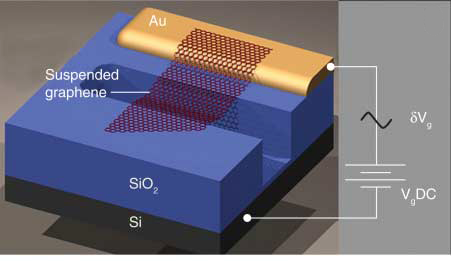}
      	\caption{\textbf{Schematic of a suspended graphene resonator for electrical actuation}  \cite{bunch07}.}
  	\label{fig:resonator1}
\end{figure}

The mechanical resonators are exfoliated graphene sheets that are suspended over trenches on SiO$_2$/Si substrates. These micron-sized sheets are doubly-clamped beams which are secured to the SiO$_2$ surface via van der Waals forces. A gold electrode defined by photolithography is used to electrically actuate the graphene resonator, filling the role of the driving force $F$ from \eqnref{eq:driving_f}. All resonator measurements are performed in ultra-high vacuum at room temperature. The actuation is an applied alternating electric field used to drive the resonant motion of the beam. A capacitor is then formed between the bottom gate electrode and the contacted graphene, shown in \figref{fig:resonator1}. A voltage $V_g$ applied to the capacitor induces electric charges onto the beam. A small time-varying radio frequency (RF) voltage $\delta V_g$ at frequency $f$ is used to electrically modulate the beam on top of a constant DC voltage \cite{bunch07}:

\begin{equation}
V_g = V_g^{DC} + \delta V_g.
\end{equation}

The resulting electrostatic force is then given by:

\begin{equation}
F_{el} = \frac{1}{2}\frac{dC_g}{dz}V^2_g \approx \frac{1}{2}\frac{dC_g}{dz}(V_g^{DC})^2 + \frac{dC_g}{dz}V_g\delta V_g
\label{eq:electro}
\end{equation}

where z is the distance between the graphene and the gate electrode.

The response due to a varying RF voltage is monitored by a 632 nm He-Ne laser focused on the resonator. As a result, the light creates an interference between the suspended graphene sheets and the silicon back plane. Variations of the reflected light intensity are monitored by a photodiode.

\begin{figure}[!h]
  	\centering
      	\includegraphics[scale=0.36]{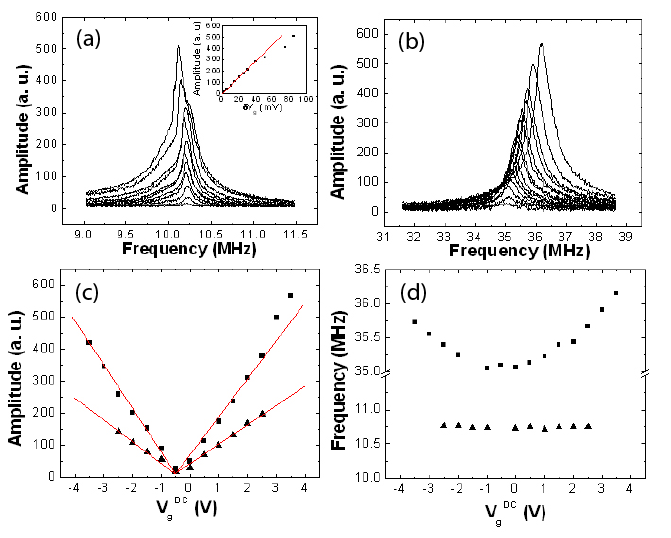}
      	\caption{\textbf{Resonance spectrum taken with an electrostatic drive.} Plot of the amplitude versus frequency of the (a) fundamental mode at 10 MHz and (b) the higher mode at 35 MHz with increasing $\delta V_g$ while $V_g^{DC}$ remains fixed. Inset: Plot of the resonant peak amplitude with increasing $\delta V_g$. Dependence of the (c) amplitude and (d) frequency as function of $V_g^{DC}$ at fixed $\delta V_g$ where the solid squares and triangles represent the fundamental and higher mode respectively \cite{bunch07}.}
  	\label{fig:data_el}
\end{figure}

Plugging \eqnref{eq:electro} into \eqnref{eq:amplitude}, one can speculate on the behavior of the resonant peak as a function of applied voltage. The data for a few-layer graphene stack is shown in \figref{fig:data_el} for electrical drive on resonance. One can see that the amplitude and frequency increase linearly with $\delta V_g$ at a fixed $V_g^{DC}$ for the higher mode (\figref{fig:data_el}B) while the frequency of the fundamental mode (\figref{fig:data_el}A) does not change significantly. In addition, if $V_g$ is increased while $\delta V_g$ is fixed (\figref{fig:data_el}C), both the amplitude of the fundamental mode and that of the higher mode increase linearly as expected from \eqnref{eq:amplitude}. As for the frequency (\figref{fig:data_el}D), it seems that there is evidence of positive tuning for the higher mode but that the fundamental mode remains unaffected. A possible explanation for the tuning may originate from an electrostatic attraction to the gate which increases the tension from stretching the graphene.

\subsection{Resonance spectrum by optical actuation}

Another technique to actuate vibrations in resonators is an optical drive. A diode laser is shined on the suspended graphene in which the graphene actuates by itself via thermal expansion and contraction. This motion is controlled by the intensity of the diode laser set to a frequency $f$. This is then monitored by a photodiode using a He-Ne laser through the procedure given in the previous section \cite{bunch07}.\\

\begin{figure}[!h]
  	\centering
      	\includegraphics[scale=0.35]{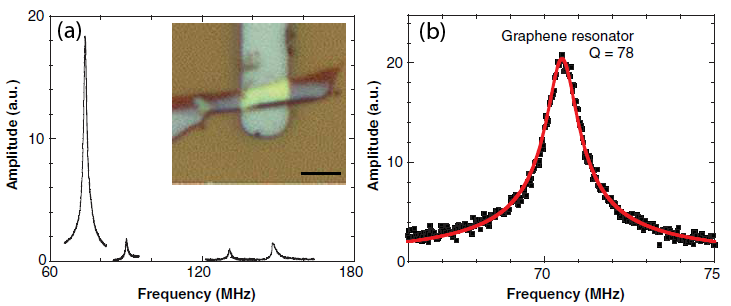}
      	\caption{\textbf{Plots showing the resonance spectrum of graphene and graphite resonators.} (a) Resonance spectrum showing the fundamental mode and higher modes for a 15 nm thick multilayer graphene sheet taken with optical drive. Inset: Optical image of the resonator. Scale bar = 5 $\mu$m. (b) Amplitude vs. frequency for a single layer graphene. The red curve represents the Lorentzian fit of the data \cite{bunch07}.}
  	\label{fig:data_opt}
\end{figure}

\begin{figure}[!h]
  	\centering
      	\includegraphics[scale=0.35]{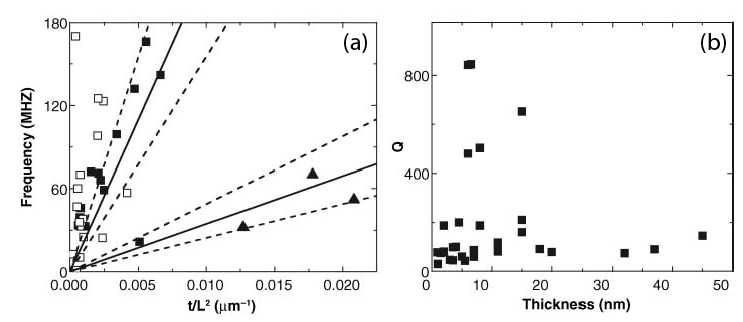}
      	\caption{\textbf{Measurements of the Young's modulus and Q factor for doubly-clamped beams.} (a) Plot showing the frequency of the fundamental mode of all the doubly-clamped beams and Si cantilevers versus $t/L^2$. The cantilevers are shown as solid triangles. The doubly clamped beams with $t > 7$ nm are indicated as solid squares while those with $t < $ 7 nm are shown as hollow squares. The solid line is the theoretical prediction with no tension. (b) Plot of the quality factor as a function of the thickness for all resonators \cite{bunch07}.}
  	\label{fig:geom_opt}
\end{figure}

\figref{fig:data_opt}A shows the resonance spectrum of a few-layer graphene sheet suspended over a SiO$_2$ trench. Several resonant peaks were measured; the first peak is the fundamental vibrational mode which will be the main focus of the discussion. The fundamental peak is the most pronounced peak, so it is easy to obtain the mechanical characteristics of the resonator. Analyzing higher modes is unnecessary as they depend on the fundamental frequency $f_0$. The most appropriate fit to extract the resonance frequency $f_0$ is a Lorentzian fit, giving $f_0$ to be 42 MHz with a quality factor Q = 210. \figref{fig:data_opt}B shows similar results for a single layer graphene, $f_0$ = 70.5 MHz and Q = 78. The same measurements were repeated for 33 resonators of different geometries with thicknesses ranging from 1 atomic layer to 75 nm. They are plotted in \figref{fig:geom_opt}. The data shows that the fundamental frequency $f_0$ largely varies from 1 MHz to 166 MHz with quality factors ranging between 20 and 850. Knowing the fundamental resonance $f_0$, the following equation for a doubly-clamped geometry is used as a reference to extract the Young's modulus of the sheets \cite{timos74}:\\

\begin{equation}
f_0 = \sqrt{\bigg(A\sqrt{\frac{E}{\rho}\frac{t}{L^2}}\bigg)^2 + \frac{0.57A^2S}{\rho L^2t}},
\label{eq:doub_clamp}
\end{equation}

where $E$ is Young's modulus, $S$ is the tension per width, $\rho$ is the mass density, $t$ and $L$ are the thickness and length of the suspended graphene sheet, and $A$ is a geometrical constant; $A$ = 1.03 for doubly-clamped beams and 0.162 for cantilevers. The parameter $S$ represents the built-in tension on the graphene sheet which may come from the fabrication process or from the van der Waals interaction between the substrate and graphene. Assuming that the tension is sufficiently small, \eqnref{eq:doub_clamp} predicts that the fundamental frequency $f_0$ scales as $t/L^2$. \figref{fig:geom_opt}A is a plot of the resonance frequency for resonators with $t > 7$ nm (solid squares) and $t < 7$ nm (hollow squares) as a function of $t/L^2$ and the slope gives the Young's modulus $E$. Taken the density for bulk graphite as $\rho$ = 2200 kg/m$^3$, the Young's modulus of the graphene sheets plotted as dashed lines correspond to values between 0.5 and 2 TPa. These values correspond well to the Young's modulus of 1 TPa of bulk graphite \cite{kelly81}. This is the highest modulus resonator to date. This is in stark contrast to the 12-300 nm thick Si cantilevers \cite{vera04} which achieve values ranging between 53 - 170 GPa, and which are plotted as solid triangles in \figref{fig:geom_opt}A. For graphene resonators with $t < 7$ nm, the data points are not as linear as for the thick resonators. To elucidate these results, Scharfenberg {\it \textit{et al.}} studied the elasticity effect of thin and thick graphene samples on a corrugated elastic substrate such as polydimethylsiloxane (PDMS). For thin graphene samples, it adheres fully to the substrate so that the graphene follows the topography of the substrate. As for thicker samples, it flattens the corrugated substrate. These observations suggest that thin graphene flakes, despite their size, are highly sensitive to tension, making them easier to deform than the thicker samples \cite{mason11}.

\begin{figure}
  	\centering
      	\includegraphics[scale=0.45]{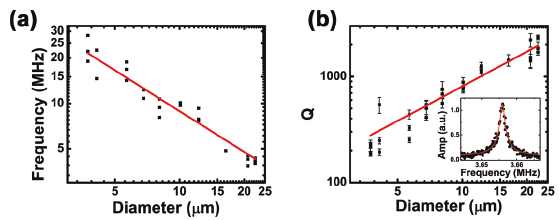}
      	\caption{\textbf{Resonance frequency and Q factor of graphene membranes.} (a) Fundamental frequency $f$ as a function of diameter $D$ for 29 graphene membranes. The red line is a fit of the data in which $f \sim D^{-0.9 \pm 0.1}$. (b) Quality factor of membranes as a function of diameter. Inset: the highest Q factor peak observed fitted with a Lorentzian indicating a value of 2400 \cite{barton11}.}
  	\label{fig:res_memb}
\end{figure}

The quality factor is an important parameter to consider when dealing with resonators. It gives a good indication of the resonator's sensitivity to external perturbations. A high Q is thus essential for most applications. A plot of Q as a function of the thickness for all graphene resonators is shown in \figref{fig:geom_opt}B. It seems that there is no clear dependence of quality factor on resonator thickness observed and the Q factor is on the order of hundreds. These Q factors are lower than diamond NEMS (Q$\simeq$2500-3000) \cite{sek02} and significantly lower than high tensile Si$_3$N$_4$ (Q$\simeq$200,000) \cite{verb06}. Thicker variants of graphene, like graphene oxide (Q$\simeq$4000) \cite{robin08} and multilayer epitaxially grown graphene (Q$\simeq$1000) \cite{shiva09} have shown higher quality factors. Despite the usability of graphene as a mechanical resonator, the source of dissipation is still poorly understood. Recent progress on graphene resonators suspended over circular holes on SiN$_x$ membranes have shown some improvements on the Q factor. Thus, \figref{fig:res_memb} demonstrates a clear dependence of quality factor on the resonator diameter. The highest Q factor is about 2400 for a resonator with a diameter of 22.5 $\mu$m \cite{barton11}.

\subsection{Measurements by atomic force microscopy}

In the previous subsections, the Young's modulus is measured by considering the tension being small, neglecting the second term in \eqnref{eq:doub_clamp} for a doubly-clamped beam. The built-in tension in the suspended graphene may indeed play a role in the mechanical measurements, a plausible explanation to the scattered data found in \figref{fig:geom_opt} for thin graphene resonators. To measure the built-in tension, an AFM tip with a calibrated spring constant is pressed onto the suspended graphene \cite{frank07}. The spring constants on the sheet are then extracted to calculate the Young's modulus and built-in tension. With an applied static force for a doubly-clamped beam and employing the relation $f_0 = (1/2\pi)\sqrt{k_s/m}$ and \eqnref{eq:doub_clamp}, the resulting effective spring constant is:

\begin{equation}
k_ s= 16.23Ew\left(\frac{t}{L}\right)^3 + 4.93\frac{T}{L}.
\label{eq:tension}
\end{equation}

where $w$ is the width of the sheet.

\begin{figure}[!h]
  	\centering
      	\includegraphics[scale=0.25]{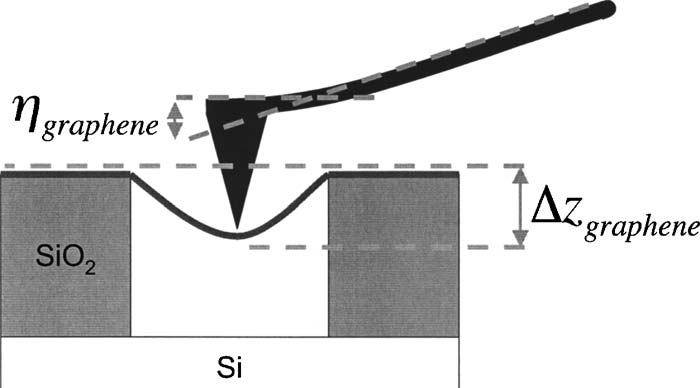}
      	\caption{\textbf{Schematic of a deflected AFM tip while pushing on the suspended graphene.} $\eta_{graphene}$ is the height that the tip is deflected by the graphene and $\Delta z_{graphene}$ is the height pushed by the tip \cite{frank07}.}
  	\label{fig:afm_meas}
\end{figure}

\begin{figure}[!h]
  	\centering
      	\includegraphics[scale=0.35]{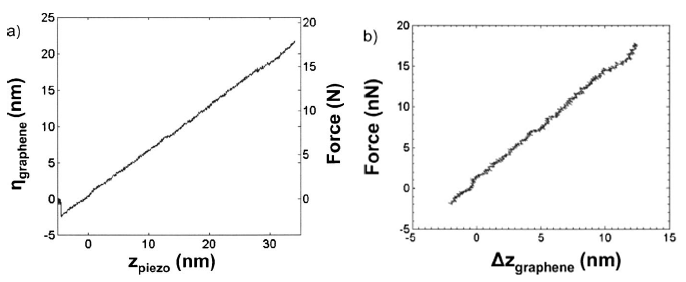}
      	\caption{\textbf{Measurement of the graphene's spring constant.} (a) Curve obtained from the deflection of the AFM tip by pushing down the suspended graphene sheet. The right axis represents the force corresponding to the tip displacement. (b) Plot of the force as a function of the displacement of the graphene sheet \cite{frank07}.}
  	\label{fig:piezo_meas}
\end{figure}

\figref{fig:afm_meas} shows the schematic of an AFM tip pushing down the suspended graphene sheet. The deflection of the tip then helps to find the effective spring constant of the suspended sheet. The chosen AFM tip has a spring constant 2 N/m which enables the graphene to be deflected by a detectable amount. The tip is pushed slowly against the sheet and the relation between tip displacement and the position of the piezo is then plotted in \figref{fig:piezo_meas}. As the AFM tip comes into contact with the suspended graphene, the cantilever is pulled down onto the surface resulting in a dip in the deflection as observed in \figref{fig:piezo_meas}a. From the tip's spring constant, a graph of the force exerted on the tip versus the displacement of the graphene sheets can be extracted as plotted in \figref{fig:piezo_meas}b. The displacement can be written as:

\begin{equation}
z_{piezo} = \eta_{graphene} + \Delta z_{graphene}
\end{equation}

where $\eta_{graphene}$ is the deflection of the tip, $z_{piezo}$ is the location of the piezo moving the tip, and $\Delta z_{graphene}$ is the deflection of the graphene sheet. From Hooke's law, the slope yields the effective spring constant of the suspended graphene sheet.\\

\begin{figure}[!h]
  	\centering
      	\includegraphics[scale=0.2]{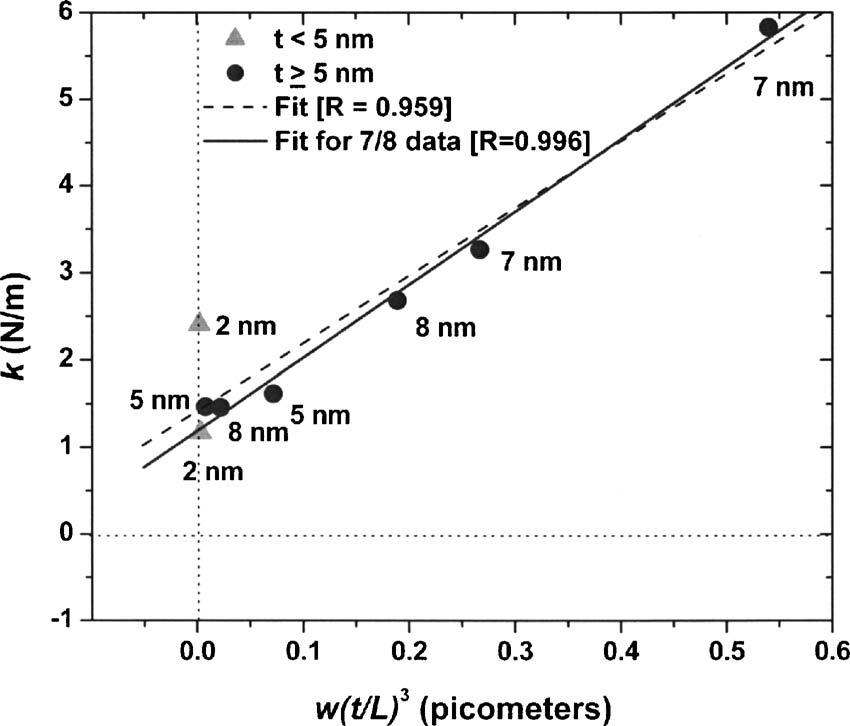}
      	\caption{\textbf{Plot of the spring constant in the center of the suspended region versus $w(t/L)^3$ for eight different samples with various thicknesses.} The linear fit provides information about the built-in tension and Young's modulus of the graphene sheets \cite{frank07}.}
  	\label{fig:spring_const}
\end{figure}

\figref{fig:spring_const} shows the plot of the spring constant as a function of the dimensions of the suspended graphene sheet for eight samples. The spring constant varies between 1-5 N/m. A linear fit is then used to extract the built-in tension and Young's modulus of the suspended graphene sheets. From \eqnref{eq:tension}, the slope suggests that a Young's modulus $E$ of 0.5 TPa, in good agreement with suspended graphene actuated optically and bulk graphite. The offset of the linear fit gives a tension of 300 nN, suggesting that the built-in tension of all suspended graphene sheets are on the order of nN.\\

Other work on AFM nanoindentation was realized on graphene suspended over micron-sized circular wells to investigate the elasticity properties of graphene. Indenting defect-free graphene with an AFM tip similar to \figref{fig:afm_meas}, one can probe the elastic stress-strain response. A non-linear elastic response model is fitted to find that the stiffness is about 340 N/m with intrinsic breaking strength of 42 N/m \cite{hone08}.

As discussed in this section, graphene as an ultra-thin membrane with extremely high bulk modulus is very promising for potential applications in mechanical systems, if the source of dissipation can be reduced substantially,  which should lead to a significant increase in Q factor. We now turn to applications of graphene as transistors.


\section{Graphene Transistors}
\label{sec:transistors}

This section provides an overview of current experimental work being pursued toward the development of graphene field effect transistors (GFET). The two most common varieties of transistor are the logic transistor and the analog transistor. The former is characterized, among other things, by a high $I_{on} / I_{off}$ ratio to ensure low energy consumption (of the off state) and to maintain a high logic interpretation yield. The latter is characterized primarily by its cutoff frequency, and is often used in high frequency applications as an amplifier. Both types of devices, as we shall see, could benefit from the remarkable electrical properties of graphene.

\subsection{Logic transistors}

For more than forty years CMOS technology has dominated the logic transistor industry with the fabrication of MOSFETs. A general trend over the years has been to reduce the length of the transistor's gate in order to achieve higher transistor densities but also increased performance produced by the higher electrical fields in the channel region. However, as the size of the device is reduced, more and more issues (commonly known as short channel effects) begin to appear \cite{Frank:2001vu}. These issues include but are not limited to such problems as hot electron effects, velocity saturation effects, and punchthrough effects. It has been suggested that graphene, due to its monoatomic thickness, would reduce these parasitic effects \cite{Schwierz:2010ix}. Graphene could be included in the channel region of transistors and thus provide a high mobility channel which would help to reduce the effect of short channel effects. One problem is that graphene does not possess a bandgap since the conduction and valence bands touch at the Dirac point. This reduces the on-off ratio of the transistor by several orders of magnitude, such that a graphene FET will typically have an on-off ratio $<10$. This ratio is unacceptable if one wants to replace CMOS technologies where ratios $>10^7$ are commonly achieved. It has been proposed by the International Technology Roadmap of Semiconductors (ITRS) that a ratio of $10^4$ would be required for logic applications. Such a ratio could be achieved in graphene only by opening a bandgap at the Dirac point in order to suppress the band-to-band tunneling \cite{Fiori:2008kl}. A bandgap of 0.4 eV could achieve the desired on-off ratio \cite{Schwierz:2010ix}. Several techniques are being envisioned for engineering a graphene bandgap. The following sections provide an overview of these methods.

\subsubsection{Bilayer graphene}

\begin{figure}[htbp]
\centering
\includegraphics[width=\columnwidth]{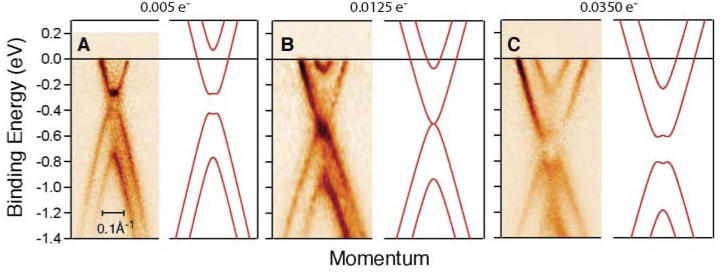}
 \caption{ \textbf{Bandgap evolution as a function of the potassium doping.} The doping level is controlled using the potassium adsorption of bilayer graphene. Shown from left to right is a diagram of the evolution of the bandgap for an increasing doping level per unit cell \cite{Ohta:2006bo}.}
 \label{fig:apesbilayer}
\end{figure}

The first technique that we shall investigate is the use of bilayer graphene. Theoretical calculations have predicted the possibility of a bandgap opening in this configuration \cite{McCann:2006ie}. Such a structure allows the opening of a bandgap by breaking the symmetry of the bilayer stack (Bernal stacking) with the application of a transverse electrical field. One of the first investigations of this approach is given in reference \cite{Ohta:2006bo}. The authors used bilayer graphene on a SiC substrate. In such a structure, light doping is provided to the bottom graphene layer by the substrate and further doping can be introduced to the top graphene layer by the adsorption of potassium atoms. This setup was designed and fabricated to facilitate band structure analysis using the ARPES technique. \figref{fig:apesbilayer} presents the experimental band structure for increasing amounts of potassium doping and a comparison to theoretical calculations. One may notice the good agreement between probed and theoretical bandstructure. A maximal bandgap of approximately 0.2 eV is observed using this technique.

\begin{figure}[htbp]
\centering
\includegraphics[width=\columnwidth]{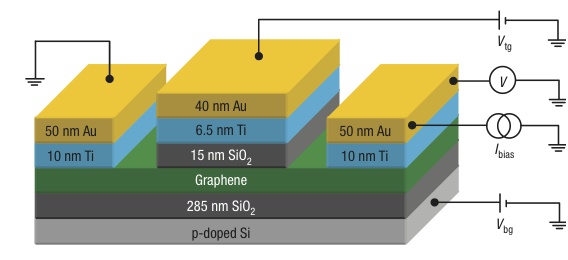}
 \caption{ \textbf{Schematic of a GFET device.} Graphene is deposited on an isolating material (e.g. SiO$_2$) after which regular photolithographic patterning can be performed \cite{Oostinga:2008ii}.}
 \label{fig:GFET}
\end{figure}

The results presented above show the potential of bilayer graphene for bandgap engineering. Several groups used this material for designing FET devices \cite{Oostinga:2008ii}. In these devices doping is induced using the field effect instead of using potassium atoms. \figref{fig:GFET} presents a schematic representation of such a device. The double gate configuration allows one to control both the Fermi position and the bandgap size.

After characterization of the device, a maximal $I_{on} / I_{off}$ ratio of approximately 100 could be achieved at 4.2 K. Decreasing performance was observed with increasing temperature. These low on-off ratios are explained by the insufficiently large bandgap. Theoretical calculations for bilayer graphene predict a maximal achievable bandgap that would be below the 0.4 eV gap needed for an on-off ration over $10^4$. These predictions are corroborated with experimental results. Therefore, bilayer graphene seems to be impractical for logic application.

\subsubsection{Graphene nanoribbons}

\begin{figure}[htbp]
\centering
\includegraphics[width=\columnwidth]{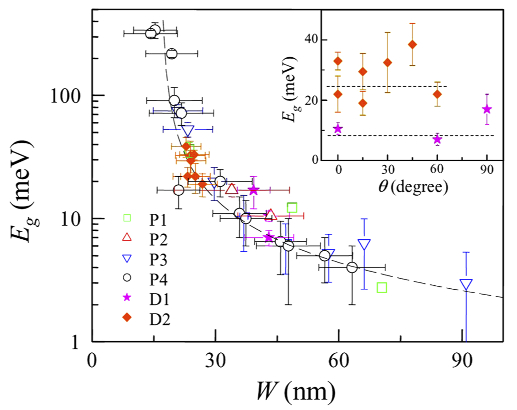}
 \caption{\textbf{Energy bandgap as a function of nanoribbon width} at $T = 4.2$ K \cite{Han:2007vx}.}
 \label{fig:ribbons_litho}
\end{figure}

A second technique used for engineering a bandgap in graphene is the use of graphene nanoribbons. The idea is to provide additional charge carrier confinement in the plane of the graphene sheet, effectively making it a one-dimensional structure. This additional confinement can lead to the opening of a transport gap. There are two common techniques for producing nanoribbons. The first technique is the lithographic patterning of graphene. On one hand, this approach makes it easy to control the position of the transistor's elements (e.g. channel, gate, electrons, etc.). On the other hand, making the width of a nanoribbon arbitrarily small is limited by the lithographic process. Furthermore, the edge quality of the patterned nanoribbons has been found to be rough, leading to increased scattering and decreased performance \cite{Han:2007vx}. \figref{fig:ribbons_litho} presents a graph of the bandgap energy ($E_g$) as a function of the width of the lithographically patterned nanoribbons at $T = 4.2$ K. Again decreased performance is correlated with increasing temperature. One can notice a maximal bandgap of $\sim$200 meV for a nanoribbon width of $\sim$15 nm. These results are similar to the ones obtained for bilayer graphene.\\

\begin{figure}[htbp]
\centering
  \subfigure[I-V curve for a 5 nm wide nanoribbon]{\includegraphics[width= 0.49\columnwidth]{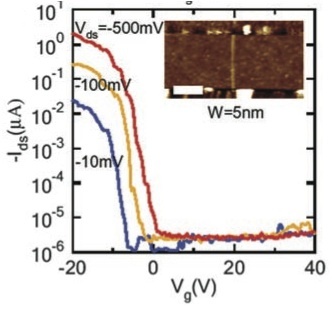}}
  \subfigure[Energy bandgap as a function of nanoribbon width at room temperature.]{\includegraphics[width= 0.49\columnwidth]{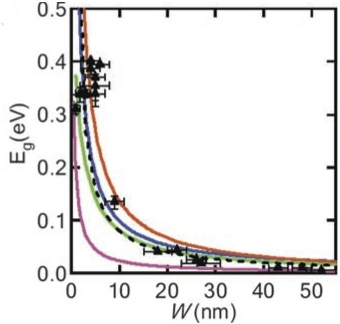}}
 \caption{\textbf{Chemically derived nanoribbons} \cite{Li:2008ht}.}
 \label{fig:chemical}
\end{figure}

A second approach for producing nanoribbons is chemical synthesis \cite{Li:2008ht}. This method, unlike lithographic patterning, produces smooth edges on the nanoribbons, while achieving widths in the sub-10 nm range. However, this method leads to a distribution of nanoribbon sizes and makes their positioning difficult. Scalability of such a technique is thus hard to achieve. \figref{fig:chemical}a presents an I-V curve for a 5 nm wide nanoribbon in a GFET device. One may notice that even at room temperature $I_{on} / I_{off}$ ratios above $10^7$ are observed. This behavior is explained by the size of the bandgap, which reaches 0.4 eV, as illustrated in \figref{fig:chemical}b. For such small widths, carrier confinement is appreciable leading to an asymptotic behavior of the bandgap as a function of width.

\subsubsection{Other techniques}

This section presents three other techniques for engineering a bandgap in graphene: strained graphene, nanomesh graphene and patterned hydrogen adsorption graphene.\\

\paragraph{Strained graphene} Several studies suggest that a bandgap would be induced in graphene by the application of strain. By keeping track of the G and 2D bands while applying strain on a graphene sheet, one can monitor the amount of strain induced. Strain of 0.8\% has been experimentally measured using this technique \cite{Ni:2008ij}. Theoretical calculations predict that such an amount of strain would lead to a 300 meV bandgap. However, experimental evidence of this gap has not been demonstrated.\\

\paragraph{Nanomesh graphene} Another novel way of inducing a bandgap in graphene is to pattern holes in a graphene sheet \cite{Bai:2010ff}. This technique led to an $I_{on} / I_{off}$ ratio of the order of 100 at room temperature for patterned holes of $\sim$7 nm in width. The on-off ratio tends to increase with decreasing neck width of the holes. A major advantage of this technique is its scalability. Work remains to be done in order to study whether an on-off ratio of $>10^4$ can be achieved.\\

\begin{figure}[htbp]
\centering
\includegraphics[width=\columnwidth]{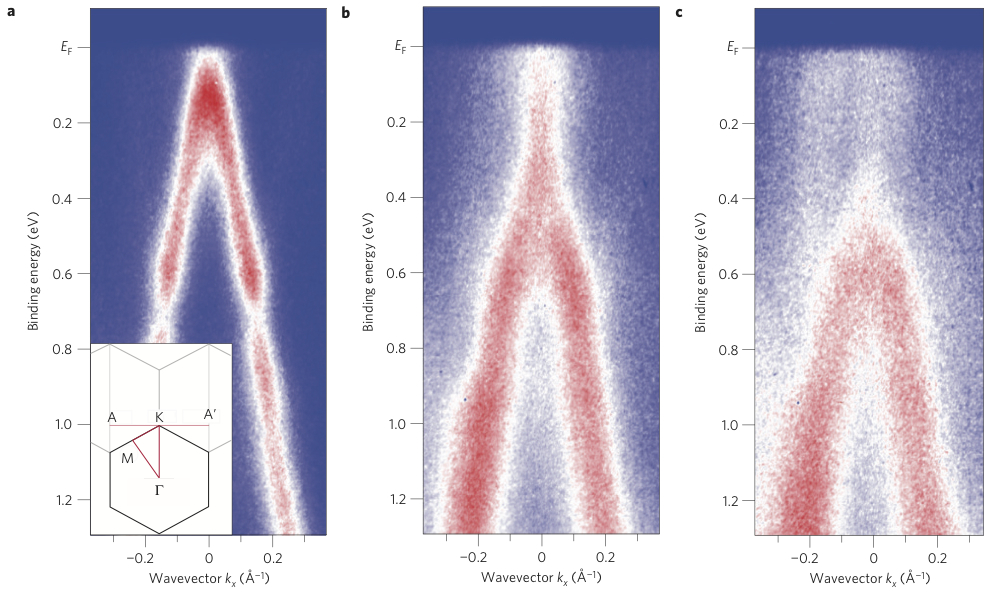}
 \caption{\textbf{Valence band evolution.} Exposition time from left to right goes from 0 sec (clean graphene) to 30 sec and finally 50 sec. A bandgap of almost 0.5 eV is observed after 50 sec exposition a result that is in good agreement with theoretical predictions \cite{Balog:2010ca}.}
 \label{fig:hydro_adsoption}
\end{figure}

\paragraph{Patterned hydrogen adsorption} This novel technique makes use of half-hydrogenated graphene on Ir(111) \cite{Balog:2010ca}. Theoretical DFT simulation predicts that this technique can open a bandgap of 0.43 eV in graphene, thus reaching our desired value of 0.4 eV. Graphene grown on Ir give rises to a superperiodic potential, also called a Moir\'e pattern. Such a pattern remains after the hydrogen adsorption and it alters the electronic properties of graphene leading to a bandgap opening. \figref{fig:hydro_adsoption} presents ARPES measurements of the hydrogen adsorbed graphene. One can easily see the the opening of a large bandgap around the Fermi level. This structure is also stable at room temperature.

\subsubsection{Techniques comparison}
Presented below is a summary table that presents achieved results for the different bandgap engineering methods presented in this section.

\begin{table}[h!]
  	\centering
\begin{tabular}{p{2cm}p{1.5cm}p{1.5cm}p{3cm}}
\hline
Method & Bandgap & $I_{on} / I_{off}$ & Ref.\\
\hline \hline
Bilayer & 0-0.25 eV & $\sim$100 & \scriptsize{\cite{Ohta:2006bo,Oostinga:2008ii}}\\
\hline
Nanoribbons & 0-0.4 eV & $>10^6$ & \scriptsize{\cite{Han:2007vx} \ \ \ \ \ \ \ \ \ \cite{Li:2008ht}}\\
\hline
Strain & 0.3 eV* & NA & \scriptsize{\cite{Ni:2008ij}}\\
\hline
Nanomesh & NA & $10-100$ & \scriptsize{\cite{Bai:2010ff}}\\
\hline
Hydrogen \ \ adsorption & $>0.4$ eV & NA & \scriptsize{\cite{Balog:2010ca}}\\
\hline
\end{tabular}
\caption{\textbf{Bandgap engineering techniques}
\scriptsize{*No experimental data available}}
\end{table}

\subsection{Analog transistors}

A second type of transistor is the analog transistor. These devices do not require a non-conducting off-state like logic transistors, and therefore do not require a bandgap. They are used in radio frequency applications and are characterized by their cutoff frequency $f_t$. The high mobility of graphene possibly leads to very high cutoff frequencies, making graphene an attractive material for designing these transistors. One major challenge for the integration of graphene in these transistors is the preservation of high mobility while integrating graphene in a device.\\

Since 2008 there has been increased research interest in the GFET analog transistor \cite{Meric:2008dw}. Some of the initial work showed that a 14.7 GHz transistor was achievable using graphene with a 500 nm device length. Following these results, a 26 GHz device was fabricated using a shorter gate length of 150 nm and mechanically exfoliated graphene \cite{Lin:2009dr}. The cutoff frequency of GFET devices has proven to follow its well-known behavior for this type of transistors given by:

\begin{equation}
  	f_t=\frac{g_m}{2 \pi C_G}
\end{equation}

Shorter gate length increases the drift velocity of charge carriers by increasing the electrical field. It also leads to a reduced distance between the source and drain contacts, which also reduces the transit time across the FET. It was observed that $f_t \propto 1 / L_G^2$ , so reducing the gate length from 500 nm to 150 nm resulted in the cutoff frequency rising from 3 GHz to 26 GHz \cite{Lin:2009dr}. For these devices the effective mobility was estimated to be 400 cm$^2$V$^{-1}$s$^{-1}$. This value is well below the 10000 cm$^2$V$^{-1}$s$^{-1}$ mobility of exfoliated graphene at room temperature that could be achieved \cite{nov04}. Indeed, the effective mobility in graphene is highly sensitive to its environment and decreases as the amount of scattering elements and defects are increased (e.g. insulating substrate, gate oxide, source/drain contacts, etc.).\\

A similar device configuration using a higher quality graphene (graphene grown on SiC) achieved a higher effective mobility of 1000-1500 cm$^2$V$^{-1}$s$^{-1}$. For this device, a 100 GHz cutoff frequency been measured with a gate length of 240 nm \cite{Lin:2010bu}.\\

\begin{figure}[htbp]
\centering
\includegraphics[width=\columnwidth]{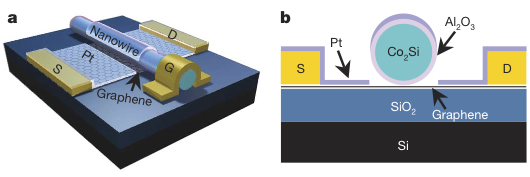}
 \caption{\textbf{GFET schematics} using a nanowire gate \cite{Liao:2010jf}.}
 \label{fig:nanowire_gate}
\end{figure}

Lately, other device configurations that minimize scattering with the graphene sheet are being engineered. \figref{fig:nanowire_gate} shows a GFET device using a nanowire gate.

This configuration minimizes scattering and defects in the graphene sheet and allows one to achieve cutoff frequencies in the 100-300 GHz with gate lengths in the 200 nm range. Further work is still required in order to achieve higher effective mobility. In the following section we shall see how the optical properties of graphene are of a great interest for optoelectronic devices.


\section{Optoelectronics}
\label{sec:optical}
\subsection{Optical properties}
\begin{widetext}

\begin{figure}[!h]
 \centering
    	\includegraphics[width=\columnwidth]{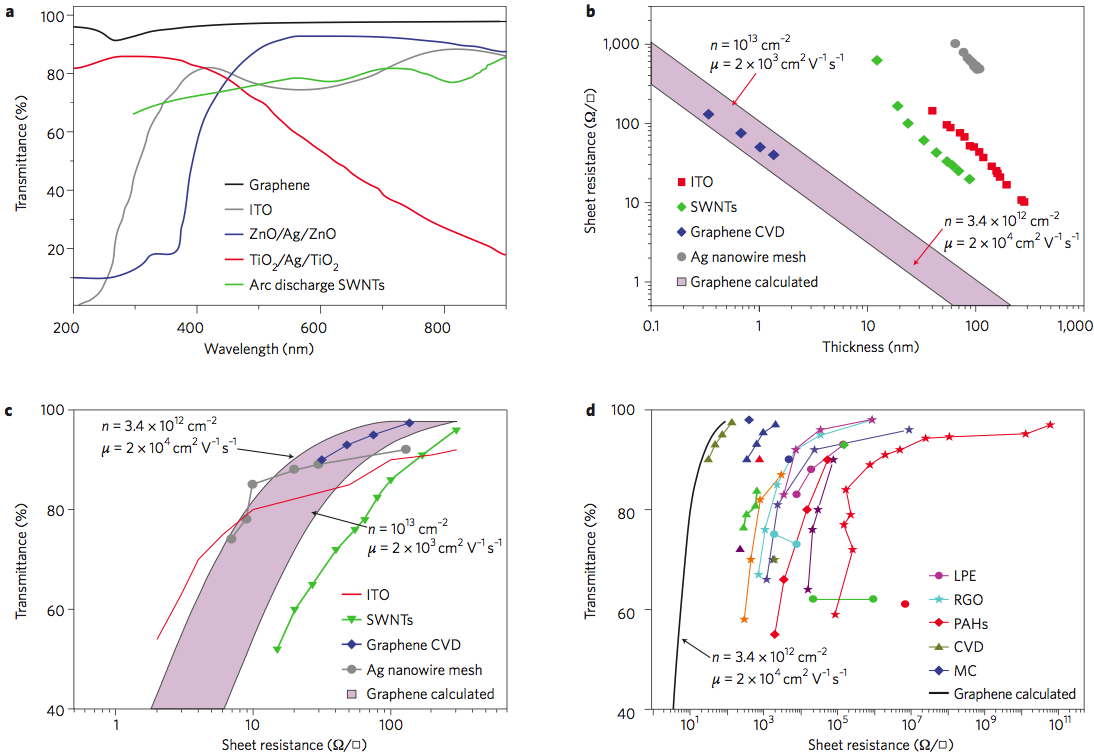}
  	\caption{\textbf{Exploring the utility of graphene for thin conducting electrodes.} Reprinted from \cite{Bonaccorso10}. Theoretical ranges are shown bound by black lines, as calculated using typical values with \eqnref{TwithSig}. (a) Transmittance as a function of wavelength for graphene, ITO and two other metal oxides, as well as SWNTs. (b) $R_{s}$ values as a function of thickness, illustrating the potential of graphene to achieve substantially lower $R_{s}$ for a given thickness. (c) Transmittance as a function of $R_{s}$ for the same materials. (d) Transmittance as a function of $R_{s}$ for a number of graphene fabrication methods: triangles, CVD; blue rhombuses, micromechanical cleavage (MC); red rhombuses, organic synthesis from polyaromatic hydrocarbons (PAHs); dots, liquid-phase exfoliation (LPE) of pristine graphene; and stars, reduced graphene oxide (rGO).}
 \label{GTCE}
\end{figure}

\end{widetext}

The properties of graphene make it an attractive choice for use in optoelectronic devices. In particular, graphene's high transparency, low reflectance, high carrier mobility and near-ballistic transport at room temperature make it a promising choice for transparent electrodes. Optically, single-layer graphene (SLG) has an unusually high absorption (given its thickness) that can be described in terms of the fine structure constant, $\alpha$ (a fundamental physical constant that describes the interaction of matter and electromagnetic fields). Light transmittance $T$ through free-standing graphene can be derived using the Fresnel equations for a thin film with a universal optical conductance of $G_{0} = e^{2}/4\hbar$:

\begin{equation}
T = (1 + 0.5\pi\alpha)^{-2} \approx 1 - \pi\alpha \approx 0.977
\end{equation}

where $\alpha = e^{2}/(4\pi\epsilon_{0}\hbar c) = G_{0}/(\pi\epsilon_{0}c)$. For the most part, each SLG will contribute an absorbance of $A = 1 - T \approx \pi\alpha \approx 2.3\%$ to visible light. Because graphene sheets behave as a 2-dimensional electron gas, they are optically almost non-interacting in superposition (though the same cannot be said of their electronic interaction \cite{Luican11}), so the absorbance of few-layer graphene (FLG) sheets is roughly proportional to the number of layers. The absorption spectrum of graphene from the ultraviolet to infrared is notably constant around 2-3\% absorption, as shown in \figref{GTCE}a, compared to some other materials. The reflectance is very low at 0.1\%, though increases to ~2\% for 10 layers \cite{Casiraghi07}. Graphene has found its way into many optoelectronic and photonic devices, a portion of which will be covered here. Other uses, including photodetectors, touch screens, smart windows, saturable absorbers and optical limiters have recently been covered in detail elsewhere \cite{Bonaccorso10}.

\subsection{Transparent conducting electrodes}
\label{TransparentConductingElectrodes}

Transparent conducting electrodes (TCEs) are critical components in a wide variety of devices, from those found in academic and industrial settings to commercial devices that define modern technology. They function primarily by injecting or collecting charge depending on the purpose of the device, and are highly transparent across some part of the electromagnetic spectrum (most often to visible light). The most widely used material for TCEs is indium tin oxide (ITO), an n-type, wide-bandgap semiconductor ($E_g$ = 3.75 eV), which typically consists of 90\% indium(III) oxide doped with 10\% tin(IV) oxide by weight. The tin atoms function as n-type donors. ITO has favorable electronic and optical properties for most TCE applications, but suffers from practical limitations. It is relatively inflexible and fragile, and a limited supply of indium makes it increasingly costly for large-scale production. Though it is highly transparent across most of the visible range and near-infrared, it becomes increasingly opaque at UV and reflects IR wavelengths, because of band-to-band absorption (excitation of an electron from the valence to the conduction band) and free carrier absorption (excitation within the conduction band), respectively \cite{Keshmiri02}. Despite the utility of ITO in optoelectronic devices, alternatives are sought due to the aforementioned reasons. One popular inorganic substitute is doped zinc oxide, which has comparable properties, though suffers from limited etching capability due to acid sensitivity, among other things. It is typically doped with aluminum, gallium or indium. The list of inorganic semiconductor materials is extensive, but ITO has become the standard to beat for many applications.

In contrast to inorganic materials, organic TCEs are characterized by low cost, flexibility and stability, though do not achieve the same level of charge mobility as inorganic materials. Many varieties of intrinsic conducting polymers (ICPs) have found use in optoelectronics. In particular, their electroluminescence, flexibility and mechanical strength makes them particularly suitable for OLEDs and touch screens, though processability is a known issue. PEDOT (poly(3,4-ethylenedioxythiophene)) is a conducting polymer that is widely used as a transparent conductive film in the form of PEDOT:polystyrene sulfonic acid (PSS) dispersions (due to the insolubility of PEDOT in water). Like other ICPs, the conductive properties of PEDOT arise from a chain of conjugated double bonds, and by controlling the extent of $\pi$-electron cloud overlap, the PEDOT band gap can be varied from 1.4 to 2.5 eV in a manner analogous to doping (earning such polymers the nickname ``organic metals'') \cite{Groenendaal00}. For organic semiconducting materials, the highest occupied molecular orbital (HOMO) and lowest unoccupied molecular orbital (LUMO) are analogous to the valence and conduction band of inorganic semiconductors.

\subsection{Graphene as TCE}

Two of the most vital parameters for TCE materials are low sheet resistance $R_{s}$ and high optical transmittance, $T$. Sheet resistance (in units of $\Omega /\Box$) is directly obtained using a four-terminal sensing measurement (see \sectionref{ElectronicProperties}), and can be thought of as resistance per aspect ratio. Under most circumstances, graphene matches or exceeds the transmittance of competing materials, though the sheet resistance has proven somewhat less predictable, and varies depending on production method, with CVD being the closest to optimal for optoelectronics (highest $T$ and lowest $R_{s}$). This is illustrated graphically in \figref{GTCE}. $T$ and $R_{s}$ in doped FLG are related fairly simply by:

\begin{equation}
T = 1 + \frac{Z_0}{2R_s}\frac{G_0}{\sigma_{2D}}
\label{TwithSig}
\end{equation}

with optical conductance $G_{0}$ as defined above, and where $Z_{0} = 1/\epsilon_{0}c = 377~\Omega$ is the free-space impedance, $\epsilon_{0}$ is the permittivity of free space and $c$ is the speed of light. Note that this relation is not directly related to the number of layers, though the 2-dimensional conductivity is given by $\sigma_{2D} = n\mu_{e}e$, where $n$ is the number of charge carriers. By taking a range of realistic values for FLG, it is possible to determine a range of $T$ and $R_{s}$ values, as shown in \figref{GTCE}. For CVD graphene, $n$ can vary from 10$^{12}$-10$^{13}$ cm$^{-2}$ and $\mu_{e}$ from 1000-20000 cm$^{2}$V$^{-1}$s$^{-1}$. Notably, by increasing the thickness, and thus the number of charge carriers, it is possible to achieve a lower $R_{s}$ value at the expense of transparency.

\subsubsection{Work function, Schottky junctions and photodetectors}

An important factor in the design of optoelectronic devices is the work function of graphene ($\phi_{G}$). A difference in the work function between materials at an interface results in charge transfer. A particular case of this is a metal/semiconductor interface, referred to as a Schottky junction, where unmatched work functions result in a potential barrier with rectifying characteristics. If the rectifying effect is absent (depending on the metal work function and band gap of the semiconductor), it is referred to as an ohmic contact. Schottky diodes differ from p-n junction diodes by lower forward voltage and faster switching speeds, allowing for high frequency signals.

Despite the absence of a band gap in graphene, it can function as the semiconductor active element for a metal/graphene photodetector \cite{Mueller10}. Normally, photoexcitation of graphene results in electron-hole pairs that quickly recombine. Near the Schottky junction, photogenerated charge is separated by the electric field of the Schottky barrier and a photocurrent results. In graphene, both electrons and holes have high mobility, providing an advantage over conventional semiconductors. In a preliminary test of an interdigitated metal-graphene-metal photodetector, no signal degradation was observed up to the frequency limit of the measurement system (40 GHz), and it is believed to be capable of up to 0.6 THz, ultimately limited by its RC constant \cite{Avouris10}. Using multiple metal types (with high and low work function) produces different doping, and by incorporating both, the photodetection area is maximized. This device was able to reliably detect optical data streams of 1.55 $\mu$m light at 10 GBits/s.

\subsubsection{Light-emitting diodes}

Light-emitting diodes (LEDs) function by radiative recombination (also known as the Lossev effect), wherein recombination of electron-hole pairs can result in the emission of photons. They offer a number of advantages over incandescent light sources including lower energy consumption and longer lifetimes. The general construction of an LED includes an electroluminescent semiconductor material (organic or inorganic) that emits light upon recombination and decay of charge injected from surrounding electrodes. The most commonly used anode material is ITO. In addition to the limitations stated above, it and other inorganic components may degrade over time, leaking ions into the active layer and compromising the proper function of the device.

A number of recent publications report gallium nitride (GaN)-based UV LED devices that use FLG as a transparent electrode \cite{Kim10,GJo10}. Graphene has better transparency to UV/blue light than ITO, and its thermal conductivity (particularly for a given thickness) is advantageous for improved heat distribution and extension of the lifetime of the device. Nitride (including gallium) LEDs are typically grown in one of two ways: epitaxially on a sapphire substrate at high temperature, for a high-quality device on a limited scale, or, for large-area or flexible devices, on glass, plastic or metal, necessitating a sacrifice in quality. Therefore a flexible, heat tolerant substrate is desired. In one design, a GaN thin film is grown on graphene (functioning as the substrate and electrode) with the assistance of ZnO nanowalls to overcome the absence of GaN nucleation on pristine graphene \cite{Chung10}. These graphene/GaN LEDs can then be transferred mechanically to other substrates.

\begin{figure}[!h]
 \centering
  	\includegraphics[width=\columnwidth]{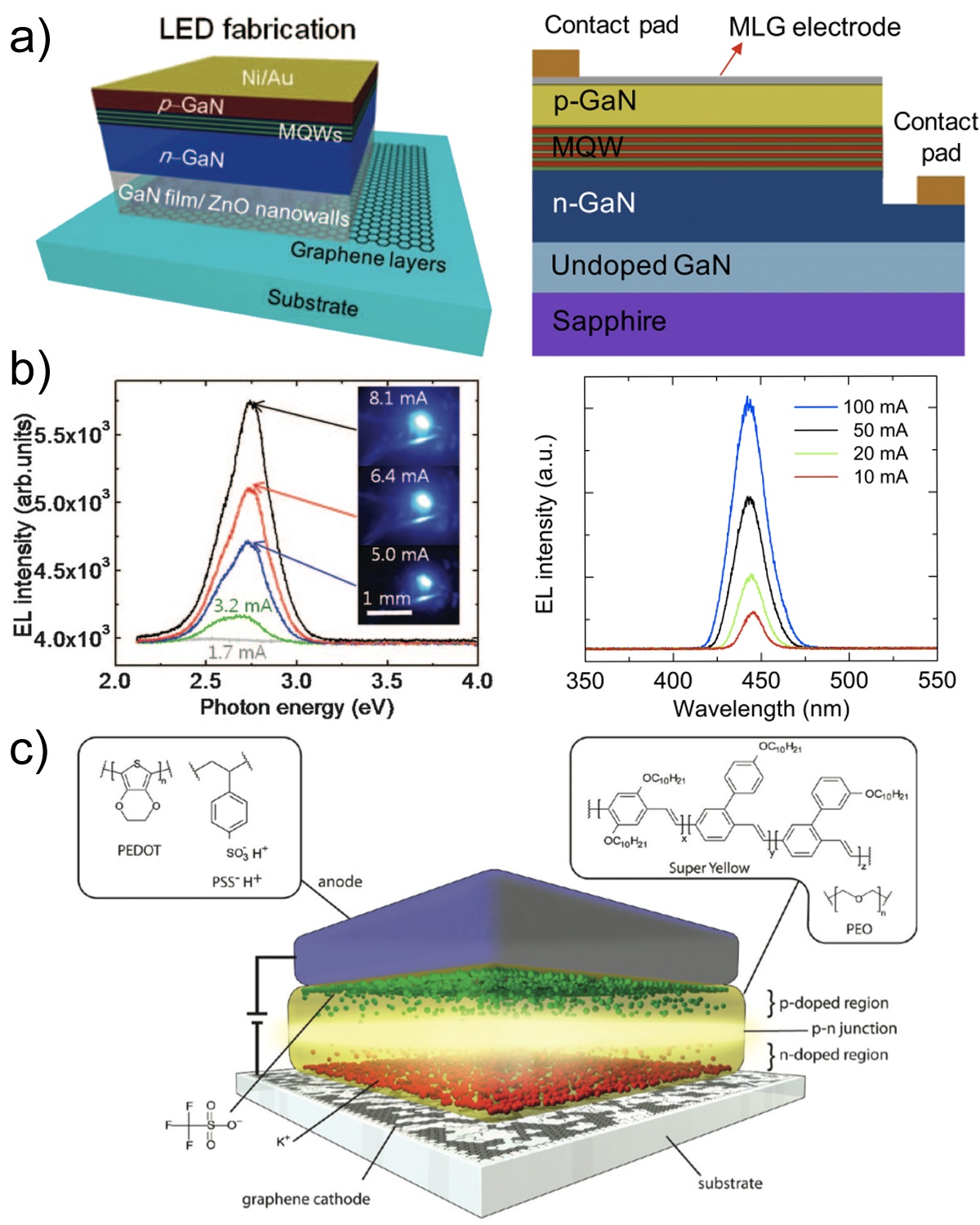}
  	\caption{\textbf{Design and function of graphene LEDs and LECs.} (a) Device schematics for inorganic LEDs incorporating graphene as a growth substrate and cathode on the left, and as an anode on the right. With graphene forming the surface in contact with the substrate, it becomes easy to move the entire device to a new substrate. Adapted from \cite{Chung10} and \cite{GJo10}. (b) Power-dependent electroluminescence of each device, showing strong blue emission spectra over a range of input currents. The left-hand device has been transferred to a plastic substrate. (c) Schematic of an LEC. Adapted from \cite{Matyba10}. Blending of the emissive material with an electrolyte solution allows the rearrangement of charge at the electrode interface, facilitating balanced electron-hole injection. As both electrodes are transparent, light is emitted through both surfaces.}
 \label{LED}
\end{figure}

In the other devices, graphene is deposited onto pre-formed GaN films. The schematics and electroluminescence intensity of two LEDs using In$_{x}$Ga$_{1-x}$N/GaN multi-quantum-wells (MQWs) as the emissive material are compared in \figref{LED}a and b. For the MLG anode device, the output power was shown to be comparable to an ITO-electrode device below 10 mA input current, becoming less efficient at higher input power. This is in part due to a larger forward voltage drop resulting from higher contact and sheet resistance of MLG compared to ITO. One way of correcting for high contact resistance is through doping. By intercalation of MLG semiconductor interfaces with bromine, it has been shown that tuning of the Schottky barrier height is possible over a wide range \cite{Tongay11}. The Schottky barrier height is the rectifying barrier for electrical conduction across a metal-semiconductor interface (here using graphene as the `metal'). Bromine intercalation serves to increase interlayer separation between graphene sheets, as well as increasing free hole carrier density and decreasing the Fermi level due to electron transfer from carbon to bromine.

Organic LEDs use polymers or small molecules for the emissive material. For efficient injection and charge conversion, the electrode work function must be well matched to the HOMO and LUMO of the emissive material. When coated with a thin and planarizing layer of PEDOT:PSS, ITO has an appropriate work function for hole injection as well as high conductivity and transparency; the latter is important so that the light generated within the active layer can be emitted from the device. By blending an electroluminescent polymer with an electrolyte solution, \cite{Matyba10} have constructed an OLED device referred to as a light-emitting electrochemical cell (LEC). It uses entirely solution-processable carbon-based materials, eliminating the need for a metal electrode. The composition uses chemically derived graphene for the transparent cathode, an active layer solution, and the conducting polymer PEDOT:PSS as the transparent anode. The active layer solution consists of a blend of a poly(para-phenylene vinylene) copolymer called ``Super Yellow'' (SY) and an electrolyte in the form of the salt KCF$_{3}$SO$_{3}$ dissolved in poly(ethylene oxide) (PEO). The design of this device is shown in \figref{LED}c. Because of the interfacial layer formed by the electrolyte, it mitigates the need for work function matching. In addition, the stability of the electrodes removes the possibility of ions from the electrode ``leaking'' into the active solution.

\subsubsection{Photovoltaics}

Photovoltaic devices (PVs, also known as solar cells) are designed to harvest solar energy by producing current from materials that exhibit the photovoltaic effect. Upon absorption of incoming light by the active material, electrodes collect the photogenerated charge carriers. Because the electrodes physically surround the active material, at least one of them is transparent to allow light to reach the active layer. In some cases, both electrodes are transparent, though typically a metal such as aluminum or silver is used. The overall energy conversion efficiency $\eta$ of the PV is determined by the product of individual efficiencies: reflectance, thermodynamic, charge carrier separation and conductive. In practice, these may be difficult to measure, so the overall efficiency is represented as:

\begin{equation}
\eta = \frac{P_{max}}{P_{inc}} = \frac{V_{OC}\times I_{SC}\times FF}{P_{inc}}
\end{equation}

where $I_{SC}$ is the maximum short circuit current, $V_{OC}$ is the maximum open-circuit voltage and FF is the fill factor, defined as the ratio of actual maximum obtainable power, given as $P_{max} = V_{max}\times I_{max}$, to the theoretical maximum, equal to $V_{OC}\times I_{SC}$. $P_{inc}$ is the total power of the incident light falling on the detector, and so $\eta$ represents the fraction of absorbed photons converted to current, and is also known as internal photocurrent efficiency. The best widely-available PVs are made from silicon, though comparable semiconductor PVs are increasingly available, such as thin-film cadmium telluride, copper indium selenide, and single-junction gallium arsenide. Silicon solar cells have achieved an $\eta$ value as high as 25\%. The FF for commercial cells can reach 70\%, with 40\% to 70\% being typical.

One instinctive direction is to integrate graphene with silicon wafers. By depositing a graphene sheet onto n-silicon, the authors of reference \cite{XLi10} have created a Schottky junction, which provides faster switching speeds and better efficiency due to a lower forward voltage drop. The graphene functions both as transparent upper electrode and antireflection coating, and through intrinsic electric field properties, aids in charge separation and hole transport. Though the work is preliminary, opportunities for optimization are apparent, such as passivation of the silicon to improve the interface - typically the deposition of an oxide layer forms a protective surface to prevent corrosion and reduce reactivity. 

One way of increasing the efficiency of devices containing graphene films is to better match the work function of the graphene to the neighboring components, and several recent efforts have been made toward engineering the graphene work function $\phi_{G}$. In designing a simple FLG/Si heterojunction device, Ihm \textit{et al.} revealed a direct relationship between the number of layers in the FLG and the resulting $V_{OC}$, with the best matching at 4 layers, despite disordered stacking \cite{Ihm10}. As the Si substrate is unchanged, the $V_{OC}$ is determined by the work function of the CVD-grown graphene layers, which was shown to better match the silicon as the number of layers increased, approaching the silicon work function $\phi_{Si}$ from below.

Organic cells made from polymers can be manufactured at a lower expense, though they have not yet achieved comparable efficiencies to silicon. A popular choice of active layer for proof-of-concept devices is an organic polymer dispersion consisting of a narrow-band electron donor, poly(3-hexylthiophene) (P3HT), and a fullerene derivative, [6,6]-phenyl-C$_{61}$-butyric acid methyl ester (PCBM), which functions as an electron acceptor. Cells with this kind of donor/acceptor blend constituting the active layer are referred to as bulk heterojunction (BHJ) cells. As we will discuss, graphene has been integrated into inorganic and BHJ cells in various roles, as summarized in \tblref{PVtable}. Most of the work on graphene PVs is at an early stage and in general they do not yet contend with the best performance obtained from similar organic solar cells made with ITO electrodes. Most of the graphene PVs do not reach an $\eta$ of 2\%, whereas those with ITO manage upwards of 6\%.

\begin{figure}[!h]
 \centering
  	\includegraphics[width=\columnwidth]{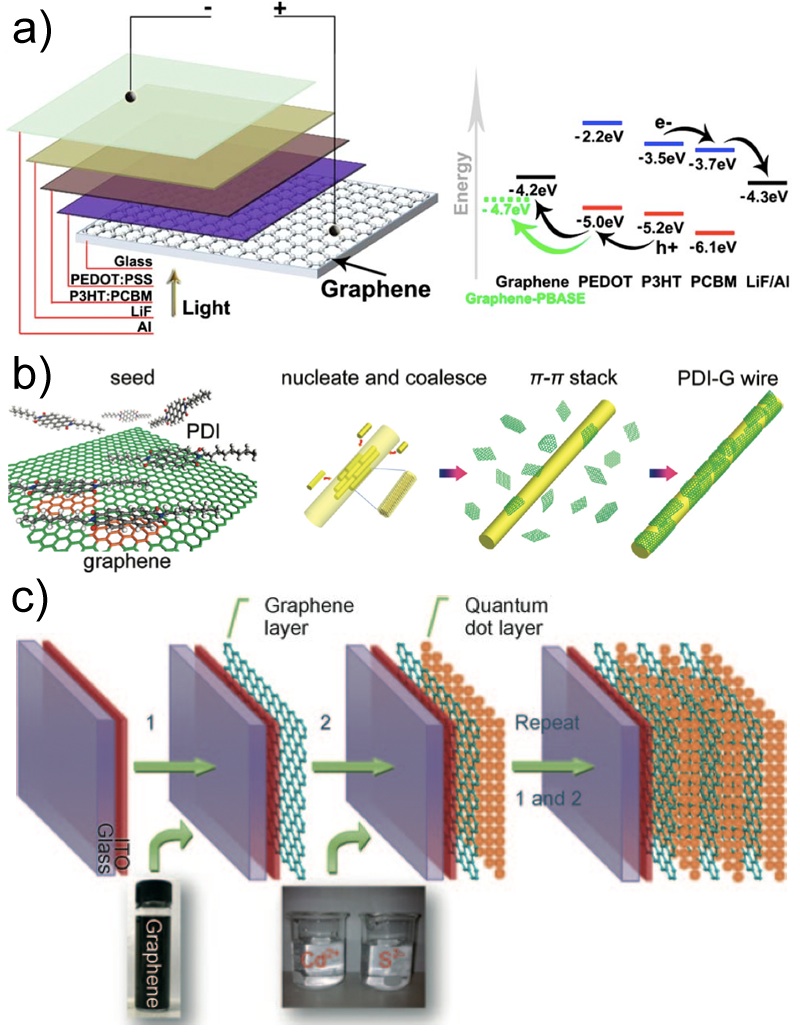}
  	\caption{\textbf{Schematics for some photovoltaic graphene devices.} (a) Schematic and energy diagram of a bulk heterojunction cell that uses graphene as a transparent anode. Adapted from reference \cite{Wang09}. (b) Construction of an rGO-wrapped PDI hybrid wire, showing the interaction between graphene and PDI. Adapted from reference \cite{Wang10}. (c) Layered graphene/CdS, using graphene as the electron acceptor. By stacking several bilayers, the $I_{SC}$ and $\eta$ of the device are substantially improved. Adapted from reference \cite{Guo10}.}
 \label{PVdiagrams}
\end{figure}

The recent work presented by \cite{Wang09} illustrates some of the design considerations and optimization problems associated with a simple graphene-modified BHJ cell. CVD-grown FLG films function as the anode in their device, filling the role of ITO (or other TCE). The films varied in thickness from 6-30 nm, with $R_{s}$ from 1350 to 210 $\Omega/\Box$ and $T$ from 91\% to 72\%. The complete device structure is: FLG graphene/PEDOT:PSS/P3HT:PCBM/LiF/Al, and is shown in \figref{PVdiagrams}a. The device is mounted on glass, which provides protection and structural integrity and permits light, which passes through the FLG layer to be absorbed by the P3HT:PCBM active layer, generating the charge that will be collected by the FLG and LiF/Al electrodes. With a simple substitution of unmodified FLG film for ITO, the device efficiency was reduced by approximately 30-fold (from $\eta$ of 3.10\% to 0.21\%). 

This efficiency reduction was attributed to several factors. First, the hydrophobicity of graphene precludes uniform coating of the PEDOT:PSS layer, which is essential as a planarizing buffer layer for reducing surface roughness, as well the hole injection barrier between P3HT and graphene. To address this, the graphene was modified by UV/ozone treatment, which improved wettability by introduction of OH and C=O groups to the surface. This resulted in improved $I_{SC}$ and $V_{OC}$ values, with slightly reduced FF for an overall improvement to $\eta$ of 0.74\%. The decrease in FF was attributed to increased series resistance of the cell, as disruption of the aromatic structure by surface groups results in decreased conductivity. 

To circumvent this effect, the graphene was modified instead by noncovalent functionalization with pyrene buanoic acid succidymidyl ester (PBASE). This resulted in the same improvement in $V_{OC}$ and further improvement in both $I_{SC}$ and FF, resulting in overall $\eta$ of 1.71\%, reaching slightly more than half of the ITO version of the device. The effect of PBASE modification was multifaceted - in addition to avoiding $\pi$-conjugation disruption and maintaining conductivity, the work function of graphene, $\phi_{G}$, was increased from 4.2 eV to 4.7 eV, increasing the open circuit potential as well as resulting in better matching between the Fermi level of FLG and the HOMO of PEDOT for more efficient hole collection.

Many improvements to graphene BHJ cell designs have appeared recently, including using metals for doping or blocking. By immersing CVD-grown graphene films into AuCl$_{3}$, Shi \textit{et al.} showed formation of Au particles on the graphene surface by spontaneous reduction of metal ions. The extent of the metal doping, controlled by immersion time, effectively tunes the surface potential, over a range of $\sim$0.5 eV \cite{Shi10}. Ng \textit{et al.} explored the effect of introducing graphene to TiO$_{2}$ nanonstructed films, using reduced graphene oxide (rGO, see \sectionref{GOsynth}) scaffolds to harvest photogenerated charge, and report up to a 90\% enhancement in photocurrent over pristine TiO$_{2}$ under UV illumination \cite{Ng10}. While pristine TiO$_{2}$ films yield a maximum $\eta$ of 7.4\%, the rGO-TiO$_{2}$ nanocomposite films yield maximum $\eta$ of 13.9\% for \emph{in situ} rGO (photo-catalytic reduction) and 11.4\% for hydrazine rGO (chemical reduction using hydrazine). In these devices, graphene serves as an efficient collector of generated electrons, and prevents recombination. It has also been shown that the introduction of a thin sol-gel processed titanium sub-oxide layer (TiO$_{x}$) for hole-blocking can improve device efficiency ~2-fold \cite{Choe10}. In this CVD graphene/PEDOT:PSS/P3HT:PCBM/TiO$_x$/Al device, the $I_{SC}$ and $V_{OC}$ values closely match those for ITO, with only the lower FF preventing the overall efficiency from being comparable.

An interfacial dipole layer (consisting of WPF-6-oxy-F, a synthesized polymer) in an inverted-structure organic solar cell can reduce the work function, increase built-in potential and improve charge extraction \cite{Jo10}. CVD-grown graphene films were used as the device cathode for design flexibility and the ability to build stacked devices. The ionic or polar groups of the WPF-6-oxy-F form interface dipoles, better matching $\phi_{G}$ to the LUMO of PCBM (4.2 eV). Untreated MLG film has $\phi_G = 4.58\pm0.08$ eV, close to HOPG at 4.5 eV. It was reduced by 0.05$\pm$0.03 eV using poly(ethylene oxide) (PEO), 0.22$\pm$0.05 eV with Cs$_{2}$CO$_{3}$, and 0.33$\pm$0.03 eV with WPF-6-oxy-F.

Though graphene finds a natural function as a TCE, it also presents potential for additional utility by virtue of its chemical structure. In particular, its aromatic skeleton can function as a nucleating agent for the self-assembly of organic nanostructures. Using rGO flakes as a seeding agent for planar aromatic molecules (which tend to self-assemble), ``hybrid'' wires were formed through noncovalent hydrogen bonding and $\pi-\pi$ stacking interactions \cite{Wang10}. By choosing a photoluminescent molecule (in this case N,N'-dioctyl-3,4,9,10-perylenedicarboximide, or PDI) which is effectively ``wrapped'' in graphene, a photovoltaic active material can be produced, as shown schematically in \figref{PVdiagrams}b. Hydrothermal reduction forms PDI-graphene hybrid wires (PDI-G) which can be integrated into a PV device, with properties as listed in \tblref{PVtable}.

\begin{widetext}

\begin{table}
\begin{tabular}{ l c c c c c c }

\hline

Device construction & $I_{SC}$ & $V_{OC}$ & FF & $\eta$ & Role of & Reference \\
& (mA/cm$^2$) & (V) & ($\%$) & ($\%$) & graphene & \\

\hline\hline

ITO/PEDOT:PSS/P3HT:PCBM/LiF/Al & 9.03 & 0.56 & 61.1 & 3.10 & Transparent anode & \cite{Wang09} \\ \hline

\hline\hline

Graphene/n-Si/Ti/Pd/Ag & 6.5 & 0.48 & 55 & 1.65 & Transparent cathode & \cite{XLi10} \\ \hline
CVD graphene/PEDOT/CuPc/C$_{60}$/BCP/Al & 4.73 & 0.48 & 52 & 1.18 & Transparent anode & \cite{DeArco10} \\ \hline
ITO/PEDOT:PSS/P3HT:graphene/LiF/Al & 4.0 & 0.72 & 38 & 1.1 & $e^-$ acceptor & \cite{Liu09} \\ \hline
ITO/PEDOT:PSS/P3HT:PCBM-graphene/Al & 5.3 & 0.64 & 41 & 1.4 & Dopant & \cite{Liu10} \\ \hline
Graphene/PEDOT:PSS/P3HT:PCBM/LiF/Al & 6.05 & 0.55 & 51.3 & 1.71 & Transparent anode & \cite{Wang09} \\ \hline
MLG/PEDOT:PSS/P3HT:PCBM/TiO$_x$/Al & 9.03 & 0.60 & 48 & 2.60 & Transparent anode & \cite{Choe10} \\ \hline
Al/PEDOT:PSS/P3HT:PCBM/WPF-6-oxy-F/MLG & 6.61 & 0.57 & 33 & 1.23 & Transparent cathode & \cite{Jo10} \\ \hline
ITO/PEDOT:PSS/P3HT:GQDs/Al & 6.33 & 0.67 & 30 & 1.28 & $e^-$ acceptor & \cite{Li11} \\ \hline
ITO/PEDOT/P3HT:PDI-G/LiF/Al & 3.85 & 0.78 & 35 & 1.04 & $e^-$ acceptor & \cite{Wang10} \\ \hline
Layered graphene/CdS QDs on ITO glass* & 1.08 & 0.68 & - & 16 & $e^-$ acceptor & \cite{Guo10} \\ \hline

\end{tabular}
\caption{\textbf{Function of graphene photovoltaic devices.} A summary of some recently developed photovoltaic devices that have incorporated graphene in various capacities, compared to an ITO device (shown at the top).
\\$^\ast$For 8 bilayers.}
\label{PVtable}
\end{table}

\end{widetext}

\subsubsection{Graphene quantum dots (GQDs) and graphene-QD nanocomposites}

Another interesting development is the formation of 0D graphene nanoparticles or nanowells from isolated 2D SLG. Quantum dots are described as semiconductors (typically nanoparticles) with excitons confined in 3 dimensions, which results in interesting properties that fall somewhere between bulk semiconductors and discrete molecules. Most inorganic QDs (CdSe, CdTe, InP) are fluorescent, with a large Stokes shift and narrow, size-dependent emission peaks, making them useful in fluorescence applications.

Recently, a straightforward electrochemical method of producing GQDs from freestanding graphene film was reported \cite{Li11}. By treating with O$_{2}$ and using the graphene as the working electrode in a cyclic voltammetry cell, they were able to produce hydrophilic, monodisperse GQDs 3-5 nm in diameter. Interestingly, they show an excitation-dependent emission wavelength and intensity, distinct from other QD compositions. With high-resolution XPS, it was determined that the as-prepared particles' surface contained hydroxyl, carbonyl and carboxylic acid groups, both making them water-soluble, as well as paving the way for potential conjugation to other molecules. They are also soluble in some organic solvents for use in the non-aqueous phase of devices. With this in mind, they were integrated into polymer PV cells, with a composition of ITO/PEDOT:PSS/P3HT:GQDs/Al to form a metal-insulator-metal (MIM) device, giving $\eta$ = 1.28\% after some optimization, outperforming the GQD-free assembly.

\begin{figure}[!h]
 \centering
  	\includegraphics[width=\columnwidth]{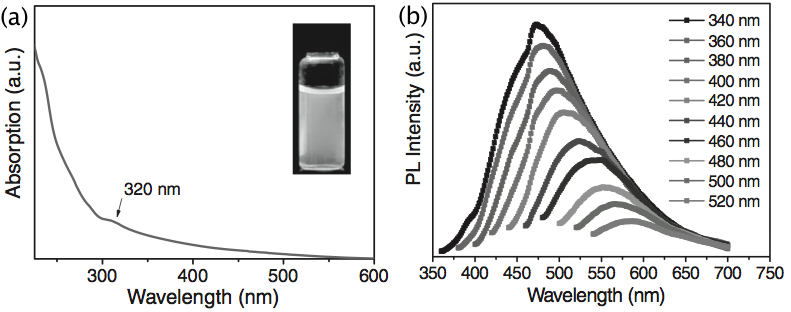}
  	\caption{\textbf{Absorption and photoluminescence of graphene quantum dots in water.} (a) Absorbance spectrum of GQDs, showing a broad UV absorption. Inset shows a vial of GQDs in water. (b) Excitation-dependent emission of GQDs. As the excitation wavelength is red-shifted, the emission is also red-shifted to a lesser extent, along with diminished intensity.}
 \label{GQDemit}
\end{figure}

Graphene sheets have also been integrated into QD nanocomposites. Such a device been developed with CdS QDs and graphene, harnessing the electrical conductivity of graphene to form an electron-transport matrix \cite{Cao10}. The fabrication process uses graphene oxide in dimethyl sulfoxide (DMSO), allowing GO reduction and CdS deposition simultaneously. They report picosecond electron transfer from the excited QDs to graphene using time-resolved fluorescence microscopy. Another promising PV device consists of layers of graphene and CdS QDs built upon an ITO glass support \cite{Guo10}, as shown in \figref{PVdiagrams}c. The effect of the number of graphene/QD bilayers on the photocurrent and voltage was investigated, and a substantial increase in the $I_{SC}$ up to 8 layers was shown, while the $V_{OC}$ held around 0.68 V for 2 or more layers. An $\eta$ value of 16\% is reported, considerably surpassing previous versions of carbon/QD solar cells ($\eta \leq 5\%$). This is attributed to the thin-layered structure being effective at charge collection and transport.


\section{Graphene Sensors}

In addition to the optoelectronic applications discussed above, important research has also been conducted on graphene sensors. After the isolation of graphene in 2004 \cite{nov04}, graphene-based nanomaterials have for the most part replaced CNTs in sensor applications \cite{geim07,DL2008}, due to higher sensitivity to adsorbed materials \cite{Pumera2009,Lilang2009}. 

\subsection{Electrochemical sensors}

\begin{figure}[H]
  	\centering
      	\includegraphics[width=8.5cm]{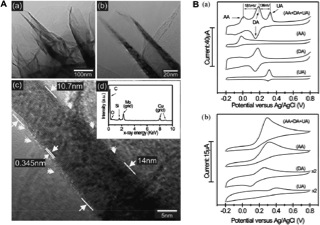}
      	\caption{\textbf{Graphene platelets (GNPs) for electrochemical sensing} (A) (a) and (b) Transmission electron microscopy (TEM) images of GNPs at different magnifications; (c) high-resolution TEM image of GNPs showing a nanoflake with a knife-edge or conical structure with open graphitic planes; (d) energy-dispersive X-ray spectrum, showing the chemical composition of GNP films. (B) (a) and (b) Cyclic voltammetry profiles of GNPs and bare GCEs, respectively, in a solution of 50 mM PBS buffer at pH 7, containing 1 mM ascorbic acid, 0.1 mM dopamine, and 0.1 mM uric acid \cite{Shang2008}.}
  	\label{fig:GNP}
\end{figure}

Graphene exhibits improved electrochemical response when compared with other electrodes such as glassy carbon electrodes \cite{Shang2008}, graphite \cite{Shan2009}, and CNTs \cite{Alwarappan2009,Wang2009}. The authors of \cite{Zhou2009} have shown that graphene exhibits a wide electrochemical potential window of ca. 2.5 V in 0.1 M phosphate-buffered saline solution (PBS) at pH 7.0. In \figref{fig:GNP}, graphene platelets (GNPs) were shown to simultaneously detect the small organic molecules dopamine, ascorbic acid, and uric acid. Other groups have explored the electrochemical properties of reduced graphene oxide (rGO) and graphene-based nanomaterials \cite{Wang2009,Zhou2009}. These studies used different probes such as nucleic acids, potassium ferricyanide, dopamine, and acetaminophen.

CNTs have shown the ability to detect gases when decorated with nanoparticles \cite{Kong2000,Sippel2005,Star2004}. Graphene based sensors also exhibit an electrical response to gaseous CO, H$_{2}$O, NO$_2$, and NH$_3$. Sensors based on graphene operate by measuring a change in resistivity resulting from the adsorption of gas molecules \cite{Schedin2007,Qazi2007,Robinson2008,Arsat2009,Fowler2009,Lu2009,Dan2009}. The authors of \cite{Robinson2008} demonstrated a molecular sensor based on rGO capable of detecting toxic gases with ppb sensitivity. In comparison to CNT based sensors, RGO based sensors show similar performance with greatly reduced noise. The detection of NO$_2$ as low as 60 ppb as been demonstrated by measuring the changes in conductivity of flakes in thin graphite \cite{Qazi2007}. Detection of organic vapors like nonanol, octanic acid, and trimethyl amine was reported in reference \cite{Dan2009}. Furthermore, the authors of reference \cite{Schedin2007} achieved single molecule detection with an electrical sensor made of few-layered graphene.

\begin{figure}[H]
  	\centering
      	\includegraphics[width=9cm]{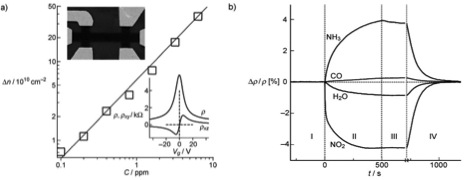}
      	\caption{\textbf{Electrical response of graphene to various gases} (a) Concentration $\Delta$n of chemically induced charge carriers in single-layer graphene exposed to different concentrations $C$ of NO$_2$. Upper inset: SEM micrograph of this device; the width of the Hall bar is 1 $\mu$m. Lower inset: the changes in resistivity $\rho$ and Hall resistivity $\rho_{xy}$ of the device with gate voltage $V_g$. The ambipolar field effect is clearly illustrated. (b) Changes in resistivity at zero magnetic field as caused by exposure of graphene to different gases diluted in a carrier gas of either He or N$_2$ (final concentration 1 ppm). Region I: device is in vacuum; II: analyte exposure; III: analyte evacuation; and IV: annealing at 1500$^{\circ}$C \cite{Schedin2007}.}
  	\label{fig:gas}
\end{figure}

\figref {fig:gas} shows the electrical response of graphene to various gases. Here the donation of electrons to graphene increases the resistance of the device, whereas scavenging of electrons decreases the overall resistance. In this case, Schedin and co-workers managed single molecule detection with pure graphene due to an unintentional functionalization of their device with a residual polymer layer from lithographic resist. The response of the sensor was measured before and after the removal of polymer resist \cite{Schedin2007}. Without the resist, the sensitivity of the device dropped by one to two orders of magnitude \cite{Dan2009}. Impurities and defects were also shown to affect the response to gas adsorption onto graphene \cite{Ao2008,Zhang2009}.

\subsection{Graphene as a biosensor}
The same mechanism allows graphene and its derivatives to be used in biosensors. The electrochemical response of hydrogen peroxide  (H$_{2}$O$_{2}$) on an rGO electrode has been studied which resulted in improved performance compared to glassy carbon, graphite and CNTs \cite{Zhou2009}. A similar effect was observed in the electrochemical behavior of the small molecule reducing agent nicotinamide adenine dinucleotide (NADH) on graphene-modified electrodes \cite{Tang2009}. In both the cases, the superior electrolytic activity of graphene is attributed to the high density of edge-plane-like defective sites on graphene. The enhanced oxidation of NADH on graphene-modified electrodes is strongly confirmed when compared with bare edge plane pyrolytic graphite electrode (EPPGE). High density of the edge-plane-like defective sites contributes to enhanced oxidation/reduction of biomolecules \cite{Banks2004}. A multilayer graphene nanoflake film electrode demonstrated highly resolved simultaneous detection of uric acid, ascorbic acid and dopamine with the detection limit as low as 0.17 $\mu$M \cite{Zhou2009}. It was reported that graphene exhibited better sensing capability of dopamine than CNTs in resolving ascorbic acid, dopamine and serotonin \cite{Alwarappan2009}. In a similar study, graphene exhibited high sensitivity to dopamine with a linear range of 5-200 $\mu$M and a better performance compared to multi-walled CNTs \cite{Wang2009}. This performance is attributed to high conductivity, large surface area and $\pi-\pi$ bond interaction between dopamine and graphene.

An electrochemical sensor based on functionalized graphene as been designed for the sensitive detection of acetominophen \cite{Kang2010}. The electrochemical behavior of acetominophen on graphene modified GCEs was studied with the help of cyclic voltammetry and square wave voltammetry. Chen \textit{et al.} designed a glucose biosensor using graphite platelets based on a composite material including GNPs, Nafion binder, and glucose oxidase \cite{Fu2009}. GO has been successfully shown to support electrical writing of the redox centers of various metalloproteins, without altering their structural integrity or biological activity \cite{Fu2009}.

\begin{figure}[h]
  	\centering
      	\includegraphics[width=8.5cm]{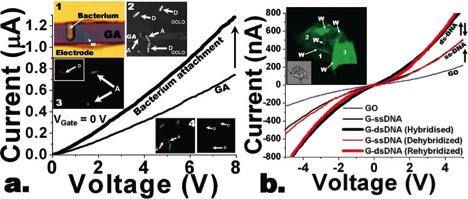}
      	\caption{\textbf{Negatively charged bacteria adsorbed on positively charged graphene amine} (a) Electrical response of graphene amine (GA) device upon attachment of a single bacterium on the surface of GA (inset 1). LIVE/DEAD confocal microscopy test on the bacteria deposited on GA confirmed that most of the bacteria were alive after the electrostatic deposition (inset 3). a) Alive and d) Dead. (b) DNA transistor: ssDNA tethering on GO increases the conductivity of the device. Successive hybridization and dehybridization of DNA on the G-DNA device results in completely reversible increase and restoration of conductivity. Inset shows a G-DNA(ds) sheet with wrinkles and folds clearly visible \cite{Mohanty2008}.}
  	\label{fig:DNA}
\end{figure}

Biosensors produced from GO and from graphene amine made by treating GO with nitrogenous plasma or ethylenediamine were developed in reference \cite{Mohanty2008}. Once treated, graphene layers were electrostatically adsorbed on to a silica substrate. Electrical measurements showed that rGO sheets acted as p-type semiconductors with high resistance and with extremely low carrier mobilities. \figref{fig:DNA} shows negatively charged bacteria adsorbed on positively charged graphene amine. This adsorption of single bacteria is accompanied by a 42$\%$ increase in conductivity. In another experiment, single-stranded DNA (ssDNA) was chemically grafted onto a graphene surface. Hybridization of the tethered strands by fluorescently-tagged complementary strands was observed. The authors further observed that the tethering of ssDNA on graphene was preferred on thicker sheets and on wrinkled surfaces of rGO rather than on flat surfaces. Initial tethering of ssDNA more than doubled the composite's conductivity, implying an increase in hole density at the graphene layer on the order of $5.61\times{10^{12}}$ cm$^{-2}$. Monolayers of ssDNA molecules were adsorbed on both sides of graphene sheets by $\pi-\pi$ stacking \cite{Patil2009}. The experimental procedure used to produce ssDNA coated graphene sheets involved vigorous sonication of a suspension of ssDNA and hydrophilized graphene.

Graphene nanomaterials have also been incorporated into functionalized biosystems integrated with DNA, peptides, proteins, aptamers, and cells. Ohno \textit{et al.} reported the use of a graphene-electrolyte field effect transistor to detect dissolved bovine serum albumin (BSA), which is an important protein for biochemical applications \cite{Ohno2009}. The sensor was made of mechanically exfoliated single layer graphene supported on a silicon dioxide wafer. An electrolyte solution and reference electrode completed the circuit. The authors observed the electrical response of their device with adsorption of BSA from standard solutions. The detection limit of BSA was as low as 0.3 nM. The same research group fabricated an aptamer-modified graphene-FET immunosensor. Functionalization of the G-FET with aptamers was confirmed by atomic force microscope \cite{Ohno2010}.

\section{Conclusion}

Graphene's journey is quite remarkable considering the tremendous scientific and technological impact this material has had on the scientific community. From the first study of graphene's band structure by Wallace in 1947 at McGill University, in the footsteps of whom the authors of this review have followed, to the pioneering experimental studies that were awarded the Nobel Prize in Physics in 2010. The former \cite{Wallace1947} led to the theoretical foundation of graphene, the unique and simple dispersion described by the Dirac Hamiltonian, while the latter work culminated with the more recent isolation of single graphene crystals in 2004.  This last discovery ignited an explosion of theoretical and experimental work on graphene and inspired the writing of this review of graphene's remarkable experimental properties with a practical emphasis. The decision was made to focus on the experimental properties, since we believe that they make for an excellent introduction on many different experimental techniques which are also relevant to other fields in condensed matter physics, materials engineering, chemistry and biophysics.

Writing a review is necessarily selective and the goal was not to be fully comprehensive, but to include the topics the authors believe to be the most interesting for the widest possible range of new researchers in this field. This naturally includes particular highlights such as the ambipolar effect for transistors, the anomalous quantum Hall effect, which brought this field to room temperature, the remarkable optical properties, while remaining a good conductor, the fascinating mechanical properties and selective sensitivity to the environment, with great outlook as sensor material. All of this would be useless without the possibility to synthesize and fabricate graphene, which is extensively reviewed here.

History has shown that predicting the future of the field and all its possible applications is a difficult and risky endeavor.  As the experience of the authors is very modest in that respect, we shall only say that a lot remains to be done and that graphene will be an important topic for many years to come.

\bibliography{References_final}
\end{document}